\documentclass[aps,twocolumn,
 superscriptaddress,amsmath,amssymb,nofootinbib]{revtex4-2}
\usepackage{amsfonts}
\usepackage{amsmath}
\usepackage{bbm}
\usepackage{amssymb}
\usepackage{graphicx}%
\usepackage{footmisc}
\usepackage{bm}
\usepackage{amsthm}

\usepackage{braket}
\usepackage{xeCJK}
\usepackage{hyperref}
\usepackage{cleveref}
\usepackage{longtable,booktabs}
\usepackage{subfigure}

\def\XXint#1#2#3{{\setbox0=\hbox{$#1{#2#3}{\int}$}
    \vcenter{\hbox{$#2#3$}}\kern-.5\wd0}}

\newcommand{\abs}[1]{\lvert#1\rvert}
\newcommand{\norm}[1]{\lVert#1\rVert}
\newcommand{\Tr}{\text{Tr}}

\begin{document}
\title{On the generic generation of steady-state entropy in isolated quantum systems}

\author{Zhiqiang Huang}
\affiliation{School of Physics, Hubei University, Wuhan 430062, China.}
\author{Qing-yu Cai}
\email{qycai@hainanu.edu.cn}
\affiliation{Center for Theoretical Physics, Hainan University, Haikou 570228, China}
\affiliation{School of Information and Communication Engineering, Hainan University, Haikou 570228, China}

\date{\today}
\begin{abstract}
	This study establishes a self-consistent mechanism for steady-state entropy generation in isolated quantum systems governed by interactions with statistically independent phases.
	Building on the nonperturbative resolvent framework developed recently, which respects the global analytic structure of many-body systems unlike conventional perturbation theory, we derive self-consistent equations for the probability distribution of eigenstate overlaps and solve them through a hierarchical ansatz strategy.
	A Lorentzian ansatz captures the bulk region, yielding coupled equations for the broadening and shift parameters and an additive logarithmic contribution \(\ln\chi\) to the entropy, on top of the extensive density-of-states term.
	A Gaussian ansatz describes the tail behavior, and a unified Lorentzian-Gaussian hybrid profile provides a refined description of the full distribution.
	Numerical simulations of a nonintegrable Ising spin chain confirm the predicted entropy scaling and validate the self-consistent equations, with the hybrid ansatz yielding entropy estimates in close agreement with observational entropy.
	Extending the analysis to the covariance (two-resolvent) sector of the same hierarchy, we relate the intra-shell intensity fluctuations---which separate the smoothed envelope entropy \(S_{\rm env}\) from the von Neumann entropy of the decohered state---to the diagonal ridge of the resolvent covariance: in the Gaussian-amplitude limit the correction approaches the symmetry-dependent Gaussian benchmark, \(S_{\rm env}-S_{\rm vN}=2-\gamma-\ln2\approx0.73\) for real Hamiltonians, with \(\gamma\) the Euler--Mascheroni constant and non-Gaussian corrections organized by the irreducible vertex hierarchy. Exact diagonalization of the nonintegrable Ising chain (\(N=12\)--\(15\) spins) reveals mildly non-Gaussian intra-shell statistics, \(q(\lambda)\approx4.8\) at \(N=15\), and a measured correction \(\Delta S=S_{\rm vN}-S_{\rm env}\approx-1.08\), consistent with the exact decomposition and markedly smaller in magnitude than the second-order estimate \(-(q-1)/2\approx-1.9\), which overshoots it by \(\sim80\%\); over \(N=12\)--\(15\) both \(q(N)\) and the correction drift monotonically toward the Gaussian benchmark---evidence of a finite-size flow whose asymptotic form the available sizes do not determine---while the correction is captured to \(4\%\)--\(9\%\) by a one-parameter Gamma closure constrained by the measured \(q\).
	Our framework thereby connects the smooth entropy envelope to the microscopic fluctuation sector of the resolvent hierarchy, providing a hierarchical route from overlap statistics to the von Neumann entropy, while leaving the microscopic closure of the irreducible covariance ridge as an open problem.
\end{abstract}

\maketitle

\section{Introduction}\label{INTRO}

We develop a self-consistent framework for steady-state entropy in isolated quantum systems with statistically independent interaction phases. While classical systems achieve equilibrium through microscopic chaos and ergodicity, the unitary evolution of quantum systems imposes strict constraints on entropy production---a contradiction that modern frameworks often address through statistical frameworks such as eigenstate thermalization hypothesis (ETH)~\cite{Deu91,Sre94,GE16,DKPR16} and related typicality arguments.
 Over the past decade, significant progress has been made in understanding how quantum many-body systems reconcile deterministic dynamics with emergent statistical behavior. 
 However, a critical gap persists: a microscopic framework for the origin and growth of entropy in the full system---applicable beyond a specific thermalization ansatz such as ETH---remains elusive, especially under strong coupling and finite-size conditions.

A sharper formulation of the question, which organizes the present work, is the following: what microscopic information is lost when a many-body overlap distribution is replaced by its smooth resolvent envelope, and how does that lost information contribute to the entropy? The answer, developed below, has a two-layer structure: the single-resolvent sector supplies the smooth envelope (and thereby the observational entropy), while the covariance sector supplies the intra-shell intensity statistics (and thereby the correction down to the von Neumann entropy); the non-Gaussian content of that statistics is carried by the higher vertices of the same hierarchy.

Recent theoretical advances have identified observational entropy---a measure quantifying information loss from coarse-grained measurements---as a key framework bridging quantum dynamics and thermodynamics~\cite{SDA19,SASD21,NKWB24}. A parallel line of research has focused on temporal coarse-graining, where averaging observables over long timescales mimics thermodynamic irreversibility. This approach, formalized in the steady-state hypothesis~\cite{S11,FMP20,IKS22}, posits that time-averaged states of closed systems exhibit thermal properties despite unitary constraints. These developments underscore the need to unify disparate entropy-generation mechanisms into a coherent predictive framework.

While these frameworks have significantly advanced thermodynamic descriptions, several open questions remain. First, the relationship between interaction strength and entropy production is still inadequately characterized, particularly near the threshold of integrability. Second, the universality of entropy growth---whether it is governed solely by the density of states or requires finely tuned interactions---remains unsettled. Moreover, the emergence of deep thermalization~\cite{IH22,MSESC24} has expanded the scope of thermalization criteria, demanding refined entropy measures.

In this work, we address these challenges by combining two complementary ingredients. The first is a general statistical condition: we consider systems where interaction matrix elements exhibit statistically independent phases that dominate systematic biases. Under this condition, complicated cross-correlated terms average out at the ensemble level, enabling a closed mean-field description, while their fluctuations encode the finer structure of the distribution---its tails and branch splitting. The second ingredient is the nonperturbative resolvent hierarchy recently developed in Ref.~\cite{HC26}. This framework provides an exact recursive reorganization of many-body expansions in resolvent space. Starting from exact projection identities, it expresses the diagonal resolvent through a self-energy that decomposes into a mean-field part and a systematically expandable hierarchy of cross-correlated contributions expressed in terms of products of diagonal resolvents. Under the random-phase condition, the mean-field part dominates, yielding self-consistent equations for the probability distribution of eigenstate overlaps; for a typical realization we assume the residual cross-correlated contribution to be parametrically smaller than the mean-field part, consistent with the random-phase scaling and with the numerical tests below. The full hierarchy further controls distribution tails, branch splitting, and spectral skewness.

The present work applies this resolvent framework to the problem of steady-state entropy. We derive self-consistent equations for the probability distribution and solve them through a hierarchical ansatz strategy~\cite{HC26}: a Lorentzian ansatz for the bulk region, a Gaussian ansatz for the tails, and a unified Lorentzian-Gaussian hybrid ansatz for a complete description. The Lorentzian ansatz yields an explicit additive logarithmic contribution \(S\supset\ln\chi\) to the entropy, where \(\chi\) is the interaction-induced broadening, on top of the extensive density-of-states term. The hybrid ansatz refines this estimate, bringing the theoretical prediction into close quantitative agreement with numerically computed observational entropy. Going one level deeper, we employ the covariance (two-resolvent) sector of the hierarchy, whose exact structure is established in Ref.~\cite{HOETH}, to relate the intra-shell fluctuations that separate the smoothed entropy from the von Neumann entropy of the decohered state to the diagonal ridge of the resolvent covariance: in the Gaussian-amplitude limit the difference approaches the symmetry-dependent Gaussian benchmark, \(S_{\rm vN}\simeq S_{\rm env}-(2-\gamma-\ln2)\approx S_{\rm env}-0.73\) for real Hamiltonians (\(\gamma\) the Euler--Mascheroni constant), while the corrections beyond Gaussianity are organized by the irreducible vertex hierarchy and quantified by the energy-resolved Porter--Thomas factor \(q(\lambda)\). Exact diagonalization of the nonintegrable Ising chain gives \(q(\lambda)\approx4.8\) and \(S_{\rm vN}-S_{\rm env}\approx-1.08\) at \(N=15\), confirming the exact decomposition and the predicted breakdown of the second-order truncation. The smoothed envelope and the microscopic fluctuations are thereby unified as the one- and two-resolvent sectors of a single entropy-generating hierarchy.

Through analytical arguments and exact-diagonalization calculations of a nonintegrable Ising spin chain, we show that steady-state entropy emerges from interaction-induced delocalization of quantum 
information across the energy spectrum~\cite{DKPR16}, combined with the temporal 
coarse-graining inherent in the construction of the decohered steady state. This study bridges critical gaps between abstract thermalization theory and practical entropy quantification. By demonstrating that systems satisfying the random-phase condition exhibit a logarithmic \(\ln\chi\) contribution to the entropy from the interaction-induced broadening, we provide a self-consistent framework for relating interaction-induced spectral spreading to entropy generation.

The remainder of this paper is organized as follows. Section~II defines the steady state of a closed composite system and the associated Shannon entropy. Section~III presents the self-consistent resolvent analysis: we summarize the framework of Ref.~\cite{HC26}, introduce the model and numerical methodology, derive the Lorentzian, Gaussian, and hybrid Lorentzian-Gaussian self-consistent equations, extract the entropy scaling, and derive the intra-shell fluctuation correction that connects the smoothed entropy to the von Neumann entropy. Section~IV discusses oscillations around the steady state. Section~V concludes with a summary and outlook. Several appendices provide detailed derivations of the branch-splitting mechanism, the cross-correlated hierarchy, the Hilbert transform of the hybrid ansatz, the effective self-energy representation, the approximate entropy contribution, and the entropy correction from the resolvent covariance hierarchy, including its exact-diagonalization verification on the interacting chain. 

\section{Steady State of a Closed System}\label{SSCS}

Consider a system \( S \) and a bath \( B \) that are initially independent, with the unperturbed Hamiltonian \( H_0 = H_S + H_B \). Assume their initial states are energy eigenstates \( \ket{\phi_{\mu i}} = \ket{\phi^{S}_{i}} \otimes \ket{\phi^{B}_{\mu}} \), satisfying \( H_0 \ket{\phi_{\mu i}} = a_{\mu i} \ket{\phi_{\mu i}} \), where \( a_{\mu i} = E_i + \epsilon_\mu \). When an interaction \( V \) is introduced, the total Hamiltonian becomes \( H = H_0 + V \), with eigenstates \( \ket{\psi_n} \) obeying \( H \ket{\psi_n} = \lambda_n \ket{\psi_n} \).

We analyze the time evolution of the initially uncorrelated state \( \ket{\phi_{\mu i}} \) under the interacting Hamiltonian \( H \). In this scenario, entanglement and correlations develop between the system and bath. The corresponding density matrix is \( \rho(t) = U(t) \ket{\phi_{\mu i}} \bra{\phi_{\mu i}} U^\dagger(t) \). Under long-time evolution (i.e., over timescales exceeding the decoherence time), the composite system-bath approaches a steady state defined by:
\begin{equation}
	\omega := \lim_{t\to \infty} \frac{1}{t} \int_0^{t} \rho(\tau)  d\tau.
\end{equation}

For non-degenerate systems, this steady state corresponds to the decohered state in the energy eigenbasis:
\begin{equation}\label{decsta}
	\omega = \sum_n \Pi_n \rho(0) \Pi_n = \sum_n \Pi_n \abs{\braket{\psi_n|\phi_{\mu i}}}^2,
\end{equation}
where \( \Pi_n = \ket{\psi_n}\bra{\psi_n} \). Letting \( \abs{\braket{\psi_n|\phi_{\mu i}}}^2 = p^{\mu i}_n \), the Shannon entropy of the decohered state becomes:
\begin{equation}
	S(\omega) = -\sum_n p^{\mu i}_n \ln p^{\mu i}_n.
\end{equation}
As a special case, when no interaction is introduced (\( V = 0 \)), the density matrix remains in a pure state \( \omega = \ket{\phi_{\mu i}} \bra{\phi_{\mu i}} \). In this scenario, its entropy remains zero with no entropy production. Although the composite system is initially in a pure state and its von Neumann 
entropy remains zero under unitary evolution, the decohered steady state 
\(\omega\), which obtained by discarding off-diagonal coherence in the energy 
eigenbasis, acquires a nonzero Shannon entropy \(S(\omega)\). This entropy 
quantifies the effective information loss induced by interaction-driven 
delocalization and the temporal coarse-graining inherent in the long-time 
limit.  Throughout this paper, ``entropy generation'' refers to this entropy
of the dephased steady state \(\omega\); the von Neumann entropy of the
instantaneous pure state remains zero under unitary evolution.

For the specific system considered in this work, the reflection symmetry of the closed chain has a mild selection effect: for a given initial state \(|\phi_{\mu i}\rangle\), the eigenstates of the full Hamiltonian whose reflection parity is definite and opposite to that of \(|\phi_{\mu i}\rangle\) have exactly vanishing overlap \(p^{\mu i}_n=0\). These parity-definite states amount to only a fraction \(\sim 1/(2N)\) of the spectrum (the remaining eigenstates belong to nearly degenerate reflection pairs, whose stored eigenvectors mix the two parities), so the support of \(p^{\mu i}_n\) is essentially the full spectrum, and no dimension correction enters the entropy formulas below; the full spectrum is used throughout.  In particular, the observational entropy is evaluated in the standard way, with the macrostate volume of each energy shell given by the full shell dimension---the total number of eigenstates within the shell (the \(d_{\mathcal M}=e^{S(\lambda)}\Delta\) of Eq.~\eqref{shellavg}), rather than any reduced effective dimension; see Appendix~\ref{fluct_entropy} for the empirical support analysis and Appendix~\ref{OMTB} for the branch structure of the tail parameters.

The resolvent-based framework for computing the probability distribution  \( p^{\mu i}_n \) is developed in Ref.~\cite{HC26}; here we summarize the essential results needed for the entropy analysis.

\section{Steady-State Entropy from the Resolvent Framework}

\subsection{Resolvent Framework (Summary of Ref.~\cite{HC26})}

We now compute the probability distribution \(p^{\mu i}_n = |\braket{\psi_n|\phi_{\mu i}}|^2\) that determines the steady-state entropy.
The analysis builds on the nonperturbative resolvent framework developed in Ref.~\cite{HC26}, whose essential elements we summarize here.

The central object is the diagonal resolvent \(\mathcal{R}_{\mu i}(z) = \bra{\phi_{\mu i}} (z - H)^{-1} \ket{\phi_{\mu i}}\).
An exact projection procedure~\cite{HC26} yields
\begin{equation}\label{CSEQ}
    \mathcal{R}_{\mu i}(z) = \frac{1}{z - a_{\mu i} - V_{\mu i} - \mathcal{G}_{\mu i}(z)},
\end{equation}
where \(V_{\mu i} = \bra{\phi_{\mu i}} V \ket{\phi_{\mu i}}\) and \(\mathcal{G}_{\mu i}(z)\) is the projected self-energy.
The self-energy decomposes exactly into a diagonal (mean-field) part and a cross-correlated part~\cite{HC26}:
\begin{equation}\label{SEdecomp}
    \mathcal{G}_{\mu i}(z) = \underbrace{\sum_{\nu j \neq \mu i} \abs{V_{\mu i,\nu j}}^2 \mathcal{R}_{\nu j}(z)}_{\mathcal{G}^{\mathrm{OD}}_{\mu i}(z)} \;+\; \mathcal{G}^{\mathrm{CC}}_{\mu i}(z).
\end{equation}

For generic nonintegrable systems, the matrix elements of local operators in the energy eigenbasis satisfy the eigenstate thermalization hypothesis (ETH)~\cite{DKPR16}:
\(\abs{\bra{\phi_{\mu}} O \ket{\phi_{\nu}}}^2 = e^{-S(\epsilon^+_{\mu\nu})} f^2(\epsilon^+_{\mu\nu}, \delta) \abs{R_{\mu\nu}}^2\),
where the random variables \(R_{\mu\nu}\) have zero mean and unit variance and are statistically independent for distinct index triples.
Consequently \(\mathbb{E}\bigl[\mathcal{G}^{\mathrm{CC}}_{\mu i}(z)\bigr] \approx 0\), yielding the mean-field approximation
\begin{equation}\label{GRE}
    \mathcal{G}_{\mu i}(z) \approx \sum_{\nu j \neq \mu i} \abs{V_{\mu i,\nu j}}^2 \mathcal{R}_{\nu j}(z),
\end{equation}
which constitutes the leading-order closure of the full resolvent hierarchy.
As shown in Ref.~\cite{HC26}, the cross-correlated terms vanish on average but are not identically zero: they admit an exact recursive expansion
\(\mathcal{G}^{\mathrm{CC}}_{\mu i} = \sum_{\ell \ge 3} \mathcal{G}^{(\ell)}_{\mu i}\)
in terms of products of diagonal resolvents. The leading term \(\mathcal{G}^{(3)}_{\mu i} \propto \sum V^{(3)} \mathcal{R}_{\nu j} \mathcal{R}_{\xi k}\)
controls spectral skewness and distribution tails---effects that lie beyond mean-field but will be referenced below.

The connection to probabilities follows from \(\mathcal{R}_{\mu i}(z) = \sum_n p^{\mu i}_n / (z - \lambda_n)\) and the Sokhotski-Plemelj identity~\cite{HC26}:
\begin{equation}\label{PAG}
    p^{\mu i}_n = \frac{1}{\pi e^{S(\lambda_n)}} \frac{\Im\, \mathcal{G}_{\mu i}(\lambda_n - \mathrm{i}0^+)}{[\Delta^{\mu i}_n - \Re\, \mathcal{G}_{\mu i}(\lambda_n)]^2 + [\Im\, \mathcal{G}_{\mu i}(\lambda_n - \mathrm{i}0^+)]^2},
\end{equation}
with \(\Delta^{\mu i}_n = \lambda_n - a_{\mu i} - V_{\mu i}\).
Under Eq.~\eqref{GRE} the real and imaginary parts are~\cite{HC26}
\begin{align}
    \frac{1}{\pi} \Im\, \mathcal{G}_{\mu i}(\lambda_n - \mathrm{i}0^+) &= \sum_{\nu j \neq \mu i} \abs{V_{\mu i,\nu j}}^2 \, p^{\nu j}_n e^{S(\lambda_n)}, \label{IMGAP}\\
    \Re\, \mathcal{G}_{\mu i}(\lambda_n) &= \sum_{\nu j \neq \mu i} \abs{V_{\mu i,\nu j}}^2 \sum_{m \neq n} \frac{p^{\nu j}_m}{\lambda_n - \lambda_m}. \label{REGAP}
\end{align}
Equations \eqref{PAG}--\eqref{REGAP} form a closed self-consistent system.

\subsection{Model and Numerical Methodology}

To validate the framework we employ a nonintegrable Ising spin chain with \(N=15\) spins:
\begin{equation}
    H = \sum_{k=1}^{N}\left(-\sigma_z^k \otimes \sigma_z^{k+1} + g \sigma_x^k + h \sigma_z^k\right),
\end{equation}
with \(g = 1.05\) and \(h = 0.1\). The composite system is constructed as:
\begin{itemize}
    \item System \(S\): a single spin, \(H_S = g \sigma_x^S + h \sigma_z^S\);
    \item Bath \(B\): the remaining \(14\) spins;
    \item Interaction: \(V = -\sigma_z^S \otimes (\sigma_z^1 + \sigma_z^{14})\).
\end{itemize}
The system energies satisfy \(E_i \approx (-1)^i 1.055\), distinguishing two energy branches.

By diagonalizing \(H_0\) and \(H\) we obtain the eigenstates and compute
\(p^{\mu i}_n = |\braket{\psi_n|\phi_{\mu i}}|^2\).
The distribution is smoothed by energy-shell averaging over intervals \(\mathcal{M}_{\lambda,\Delta} = [\lambda - \Delta/2, \lambda + \Delta/2]\) of width \(\Delta = 0.5\):
\begin{equation}\label{shellavg}
    p^{\mu i}(\lambda) = \mathbb{E}(p^{\mu i}_n) := \frac{1}{d_{\mathcal{M}}} \sum_{\lambda_m \in \mathcal{M}_{\lambda,\Delta}} p^{\mu i}_m,
\end{equation}
where \(d_{\mathcal{M}} = e^{S(\lambda)} \Delta\) and the density of states follows the binomial form~\cite{DLL18}
\(e^{S(\lambda)} = \kappa N! / [(N/2 - \kappa\lambda)! (N/2 + \kappa\lambda)!]\), with \(\kappa = \frac12(g^2 + h^2 + 1)^{-1/2} \approx 0.344\).
Under the Gaussian approximation \(e^{S(\lambda)} \approx 2^N G(\lambda; \sqrt{N}/(2\kappa))\).
\footnote{The limits used in this paper are ordered as follows: (i) \(t\to\infty\) first, defining the decohered steady state \(\omega\); (ii) the resolvent boundary value \(\eta\to0^+\) defining the spectral representation; (iii) at fixed \(N\), the shell coarse-graining is chosen with the mean level spacing much smaller than the shell width, which is in turn much smaller than the scale of macroscopic spectral variation (here \(\Delta=0.5\)); this defines the shell-averaged objects of Sec.~\ref{IntraShell} (\(p^{\rm env}\), \(q(\lambda)\), and \(\mathcal H_x\)); (iv) the large-\(N\) limit is discussed last, and only conditionally on the extensivity and regularity of the intensity statistics.}

\subsection{Mean-Field Lorentzian Ansatz and Self-Consistent Equations}

\begin{figure*}
	\centering
	\subfigure[]{\includegraphics[width=0.48\textwidth]{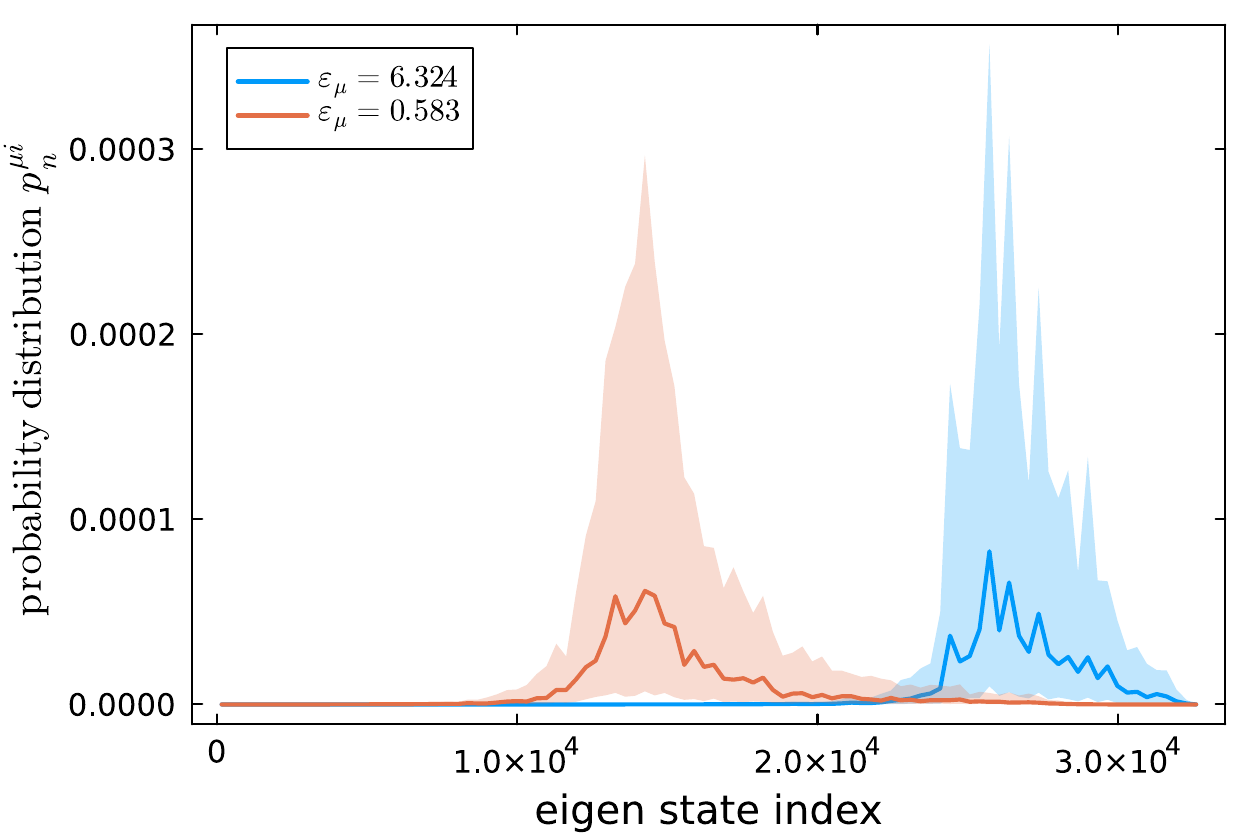}}
	\subfigure[]{\includegraphics[width=0.48\textwidth]{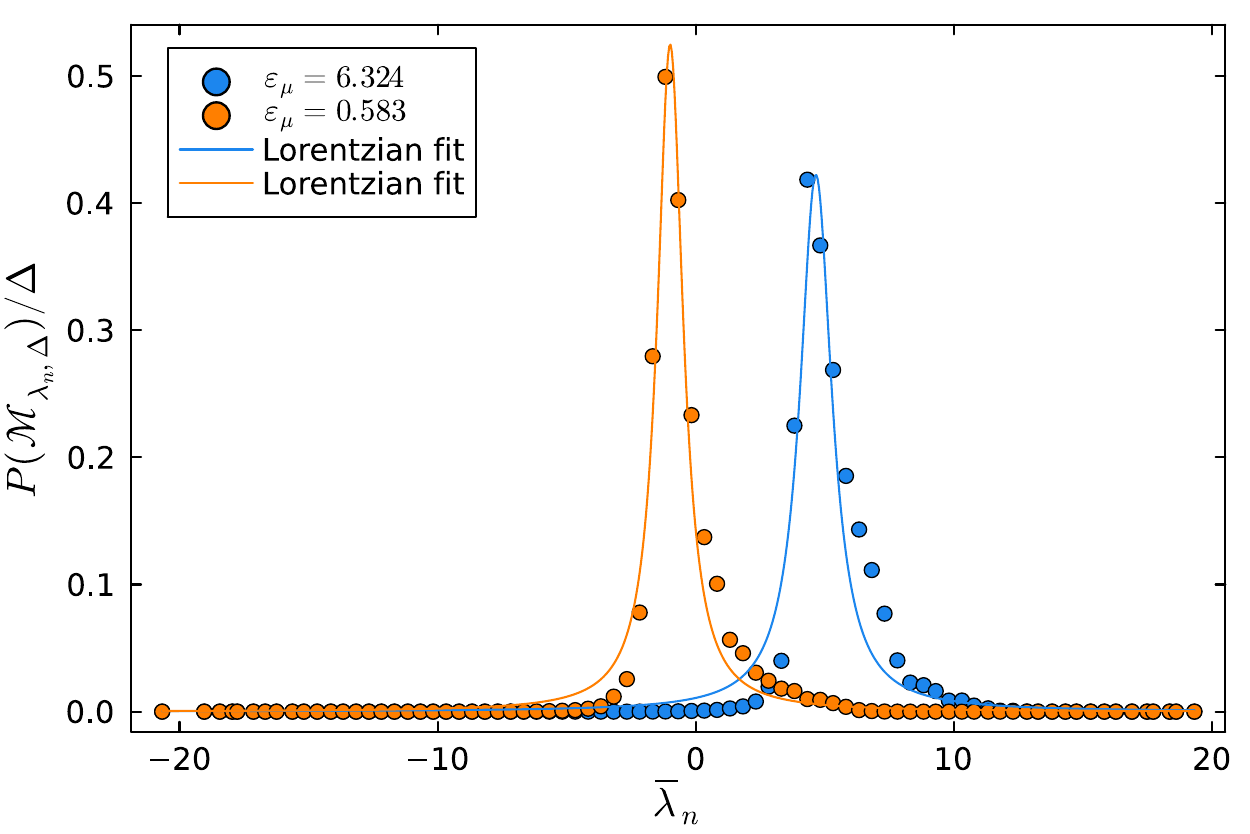}}
	\caption{
		(a) Probability distribution $p^{\mu i}_n$ ($i=1$) as a function of the eigenstate index $n$ ordered by energy, for the composite system. To suppress strong fluctuations, the data are energy-binned. The median value in each bin is shown as a solid line, while the shaded band represents the interquartile range ($25\%-75\%$), characterizing typical fluctuations within a narrow energy window. (b) Binned distribution $P(\mathcal{M}_{\lambda,\Delta})$ constructed by summing probabilities within energy intervals $\mathcal{M}_{\lambda,\Delta} $, where $\Delta = 0.5$, following Eq.~\eqref{Lorentz}. The binned probability is expressed as $P(\mathcal{M}_{\lambda,\Delta})  = \int_{\lambda-\Delta/2}^{\lambda+\Delta/2} d\lambda_m e^{S(\lambda_m)} p^{\mu i}(\lambda_m)$, with $\overline{\lambda}_n$ representing the average eigenenergy within each interval. Solid curves show Lorentzian fits to the binned distributions. Fitting parameters: For $\epsilon_\mu=6.324$, we obtain $a_{\mu i}+\Delta_{\mu i}=4.661$ and $\chi_{\mu i}=0.753$; for $\epsilon_\mu=0.583$, the parameters are $a_{\mu i}+\Delta_{\mu i}=-0.986$ and $\chi_{\mu i}=0.606$.  }
	\label{FIGs1}
\end{figure*}

\begin{figure*}[t]
    \centering
    \subfigure[Width parameter \(\chi_{\mu i}\)]{\includegraphics[width=0.48\textwidth]{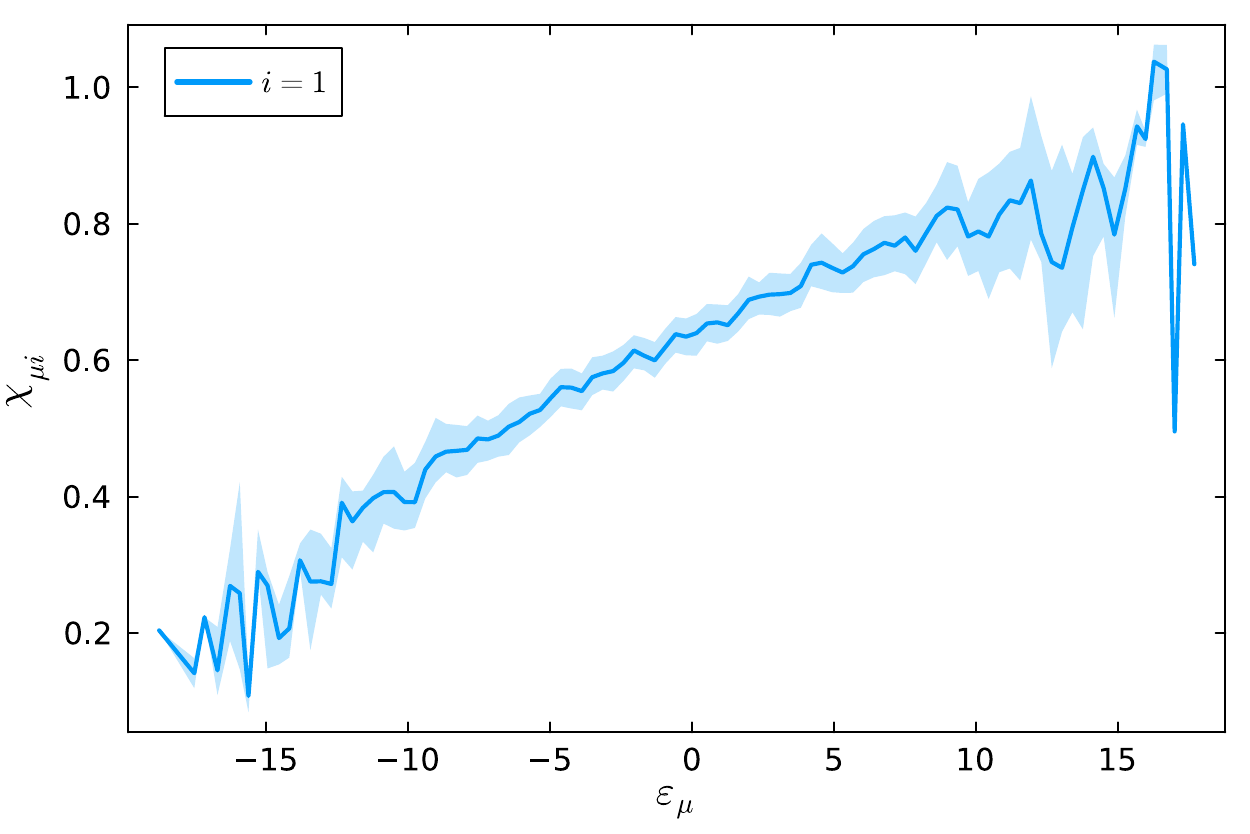}}
    \subfigure[Energy shift \(a_{\mu i} + \Delta_{\mu i}\)]{\includegraphics[width=0.48\textwidth]{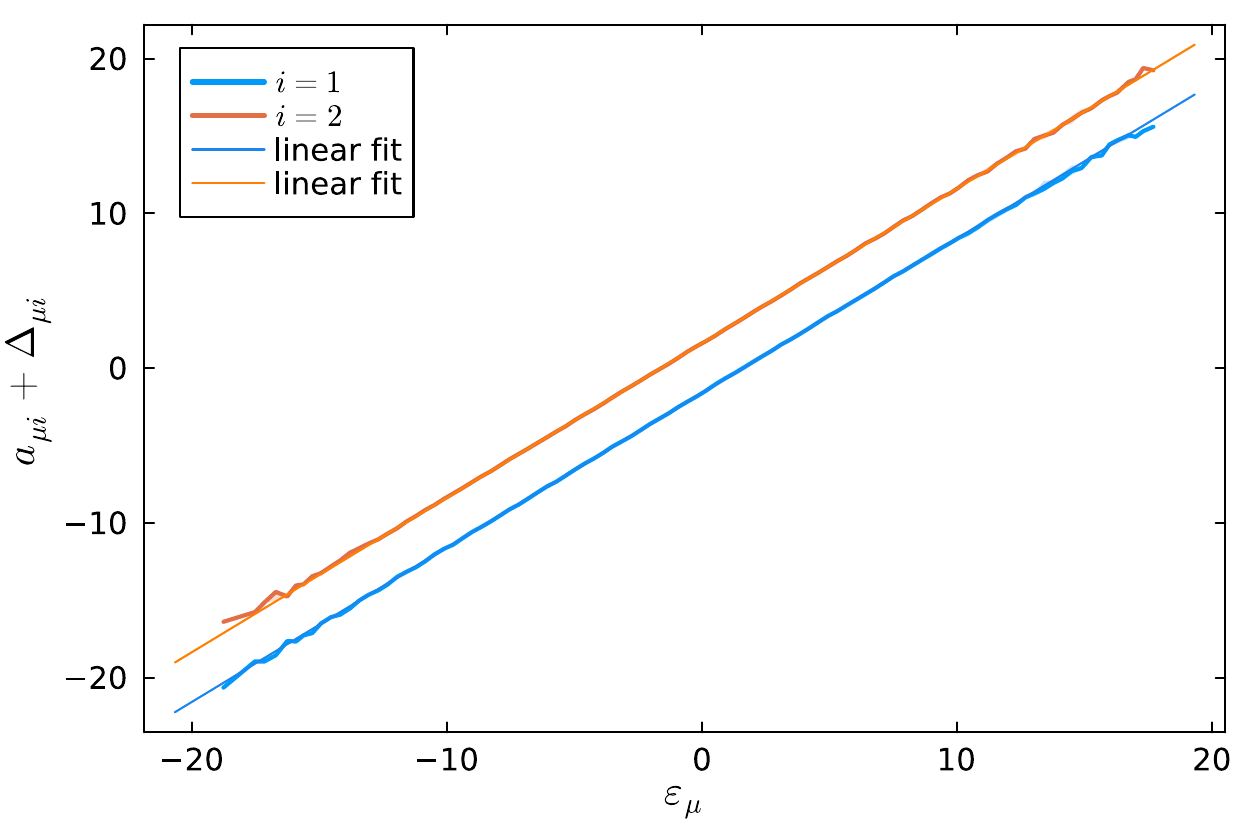}}
    \caption{(a) Lorentzian width \(\chi_{\mu i}\) versus bath energy \(\epsilon_\mu\). (b) Shifted energies \(a_{\mu 1} + \Delta_{\mu 1}\) versus \(\epsilon_\mu\).
    Linear regression: \(\braket{a_{\mu 1} + \Delta_{\mu 1}} = 0.998\epsilon_\mu - 1.610\), \(\braket{a_{\mu 2} + \Delta_{\mu 2}} = 0.998\epsilon_\mu + 1.608\).}
    \label{FIGs2}
\end{figure*}

Guided by Eq.~\eqref{PAG} and supported by numerical evidence (Fig.~\ref{FIGs1}), the minimal Lorentzian ansatz captures the bulk region~\cite{HC26}:
\begin{equation}\label{Lorentz}
    p^{\mu i}(\lambda) = \frac{1}{\pi e^{S(\lambda)}} \Im\!\left(\frac{1}{\delta \lambda_{\mu i} - \mathrm{i} \chi_{\mu i}}\right)
    = \frac{1}{e^{S(\lambda)}} L^{\mu i}(\lambda),
\end{equation}
where \(\delta \lambda_{\mu i} := \lambda - a_{\mu i} - \Delta_{\mu i}\) and \(L(x;\chi) = \chi/[\pi(\chi^2 + x^2)]\).
Inserting Eq.~\eqref{Lorentz} into Eqs.~\eqref{PAG}--\eqref{REGAP} with the decoupling \(r(\lambda)=1\)~(\cite{HC26}) yields
\begin{align}
    \chi_{\mu i}               &= \sum_{\nu j \neq \mu i} \abs{V_{\mu i,\nu j}}^2 \frac{\chi_{\nu j}}{\delta \lambda_{\nu j}^2 + \chi_{\nu j}^2}, \label{consist}\\
    \Delta_{\mu i} - V_{\mu i} &= \sum_{\nu j \neq \mu i} \abs{V_{\mu i,\nu j}}^2 \frac{\delta \lambda_{\nu j}}{\delta \lambda_{\nu j}^2 + \chi_{\nu j}^2}. \label{consist2}
\end{align}
Fitting the Lorentzian to all states \(\phi_{\mu i}\) yields the parameter evolution shown in Fig.~\ref{FIGs2}.
Linear regression gives
\begin{equation}\label{deltafit}
    \Delta_{\mu 1} \approx -0.002\epsilon_\mu - 0.555, \qquad
    \Delta_{\mu 2} \approx -0.002\epsilon_\mu + 0.553.
\end{equation}

\subsection{Interaction Characterization and Simplified Self-Consistent Equations}

\begin{figure}[htbp]
    \centering
    \includegraphics[width=0.48\textwidth]{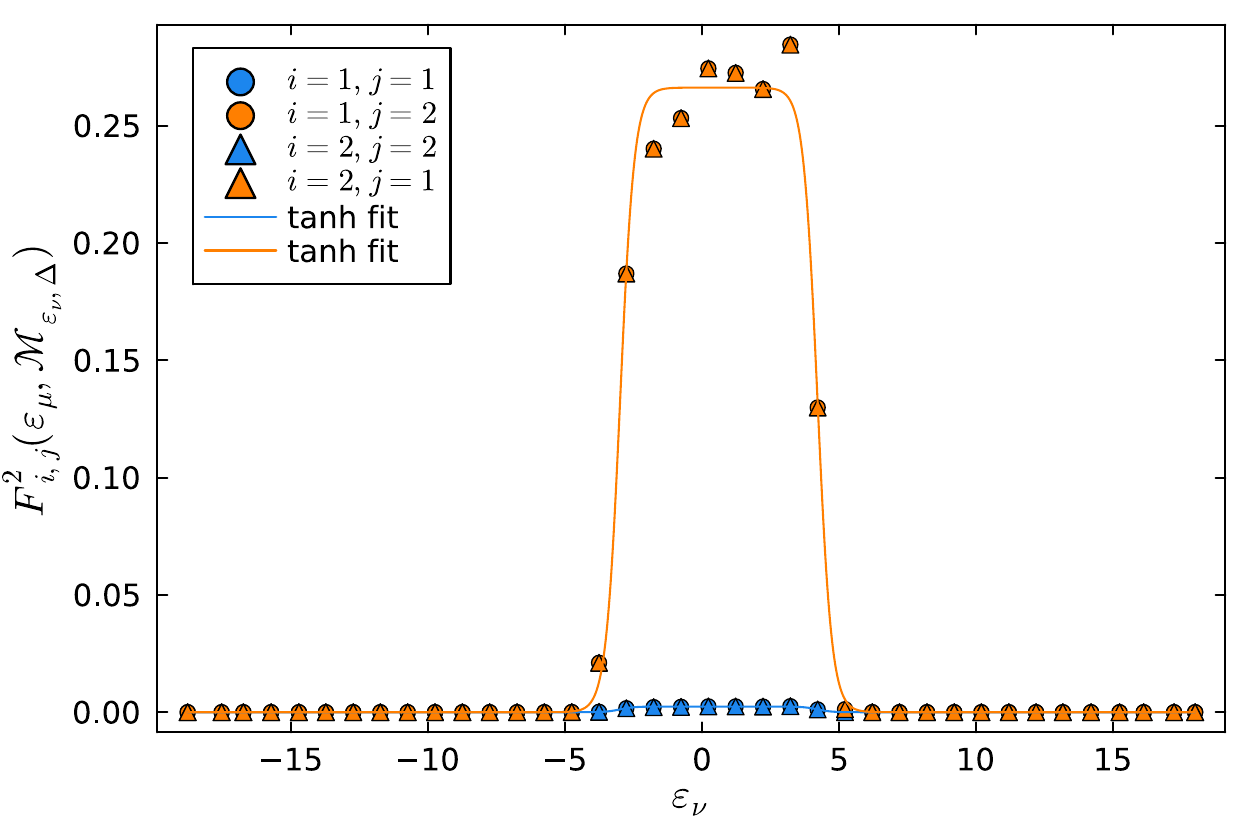}
    \caption{Interaction strength \(F^2_{i,j}(\epsilon_\mu, \mathcal{M}_{\epsilon_\nu,\Delta})\) with \(\epsilon_\mu = 0.583\), \(\Delta = 1.0\).
    Fits: \(F^2_{1,1} = F^2_{2,2} = 0.001 \times \{\tanh[1.930(x+2.985)] + \tanh[1.930(4.216-x)]\}\),
    \(F^2_{1,2} = F^2_{2,1} = 0.133 \times \{\tanh[1.930(x+2.985)] + \tanh[1.930(4.216-x)]\}\).}
    \label{FIG4}
\end{figure}

\begin{figure*}[htbp]
    \centering
    \subfigure[]{\includegraphics[width=0.48\textwidth]{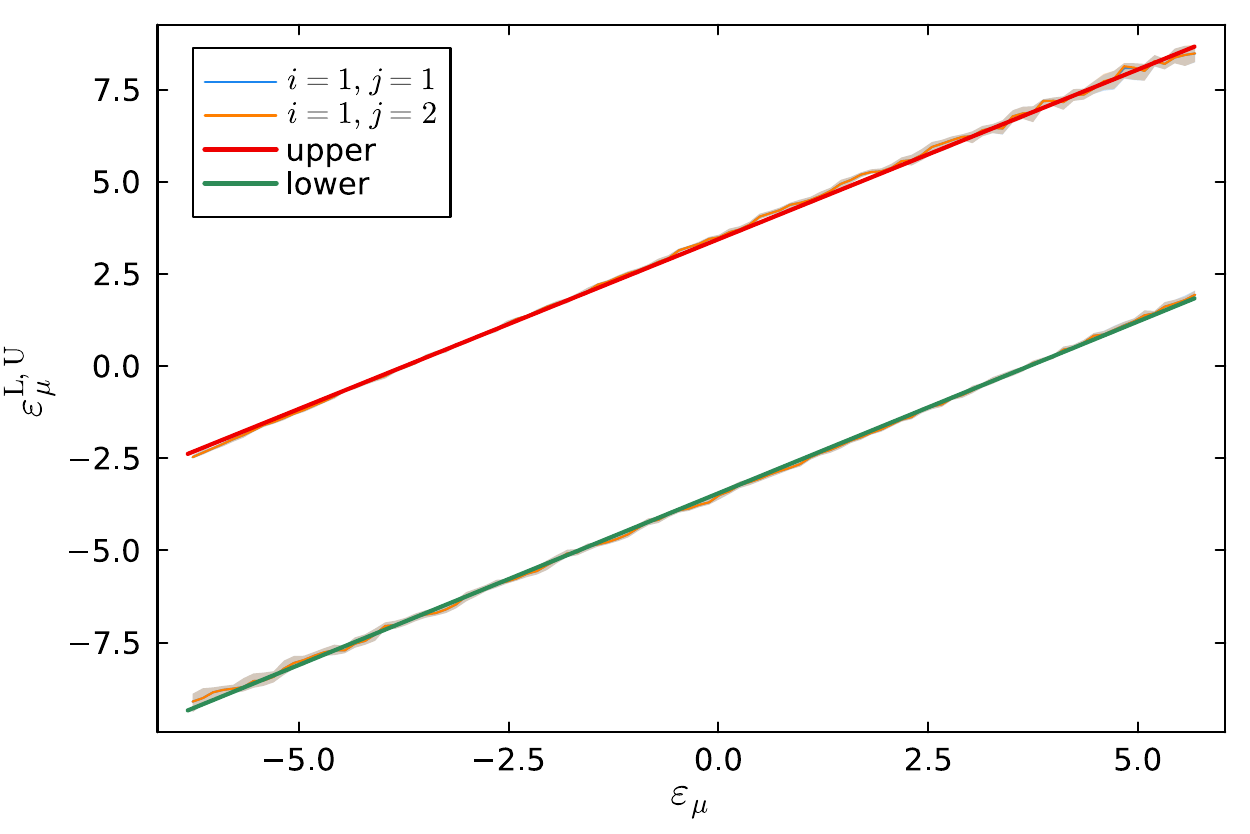}}
    \subfigure[]{\includegraphics[width=0.48\textwidth]{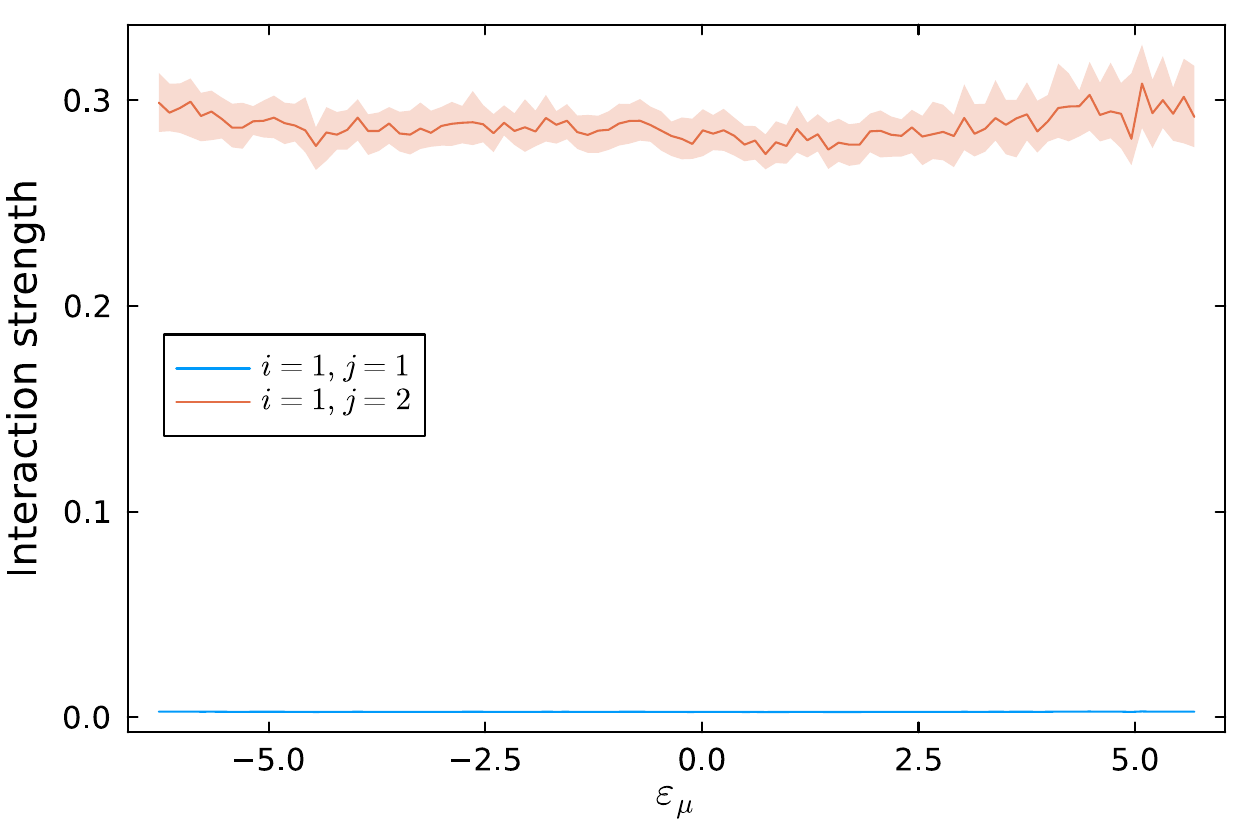}}
    \caption{Fitted interaction parameters. (a) Bounds \((\epsilon^{\mathrm{L}},\epsilon^{\mathrm{U}})\) with
    \(\braket{\epsilon^{\mathrm{U}}_{\mu}} = 0.921\epsilon_\mu + 3.440\), \(\braket{\epsilon^{\mathrm{L}}_{\mu}} = 0.931\epsilon_\mu - 3.450\).
    (b) Interaction strength \(\mathcal{F}\): \(\overline{\mathcal{F}} \approx 0.288\) for \(i \neq j\), \(\overline{\mathcal{F}} \approx 0.003\) for \(i = j\).}
    \label{interpara}
\end{figure*}

To make quantitative contact, we characterize the interaction strength.
Substituting the ETH form into Eq.~\eqref{consist} gives
\begin{equation}\label{intsmoth}
    \chi_{\mu i} = \sum_j \int d\epsilon_{\nu} \, f^2_{i,j}(\epsilon_\mu,\delta) \frac{\chi_{\nu j}}{\delta \lambda_{\nu j}^2 + \chi_{\nu j}^2}.
\end{equation}
Numerical characterization (Figs.~\ref{FIG4},~\ref{interpara}) reveals square-wave-like coupling profiles,
motivating the separation
\begin{equation}\label{cmisep}
    \chi_{\mu i} \approx \sum_j \mathcal{V}_{\mu i,j} \, \mathbb{E}_\delta\!\left( \frac{\chi_{\nu j}}{\delta \lambda_{\nu j}^2 + \chi_{\nu j}^2} \right),
\end{equation}
where \(\mathcal{V}_{\mu i,j} = \sum_\nu |V_{\mu i,\nu j}|^2 \approx \overline{\mathcal{F}}_{i,j}(\epsilon_\mu^{\mathrm{U}} - \epsilon_\mu^{\mathrm{L}})\) and
\(\mathbb{E}_\delta\) denotes averaging over the energy window \([\epsilon_\mu^{\mathrm{L}}, \epsilon_\mu^{\mathrm{U}}]\).
The dominant cross-channel coupling \(\mathcal{V}_{\mu 1,2} \approx 1.982\) overwhelms the diagonal terms
\(\mathcal{V}_{\mu i,i} \approx 0.018\), which can be neglected.

\begin{figure*}[t]
    \centering
    \subfigure[Width \(\chi_{\mu i}\)]{\includegraphics[width=0.48\textwidth]{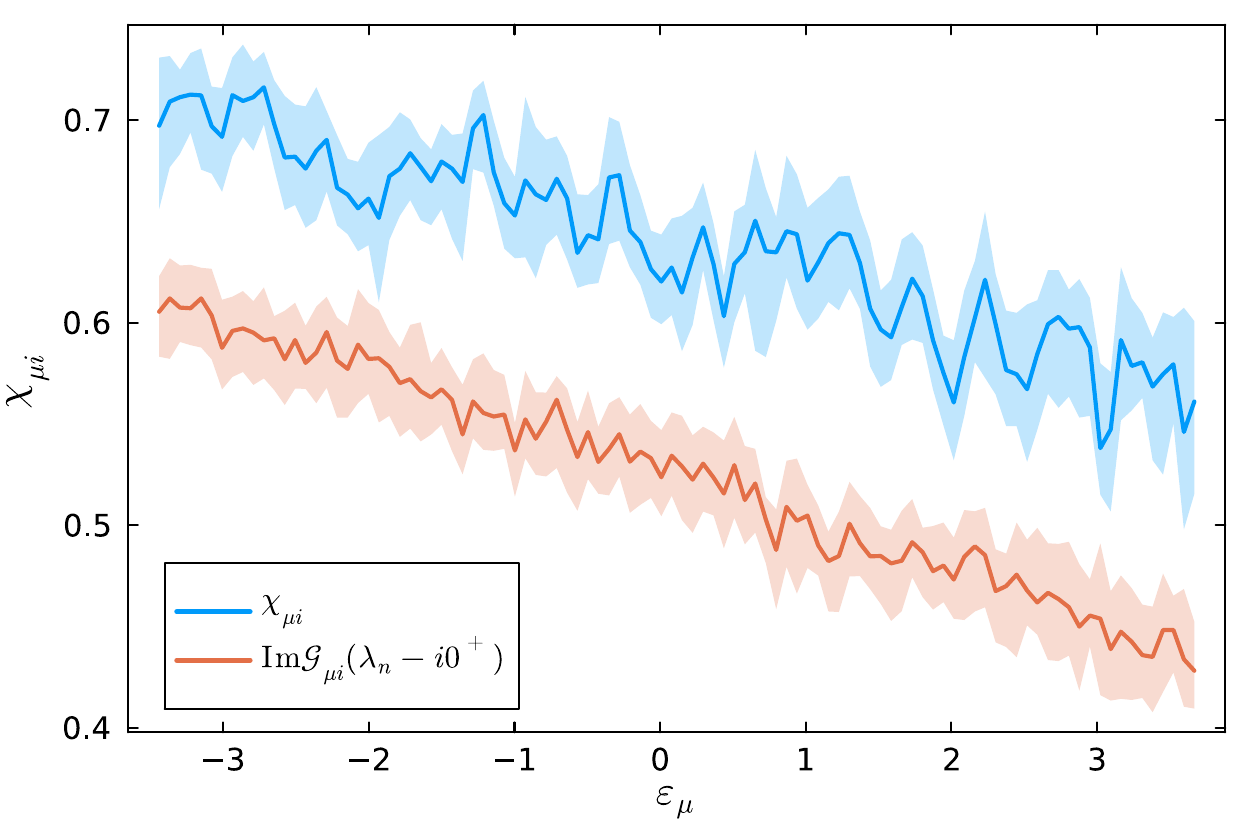}}
    \subfigure[Shift \(\Delta_{\mu i}\)]{\includegraphics[width=0.48\textwidth]{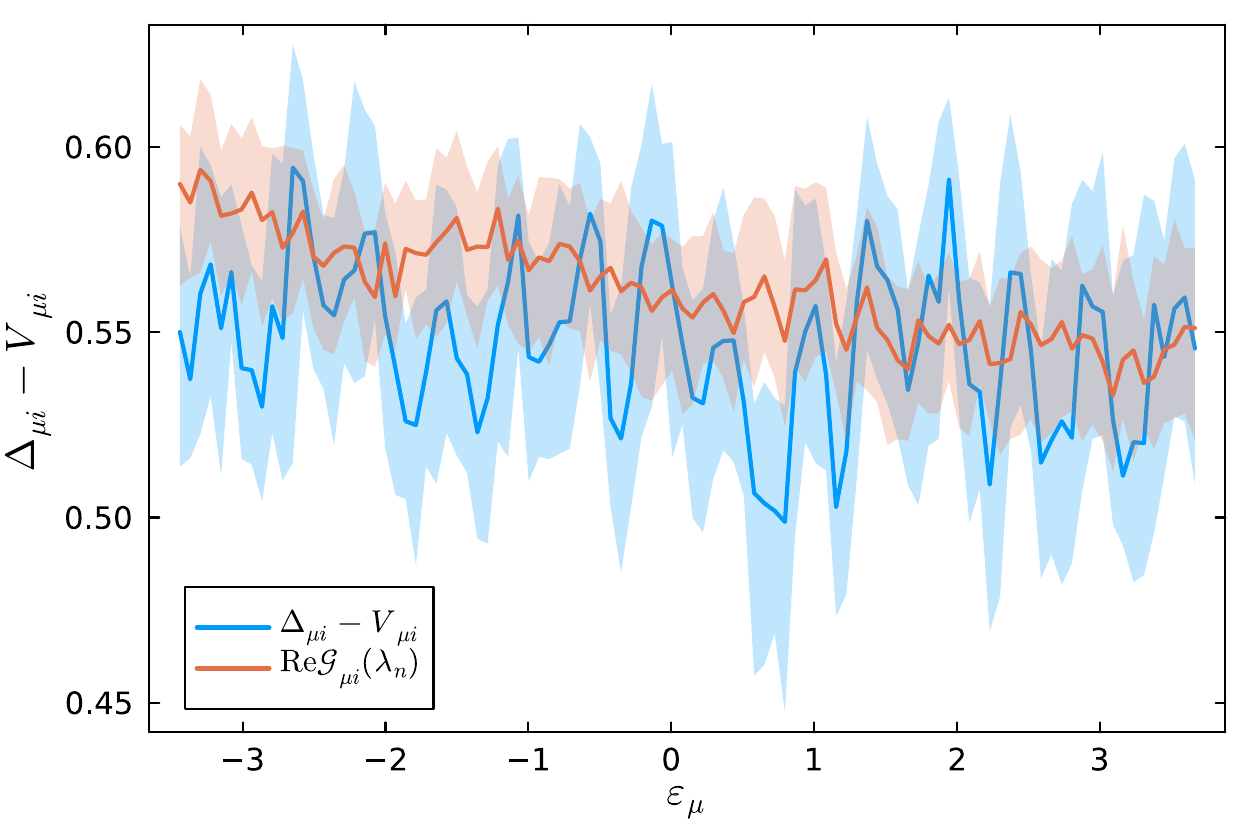}}
    \caption{Self-consistency validation for \(i=2\). Blue: directly fitted values; Orange: reconstructed via
    Eqs.~\eqref{consist},~\eqref{consist2} with parameters from Eq.~\eqref{etap}.}
    \label{FIG3}
\end{figure*}

Coarse-graining the parameters as \(\overline{\chi}_{\mu i} = \overline{\chi}_i + A_i \epsilon_\mu\) and treating the shift
\(\overline{\Delta}_i\) as approximately constant (Figs.~\ref{FIGs2},~\ref{FIG3}),
the self-consistent equations reduce to
\begin{equation}\label{simpcon}
    \overline{\Delta}_2 + \mathrm{i} \overline{\chi}_{\mu 2} = \mathcal{V}_{\mu 2,1} \, \mathbb{E}_\delta\!\left( \frac{1}{(\delta - \eta_1) - \mathrm{i}(\overline{\chi}_{\mu 1} - A_1\delta)} \right),
\end{equation}
where \(\eta_1 = \epsilon_\mu + \overline{\Delta}_1 + E_1 - \lambda\). Assuming \(\eta_i\) is constant and evaluating at \(\epsilon_\mu = 0\)
together with the slope-matching condition \(\partial_{\epsilon_\mu}(\overline{\Delta}_i + \mathrm{i}\overline{\chi}_{\mu i}) = \mathrm{i} A_i\)
gives eight real equations for \((\eta_i, \overline{\chi}_i, A_i, \overline{\Delta}_i)\).
Solving with the bounds from Fig.~\ref{interpara} yields
\begin{align}\label{etap}
    (\eta_1,\overline{\chi}_{1},A_1,\overline{\Delta}_1) &= (-3.2353, 0.5306, 0.0344, -0.6733), \notag\\
    (\eta_2,\overline{\chi}_{2},A_2,\overline{\Delta}_2) &= (3.1749, 0.4988, -0.0363, 0.6749).
\end{align}

Figure~\ref{FIG3} confirms remarkable agreement between direct parameter extraction and self-consistent solutions.
The residual discrepancy stems from the fact that the pure Lorentzian ansatz is only the leading order; refined ans\"atze introduced below substantially reduce this error.

\subsection{Entropy Scaling}

The entropy of the smoothed Lorentzian distribution evaluates to
\begin{align}\label{aproent}
    S(p^{\mu i}) &= -\int d\lambda \, e^{S(\lambda)} \, p^{\mu i}(\lambda) \ln p^{\mu i}(\lambda) \notag \\
    &= \int d\lambda \, \frac{S(\lambda)}{\pi} \frac{\chi_{\mu i}}{\delta \lambda_{\mu i}^2 + \chi_{\mu i}^2} + \ln(4\pi \chi_{\mu i}) \notag \\
    &\approx S(a_{\mu i} + \Delta_{\mu i}) + \ln(4\pi \chi_{\mu i}).
\end{align}
Using the Gaussian density of states, this becomes
\begin{align}\label{finalentro}
    S(p^{\mu i}) \approx N\ln 2 - \frac{2\kappa^2(a_{\mu i} + \Delta_{\mu i})^2}{N} \notag\\
    + \ln(4\pi \chi_{\mu i}) - \frac12\ln\!\left(\frac{\pi N}{2\kappa^2}\right).
\end{align}

\begin{figure}[htbp]
    \centering
    \includegraphics[width=0.48\textwidth]{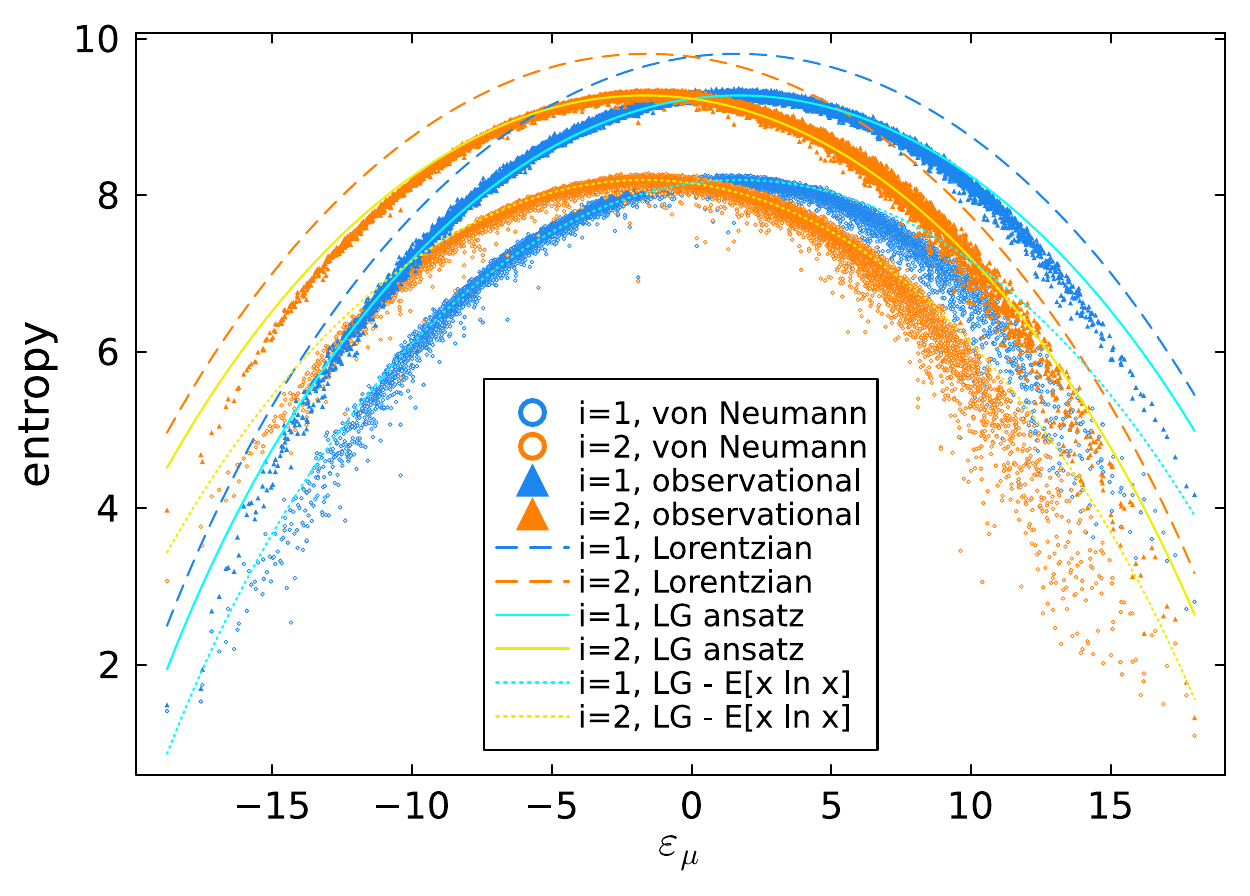}
    \caption{Entropy comparison. The observational entropy is evaluated for the
    energy-shell POVM of width \(\Delta = 0.5\), with each macrostate volume
    given by the full shell dimension---the total number of eigenstates within
    the shell (standard definition). The theoretical estimate \eqref{finalentro} (using \(\overline{\chi} \approx 0.632\))
    provides a rough estimate of this observational entropy.
    The refined LG estimate from Eq.~\eqref{LGentro} (using branch A parameters from Fig.~\ref{FIGlgp}
    and \(H(\Delta - \Delta',\sigma,\chi) \approx 1.54\)) yields improved agreement.
    The dotted curves show the fluctuation-corrected prediction \(S_{\rm vN}\simeq S_{\rm env}-\mathbb E[x\ln x]\) with the bulk-median \(\mathbb E[x\ln x]\approx1.08\) (Sec.~\ref{IntraShell}, Appendix~\ref{fluct_entropy}), which tracks the von Neumann entropy.
    The von Neumann entropy is systematically lower due to intra-shell quantum fluctuations; the measured correction \(S_{\rm vN}-S_{\rm env}\approx-1.08\) and the Gaussian-limit benchmark \(-0.73\) are derived in Sec.~\ref{IntraShell}.}
    \label{FIG5}
\end{figure}

As shown in Fig.~\ref{FIG5}, Eq.~\eqref{finalentro} aligns well with the observational entropy~\cite{NKWB24} (evaluated with the full shell dimension, Sec.~\ref{SSCS}).
The remaining overestimate of the Lorentzian-based entropy relative to the observational entropy reflects the excessive weight assigned to the broad tails, a deficiency remedied by the hybrid ansatz below.
The systematic difference from the von Neumann entropy arises from intra-shell
fluctuations; the covariance sector of the hierarchy~\cite{HOETH} converts them
into an exact additive correction whose Gaussian-limit benchmark is
\(-(2-\gamma-\ln2)\approx-0.73\), while the measured value on this model is
\(\approx-1.08\) with \(q(\lambda)\approx4.8\) (Sec.~\ref{IntraShell},
Appendix~\ref{fluct_entropy}).

\subsection{Gaussian Tail Ansatz}

The Lorentzian ansatz describes the bulk well but becomes inaccurate in the tails where
\(f^2_{i,j}(\epsilon_\mu,\delta)\) decays rapidly with \(|\delta|\), causing the actual distribution to fall faster
than the \(1/x^2\) Lorentzian form.
This motivates the Gaussian ansatz~\cite{HC26}:
\begin{equation}\label{normal}
    p^{\mu i}(\lambda) = \frac{1}{e^{S(\lambda)} \sigma_{\mu i}} \varphi\!\left(\frac{\delta \lambda'_{\mu i}}{\sigma_{\mu i}}\right),
\end{equation}
where \(\varphi(x) = e^{-x^2/2}/\sqrt{2\pi}\) and \(\delta \lambda'_{\mu i} := \lambda - a_{\mu i} - \Delta'_{\mu i}\).
Fitting all states (Fig.~\ref{FIGgp}) gives the linear regression
\begin{align}\label{deltapfit}
    a_{\mu 1} + \Delta'_{\mu 1} = 0.831\epsilon_\mu - 0.190,\notag\\ 
    a_{\mu 2} + \Delta'_{\mu 2} = 0.825\epsilon_\mu + 0.041.
\end{align}

\begin{figure}
    \centering
    \includegraphics[width=0.48\textwidth]{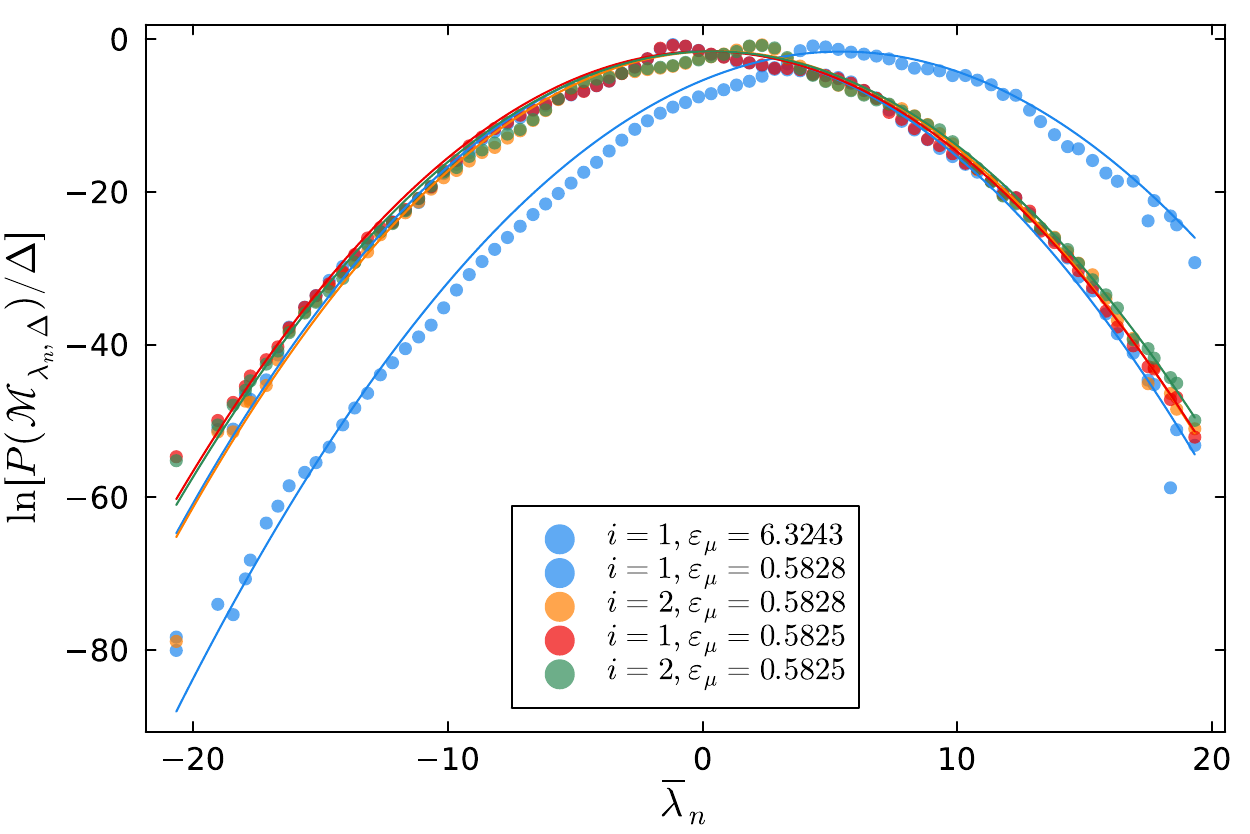}
    \caption{Binned distribution \(P(\mathcal{M}_{\lambda,\Delta})\) under the Gaussian ansatz (\(\Delta = 0.5\)).
    Solid curves: Gaussian fits. Fitted parameters:
    \((\epsilon_\mu; a_{\mu i}+\Delta'_{\mu i}, \sigma_{\mu i}) = (6.3243; 5.451, 1.986)\) for \(i=1\);
    \((0.5828; 0.220, 1.858)\) and \((0.5828; 0.529, 1.879)\);
    \((0.5825; 0.147, 1.922)\) and \((0.5825; 0.398, 1.933)\).}
    \label{FIGgf}
\end{figure}

\begin{figure*}[htbp]
    \centering
    \subfigure[Standard deviation \(\sigma_{\mu i}\)]{\includegraphics[width=0.48\textwidth]{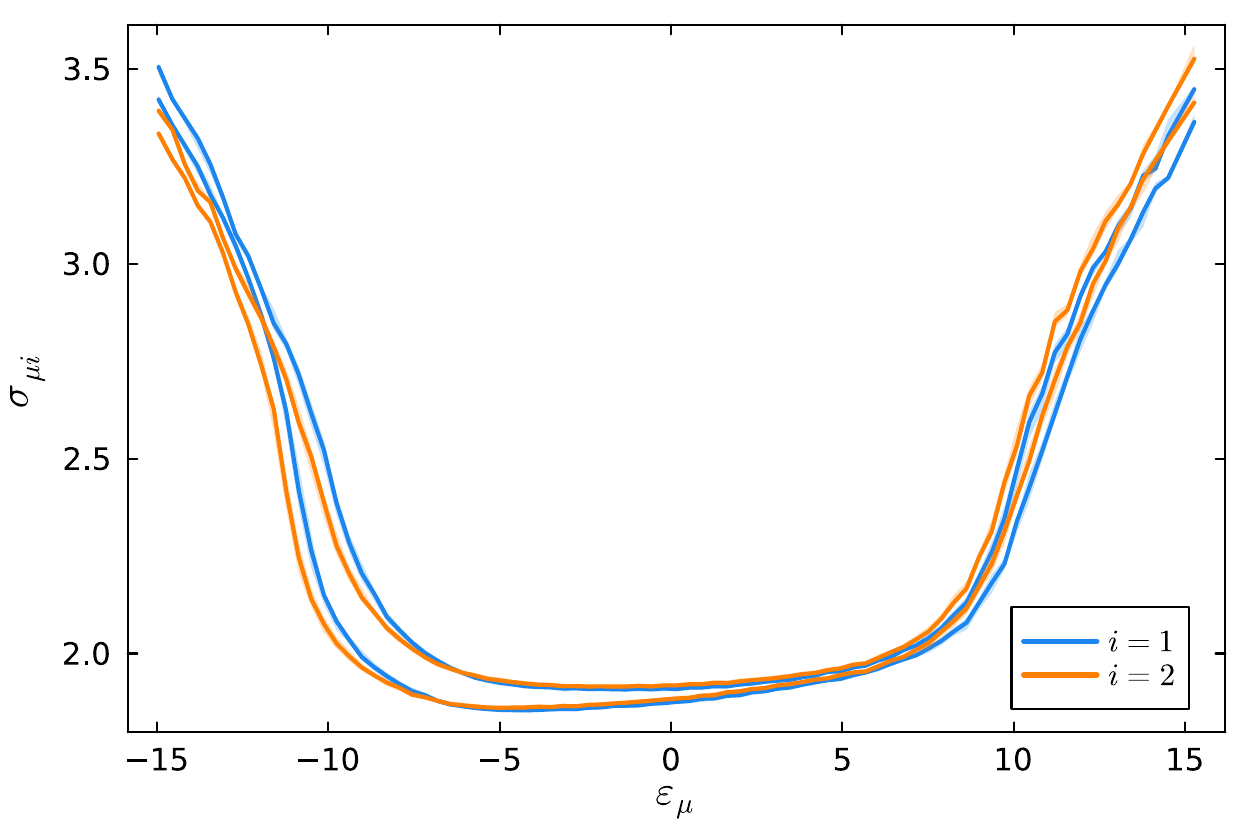}}
    \subfigure[Shift \(a_{\mu i} + \Delta'_{\mu i}\)]{\includegraphics[width=0.48\textwidth]{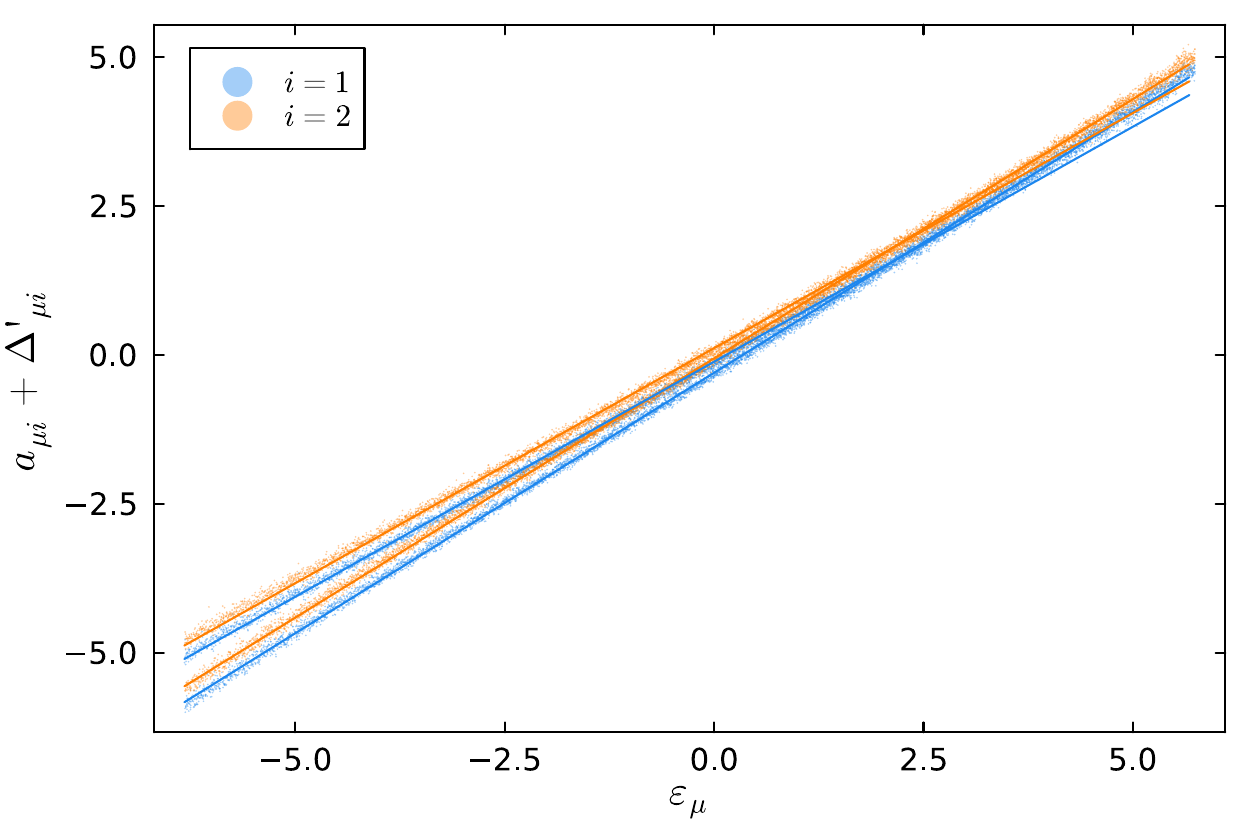}}
    \caption{(a) \(\sigma_{\mu i}\) versus \(\epsilon_\mu\). (b) \(a_{\mu i} + \Delta'_{\mu i}\) splitting into two branches.
    Branch A: \(\braket{a_{\mu 1}+\Delta'_{\mu 1}} = 0.875\epsilon_\mu - 0.301\), \(\braket{a_{\mu 2}+\Delta'_{\mu 2}} = 0.871\epsilon_\mu - 0.060\).
    Branch B: \(\braket{a_{\mu 1}+\Delta'_{\mu 1}} = 0.789\epsilon_\mu - 0.116\), \(\braket{a_{\mu 2}+\Delta'_{\mu 2}} = 0.790\epsilon_\mu + 0.113\).
    Branch A corresponds to larger \(\sigma\) values in (a).}
    \label{FIGgp}
\end{figure*}

\begin{figure}[htbp]
    \centering
    \subfigure{\includegraphics[width=0.42\textwidth]{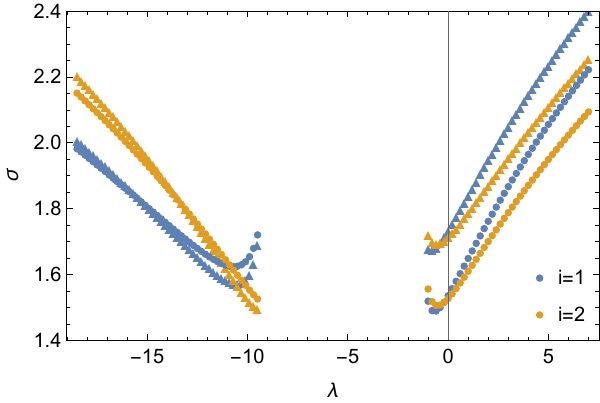}}
    \subfigure{\includegraphics[width=0.42\textwidth]{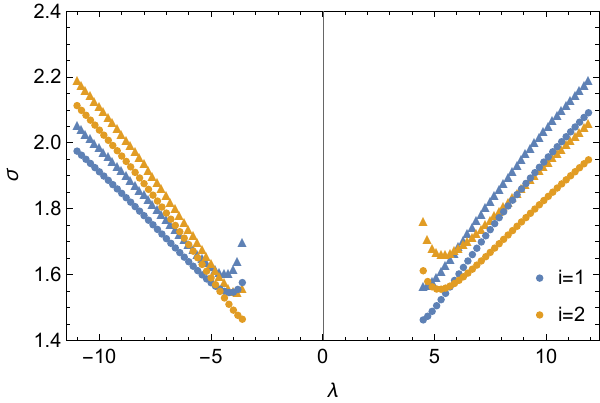}}
    \subfigure{\includegraphics[width=0.42\textwidth]{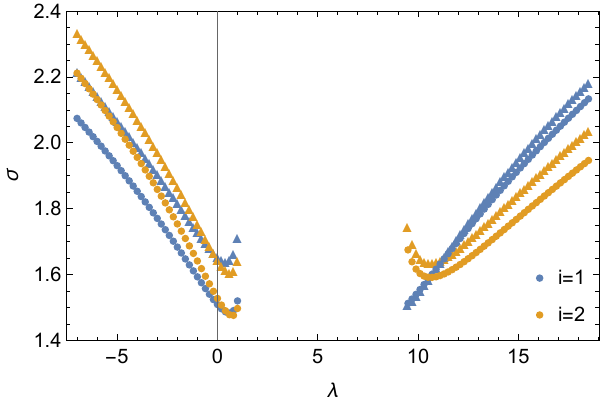}}
    \caption{\(\sigma\) versus \(\lambda\) from Eq.~\eqref{sigmaconsis} for \(\epsilon_\mu = \{-6, 0.583, 6\}\) (top, middle, and bottom panels).
    Triangles (circles): Branch A (B).}
    \label{sigdis}
\end{figure}

In the tail regime both \(\Im\,\mathcal{G}\) and \(\Re\,\mathcal{G}\) become much smaller than \(|\Delta^{\mu i}(\lambda)|\).
Substituting Eqs.~\eqref{normal} and~\eqref{IMGAP} into Eq.~\eqref{PAG} and approximating the interaction
via Eq.~\eqref{cmisep} yields the tail self-consistency condition
\begin{equation}\label{tailconsis}
    \frac{[\Delta^{\mu i}(\lambda)]^2}{\sigma_{\mu i}} \varphi\!\left(\frac{\delta\lambda'_{\mu i}}{\sigma_{\mu i}}\right)
    \approx \sum_j \mathcal{V}_{\mu i,j} \, \mathbb{E}_\delta\!\left[ \varphi\!\left(\frac{\delta\lambda'_{\nu j}}{\sigma_{\nu j}}\right) \frac{1}{\sigma_{\nu j}} \right].
\end{equation}
Assuming \(\sigma_{\nu j} \approx \sigma\) is approximately state-independent and using
\(\operatorname{erfc}(x) \approx e^{-x^2}/(\sqrt{\pi}\,x)\) for large \(x\), we obtain
\begin{align}\label{sigmaconsis}
    \frac{1}{\sigma^2} \approx \sum_j \frac{\overline{\mathcal{F}}_{i,j}}{k_j [\Delta^{\mu i}(\lambda)]^2} \,
    \frac{1}{|\lambda - \Delta'_j - k_j \epsilon^{\mathrm{sgn}_{\lambda\mu}}_\mu|}\notag\\
	\times    \exp\!\left( \frac{-(\lambda - \Delta'_j - k_j \epsilon^{\mathrm{sgn}_{\lambda\mu}}_\mu)^2 + (\Delta\lambda^{\mu i})^2}{2\sigma^2} \right).
\end{align}
Figure~\ref{sigdis} shows that \(\sigma\) varies mildly with \(\lambda\); the median values are consistent
with the fitted parameters (Fig.~\ref{FIGgp}).

A striking feature is the splitting of the Gaussian tail parameters into two distinct branches (Fig.~\ref{FIGgp}).
As analyzed in \cref{OMTB}, this originates from the interplay between the reflection symmetry of the unperturbed bath and the local system-bath coupling: the local perturbation hybridizes nearly degenerate bath eigenstates into symmetric and antisymmetric combinations that couple differently to the system.
Notably, the Lorentzian parameters \((\Delta_{\mu i}, \chi_{\mu i})\) remain largely single-valued (Fig.~\ref{FIGlgp}).
This is because the Lorentzian self-consistent equation~\eqref{consist} depends on the interaction only through the integrated spectral weight \(\mathcal{V}_{\mu i,j} = \sum_\nu |V_{\mu i,\nu j}|^2\), which is invariant under the symmetry-induced reorganization of bath states; a detailed analysis is given in \cref{OMTB}.

The Gaussian tail entropy is
\begin{equation}\label{entgau}
    S(p^{\mu i}) \approx S(a_{\mu i} + \Delta'_{\mu i}) + \ln(\sigma_{\mu i} \sqrt{2\pi e}).
\end{equation}

\subsection{Unified Hybrid Lorentzian-Gaussian Description}

\begin{figure*}[htbp]
    \centering
    \subfigure{\includegraphics[width=0.48\textwidth]{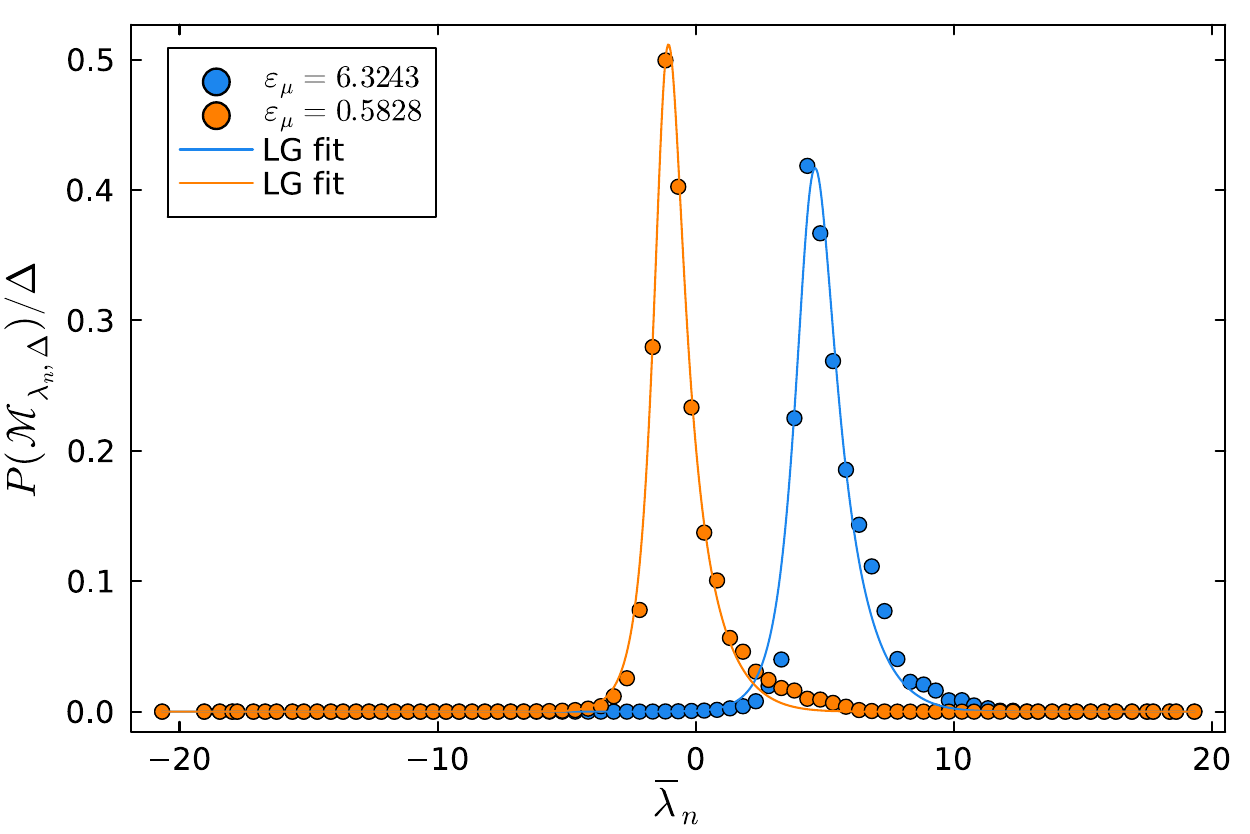}}
    \subfigure{\includegraphics[width=0.48\textwidth]{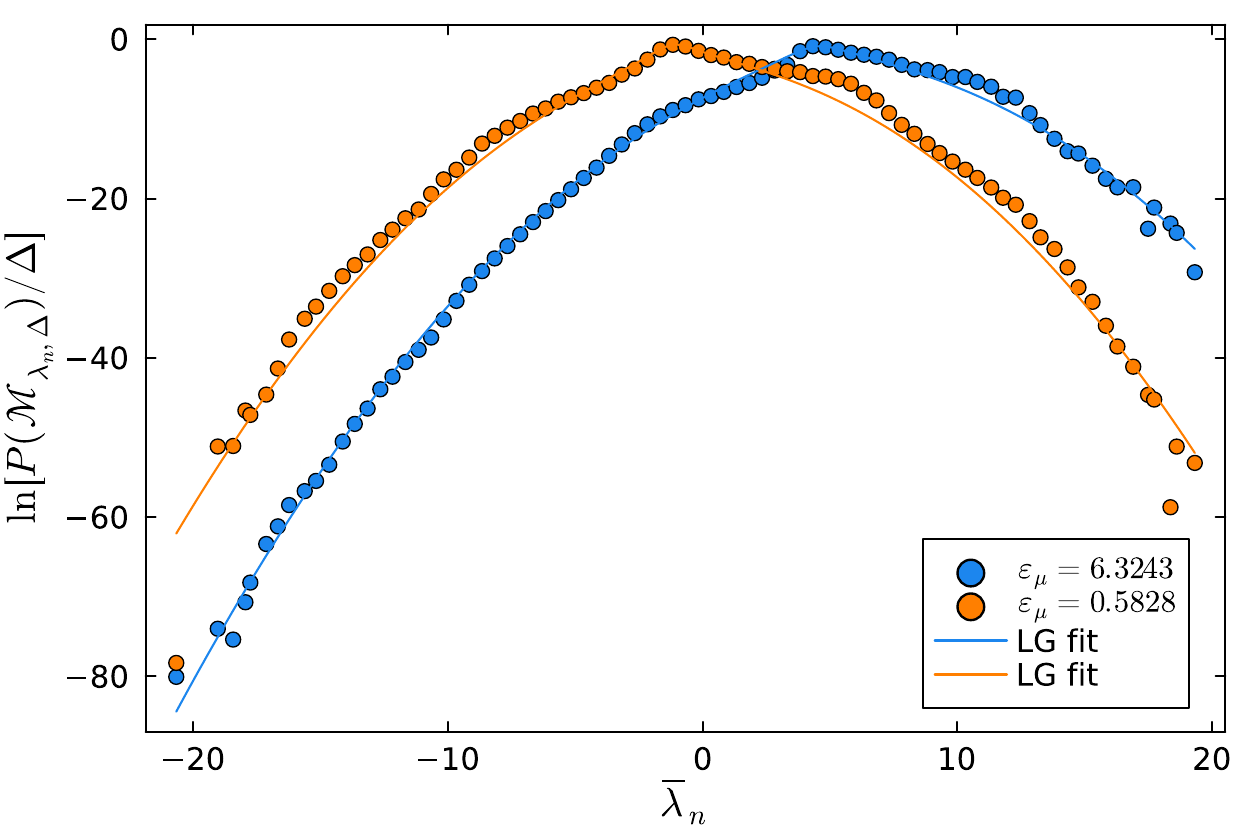}}
    \caption{Binned distribution under the LG ansatz (\(i=1\), \(\Delta = 0.5\)).
    Fitted parameters: for \(\epsilon_\mu = 6.3243\): \((4.4885, 5.7790, 1.0439, 2.1266)\);
    for \(\epsilon_\mu = 0.5828\): \((-1.1601, 0.3519, 0.7823, 2.000)\).}
    \label{FIGlgfit}
\end{figure*}

To achieve a comprehensive description, we unify both profiles into a Voigt-type hybrid~\cite{HC26}:
\begin{equation}\label{LGdistr}
    p^{\mu i}(\lambda) = \frac{G^{\mu i}(\lambda) L^{\mu i}(\lambda)}{e^{S(\lambda)} V^{\mu i}},
\end{equation}
where \(G^{\mu i}(\lambda) = G(\delta\lambda'_{\mu i}; \sigma_{\mu i})\) with \(G(x;\sigma) = e^{-x^2/(2\sigma^2)}/(\sqrt{2\pi}\sigma)\),
and the normalization factor \(V^{\mu i} = V(\Delta_{\mu i} - \Delta'_{\mu i}; \sigma_{\mu i}, \chi_{\mu i})\) is the Voigt profile
\(V(x;\sigma,\chi) = \int dx' G(x';\sigma) L(x - x'; \chi)\).
This Lorentzian-Gaussian (LG) distribution is characterized by four parameters:
\((a_{\mu i}+\Delta_{\mu i}, a_{\mu i}+\Delta'_{\mu i}, \chi_{\mu i}, \sigma_{\mu i})\).
The inequality \(\Delta'_{\mu i} \neq \Delta_{\mu i}\) breaks mirror symmetry, a feature absent in the parity-preserving
mean-field closure but naturally generated by multi-resolvent correlations~\cite{HC26}.
Figure~\ref{FIGlgfit} shows that the LG distribution provides an excellent description of both peak and tails.

\begin{figure*}[htbp]
    \centering
    \subfigure[\(\sigma_{\mu i}\)]{\includegraphics[width=0.48\textwidth]{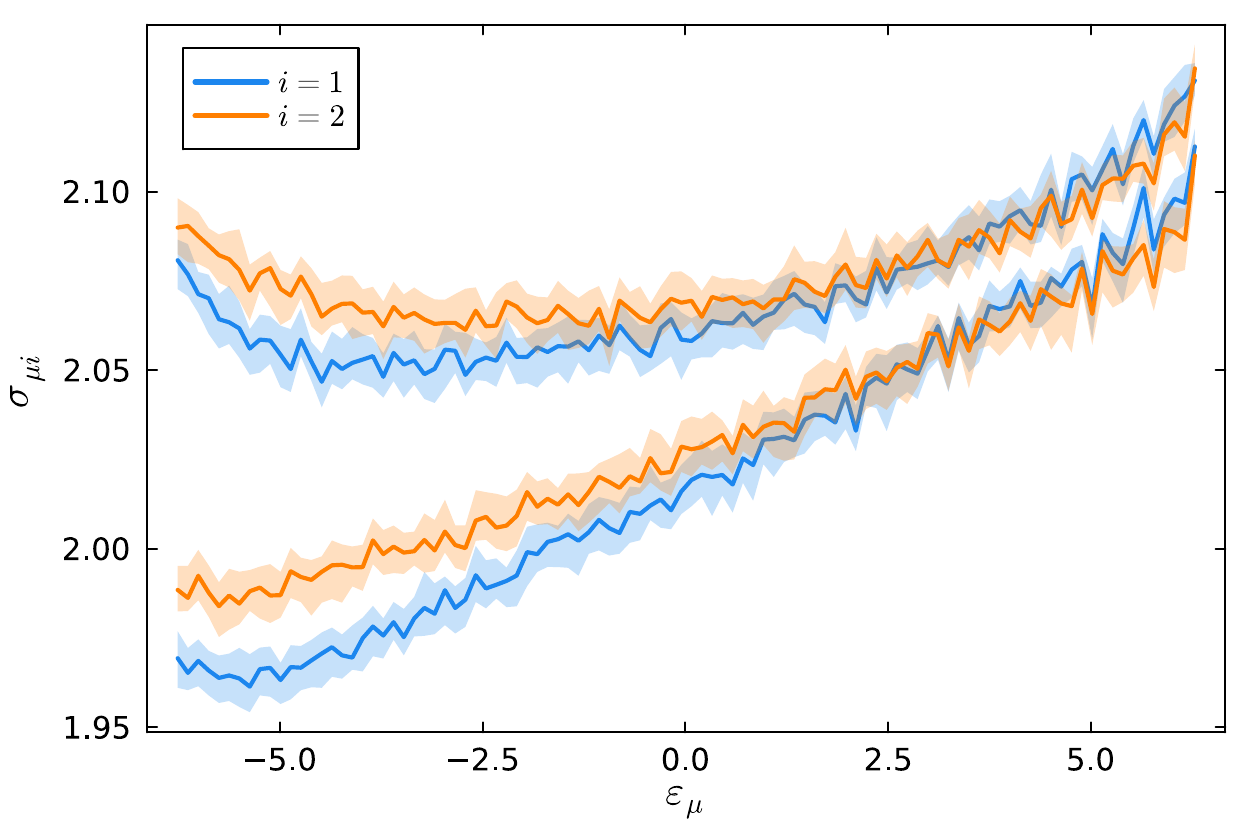}}
    \subfigure[\(a_{\mu i}+\Delta'_{\mu i}\)]{\includegraphics[width=0.48\textwidth]{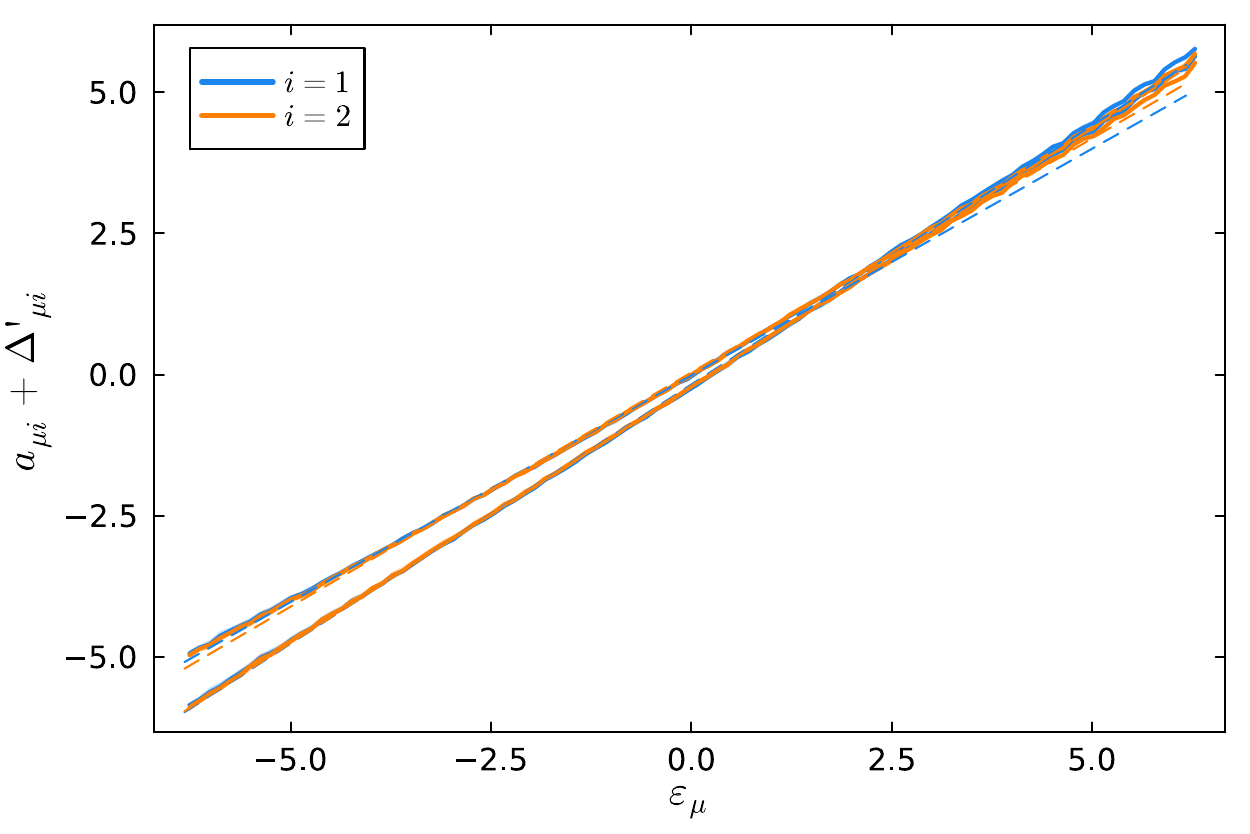}}
    \subfigure[\(\chi_{\mu i}\)]{\includegraphics[width=0.48\textwidth]{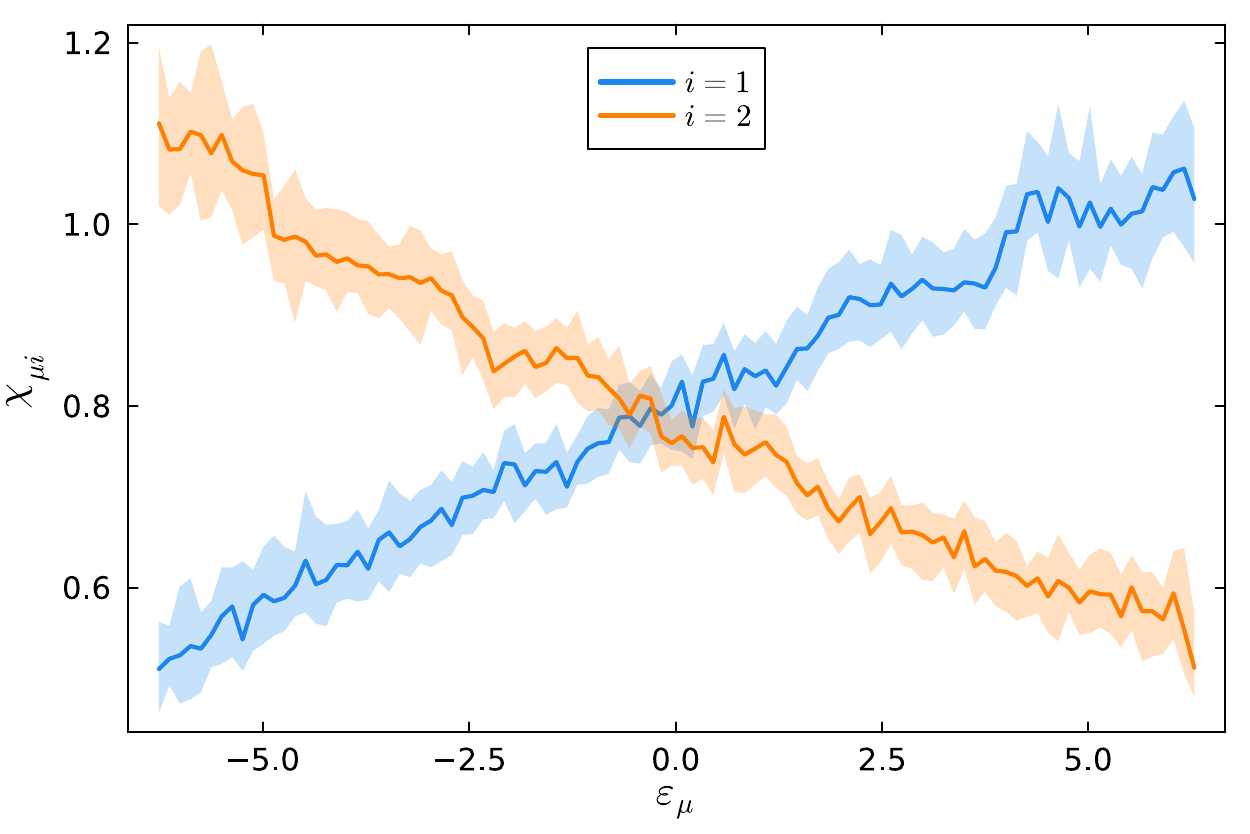}}
    \subfigure[\(a_{\mu i}+\Delta_{\mu i}\)]{\includegraphics[width=0.48\textwidth]{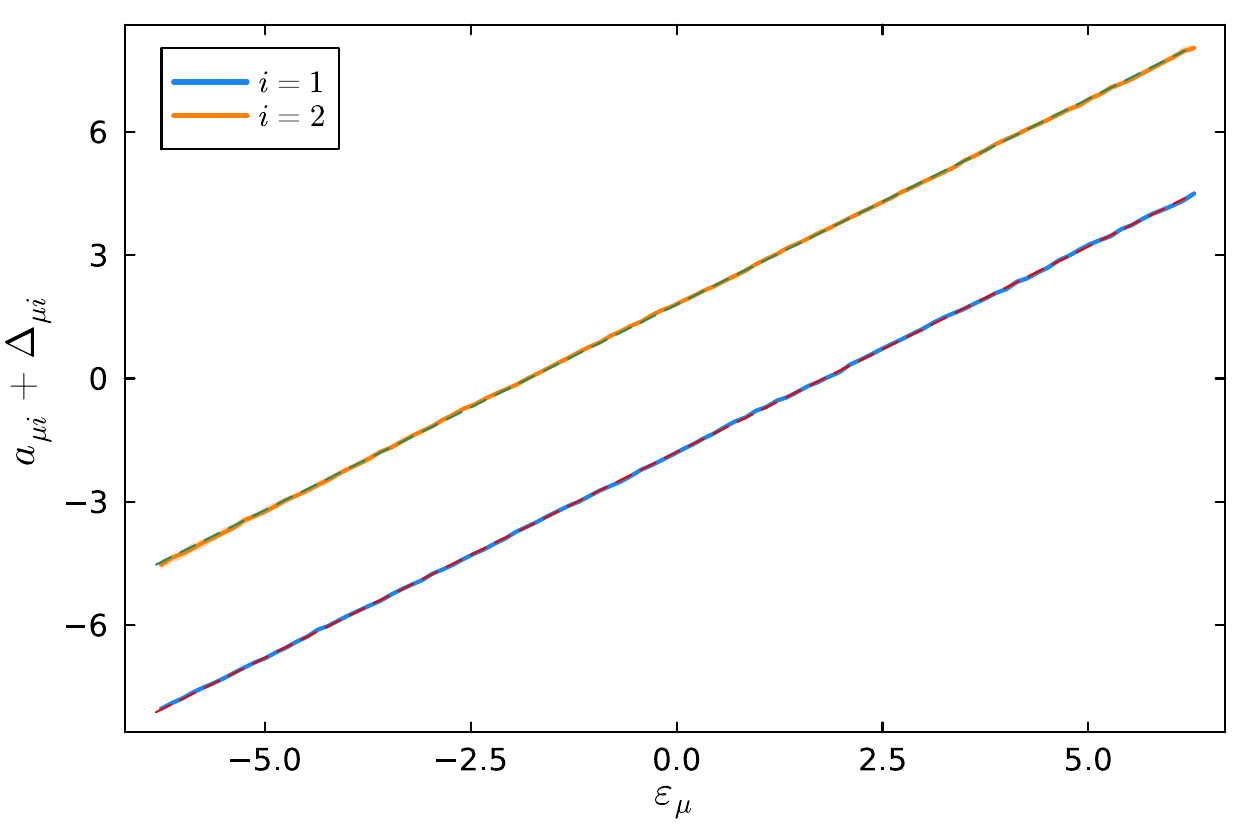}}
    \caption{LG fit parameters versus \(\epsilon_\mu\). (a) \(\sigma_{\mu i}\). (b) \(a_{\mu i}+\Delta'_{\mu i}\) showing two branches.
    Branch A: \(\braket{a_{\mu 1}+\Delta'_{\mu 1}} = 0.920\epsilon_\mu - 0.183\), \(\braket{a_{\mu 2}+\Delta'_{\mu 2}} = 0.910\epsilon_\mu - 0.192\).
    Branch B: \(\braket{a_{\mu 1}+\Delta'_{\mu 1}} = 0.816\epsilon_\mu + 0.005\), \(\braket{a_{\mu 2}+\Delta'_{\mu 2}} = 0.826\epsilon_\mu + 0.036\).
    (c) \(\chi_{\mu i}\). (d) \(a_{\mu i}+\Delta_{\mu i}\) with
    \(\braket{a_{\mu 1}+\Delta_{\mu 1}} = 1.001\epsilon_\mu - 1.791\), \(\braket{a_{\mu 2}+\Delta_{\mu 2}} = 1.001\epsilon_\mu + 1.800\).}
    \label{FIGlgp}
\end{figure*}

The self-consistent equations for the LG parameters follow from Eq.~\eqref{PAG}.
Near the center,
\begin{align}
    \chi_{\mu i} &= \Im\, \mathcal{G}_{\mu i}(\lambda - \mathrm{i}0^+) \times G^{\mu i}(\lambda)/V^{\mu i}, \label{EAFC1}\\
    \lambda - a_{\mu i} - \Delta_{\mu i} &= [\Delta^{\mu i}(\lambda) - \Re\, \mathcal{G}_{\mu i}(\lambda)] \times G^{\mu i}(\lambda)/V^{\mu i}, \label{EAFC2}
\end{align}
while far from the center
\begin{equation}\label{EAFCfar}
    [\Delta^{\mu i}(\lambda)]^2 \frac{G^{\mu i}(\lambda) L^{\mu i}(\lambda)}{V^{\mu i}} \approx \frac{1}{\pi} \Im\, \mathcal{G}_{\mu i}(\lambda - \mathrm{i}0^+).
\end{equation}

The evaluation of \(\Im\,\mathcal{G}\) and \(\Re\,\mathcal{G}\) under the LG ansatz is facilitated by the Faddeeva function
\(w(z) = e^{-z^2} \operatorname{erfc}(-\mathrm{i}z)\). Defining the dispersion profile
\(D(x;\sigma,\chi) = \frac{1}{\sqrt{2\pi}\sigma} \Im\, w\!\bigl(\frac{x + \mathrm{i}\chi}{\sqrt{2}\sigma}\bigr)\),
the real part takes the compact form (see \cref{HTEA})
\begin{equation}\label{REGui}
    \Re\, \mathcal{G}_{\mu i}(\lambda) = \sum_{\nu j \neq \mu i} \abs{V_{\mu i,\nu j}}^2 \frac{\delta\lambda_{\nu j} + \chi_{\nu j} \delta D_A}{\delta\lambda_{\nu j}^2 + \chi_{\nu j}^2},
\end{equation}
with \(\delta D_A = [D(\delta\lambda'_{\nu j}; \sigma_{\nu j}, 0) - D(\Delta_{\nu j} - \Delta'_{\nu j}; \sigma_{\nu j}, \chi_{\nu j})]/V^{\nu i}\).
Defining \(z_0^{\nu j} = (\Delta_{\nu j} - \Delta'_{\nu j} + \mathrm{i}\chi_{\nu j})/(\sqrt{2}\sigma_{\nu j})\) and
\(\delta w = [w(\delta\lambda'_{\nu j}/(\sqrt{2}\sigma_{\nu j})) - w(z_0^{\nu j})] / \Re\, w(z_0^{\nu j})\),
the resolvent boundary value decomposes elegantly as
\begin{equation}\label{KKR}
    \mathcal{R}_{\nu j}(\lambda - \mathrm{i}0^+) = \frac{1}{\delta\lambda_{\nu j} - \mathrm{i}\chi_{\nu j}} + \mathrm{i}\frac{\chi_{\nu j} \delta w^*}{|\delta\lambda_{\nu j} - \mathrm{i}\chi_{\nu j}|^2}.
\end{equation}
The second term encodes the deviation from the pure Lorentzian: when \(\sigma_{\nu j} \to \infty\), \(\delta w \to 0\)
and the single-pole SCBA structure is recovered.
When \(\Delta'_{\nu j} \neq \Delta_{\nu j}\), \(\delta w\) acquires an imaginary part that mixes
retarded and advanced sectors, generating spectral skewness~\cite{HC26}.

\begin{figure*}[t]
    \centering
    \subfigure[Width \(\chi_{\mu i}\)]{\includegraphics[width=0.48\textwidth]{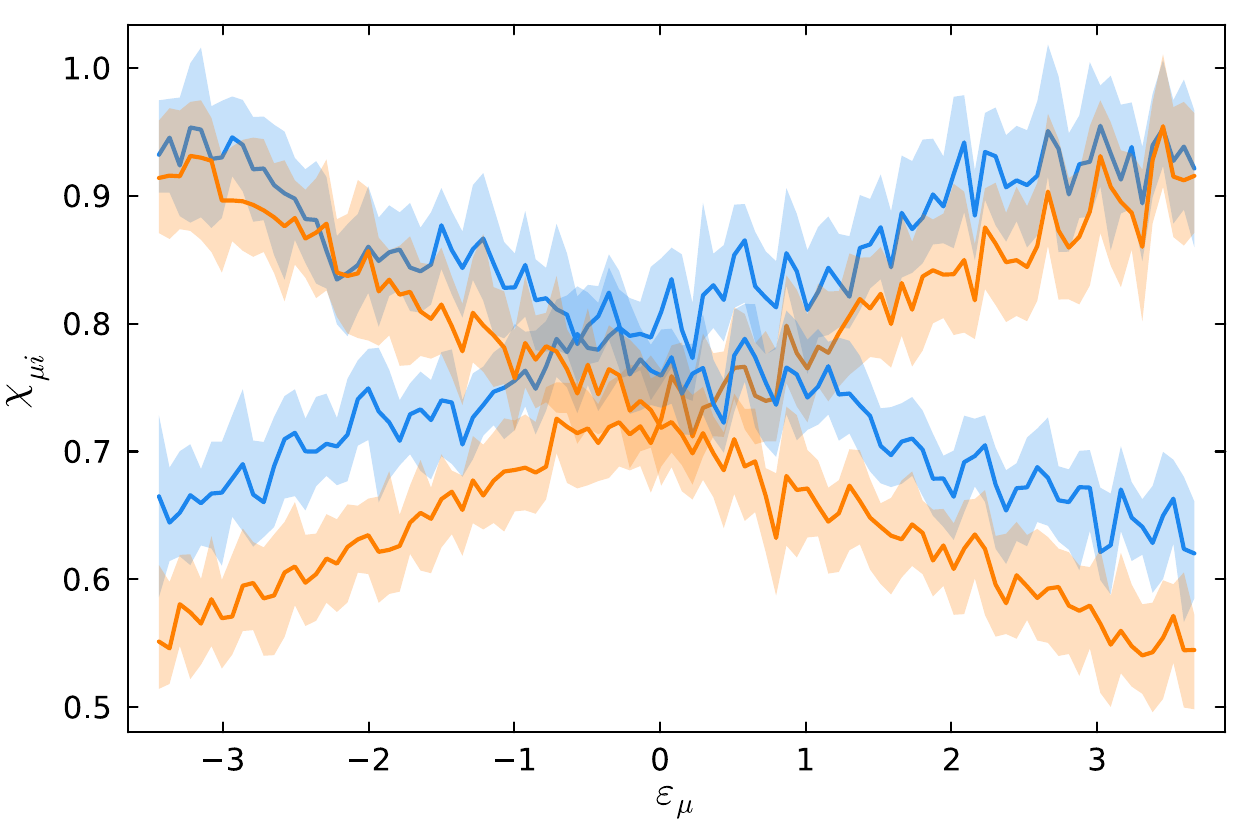}}
    \subfigure[Shift \(\Delta_{\mu i}\)]{\includegraphics[width=0.48\textwidth]{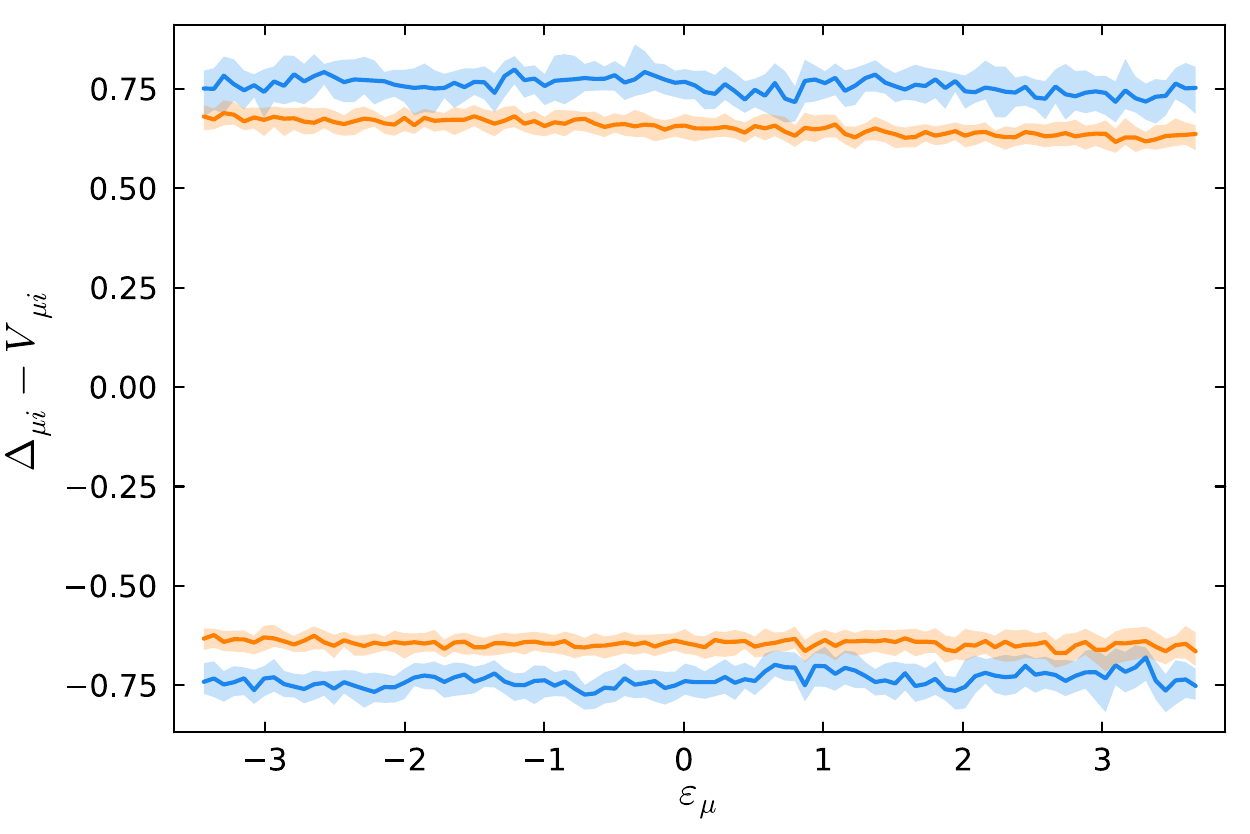}}
    \caption{LG self-consistency validation. Blue: fitted; Orange: reconstructed via Eq.~\eqref{LGsimpcon}
    with parameters from Eq.~\eqref{LGpara}.}
    \label{FIGLGconsis}
\end{figure*}

Applying the same coarse-graining as in the Lorentzian case yields the simplified LG self-consistent equation:
\begin{widetext}
	\begin{equation}\label{LGsimpcon}
    \overline{\Delta}_2 + \mathrm{i} \overline{\chi}_{\mu 2} = (E_1 - E_2 + \overline{\Delta}_1 - \eta_1)\!\left(1 - \frac{G^{\mu 2}}{V^{\mu 2}}\right)
    + \mathcal{V}_{\mu 2,1} \, \mathbb{E}_\delta\!\left[ \frac{1}{\delta - \eta_1 - \mathrm{i}(\overline{\chi}_{\mu 1} - A_1\delta)} + \mathrm{i}\frac{(\overline{\chi}_{\mu 1} - A_1\delta) \delta w^*}{(\delta - \eta_1)^2 + (\overline{\chi}_{\mu 1} - A_1\delta)^2} \right] \frac{G^{\mu 2}}{V^{\mu 2}}.
\end{equation}
Solving numerically with \(\sigma \approx 2.07\) and the fitted Gaussian parameters for branch A gives
\begin{equation}\label{LGpara}
    (\eta_1, \overline{\chi}_1, A_1, \overline{\Delta}_1) = (-3.4649, 0.6983, 0.0510, -0.7381), \quad
    (\eta_2, \overline{\chi}_2, A_2, \overline{\Delta}_2) = (3.4952, 0.6598, -0.0519, 0.7508).
\end{equation}
\end{widetext}
Compared with the pure Lorentzian parameters \eqref{etap}, Eq.~\eqref{LGpara} lies closer to the fitted data
(Fig.~\ref{FIGlgp}). The broad consistency in Fig.~\ref{FIGLGconsis} confirms the accuracy of the LG equations.

The entropy for the LG distribution, dominated by the sharp Lorentzian peak, is approximately
\begin{equation}\label{LGentro}
    S(p^{\mu i}) \approx S(\lambda_p) + H(\Delta_{\mu i} - \Delta'_{\mu i}, \sigma_{\mu i}, \chi_{\mu i}),
\end{equation}
where \(\lambda_p\) is the peak position (see \cref{ESrVN}) and \(H\) is the entropy contribution of the
\(G \times L / V\) function.
For representative parameters (\(\Delta - \Delta' = \pm 1.8\), \(\sigma = 2.0\), \(\chi = 0.75\)),
one finds \(H \approx 1.54\) (see \cref{DAE}), and the full entropy is \(S \approx S(\lambda_p) + 1.54\). 
For comparison, the pure Lorentzian estimate is
\(\ln(4\pi\chi) \approx 2.07\). This substantial reduction explains why the Lorentzian-based entropy in Fig.~\ref{FIG5}
overestimates the observational entropy: the hybrid profile suppresses the broad tails that the Lorentzian retains.

\subsection{Self-Energy Ansatz and Analyticity}

An alternative, more fundamental approach proceeds by constructing an ansatz directly for the self-energy.
As shown in Ref.~\cite{HC26}, any admissible self-energy must be analytic in the lower half-plane,
with real and imaginary parts related by the Kramers-Kronig relations.
The simplest extension of the constant self-energy that introduces a non-Lorentzian scale while preserving
causality is the Faddeeva form~\cite{HC26}
\begin{equation}\label{afg}
    \mathcal{G}_{\mu i}(\lambda) = \Delta^{\mathrm{eff}}_{\mu i} - V_{\mu i} + \mathrm{i}\chi^{\mathrm{eff}}_{\mu i} \, w\!\left(\frac{-\delta\lambda'_{\mu i}}{\sqrt{2}\sigma_{\mu i}}\right).
\end{equation}
The corresponding spectral function
\begin{equation}\label{ofag}
    p^{\mu i}_n = \frac{1}{e^{S(\lambda_n)}} \frac{1}{\pi} \Im \frac{1}{\lambda_n - a_{\mu i} - \Delta^{\mathrm{eff}}_{\mu i} - \Sigma_{\mu i}(\lambda_n)}
\end{equation}
is approximately equivalent to the LG profile \eqref{LGdistr} (see \cref{ESrVN} for the precise mapping).

\begin{figure*}[htbp]
    \centering
    \subfigure{\includegraphics[width=0.48\textwidth]{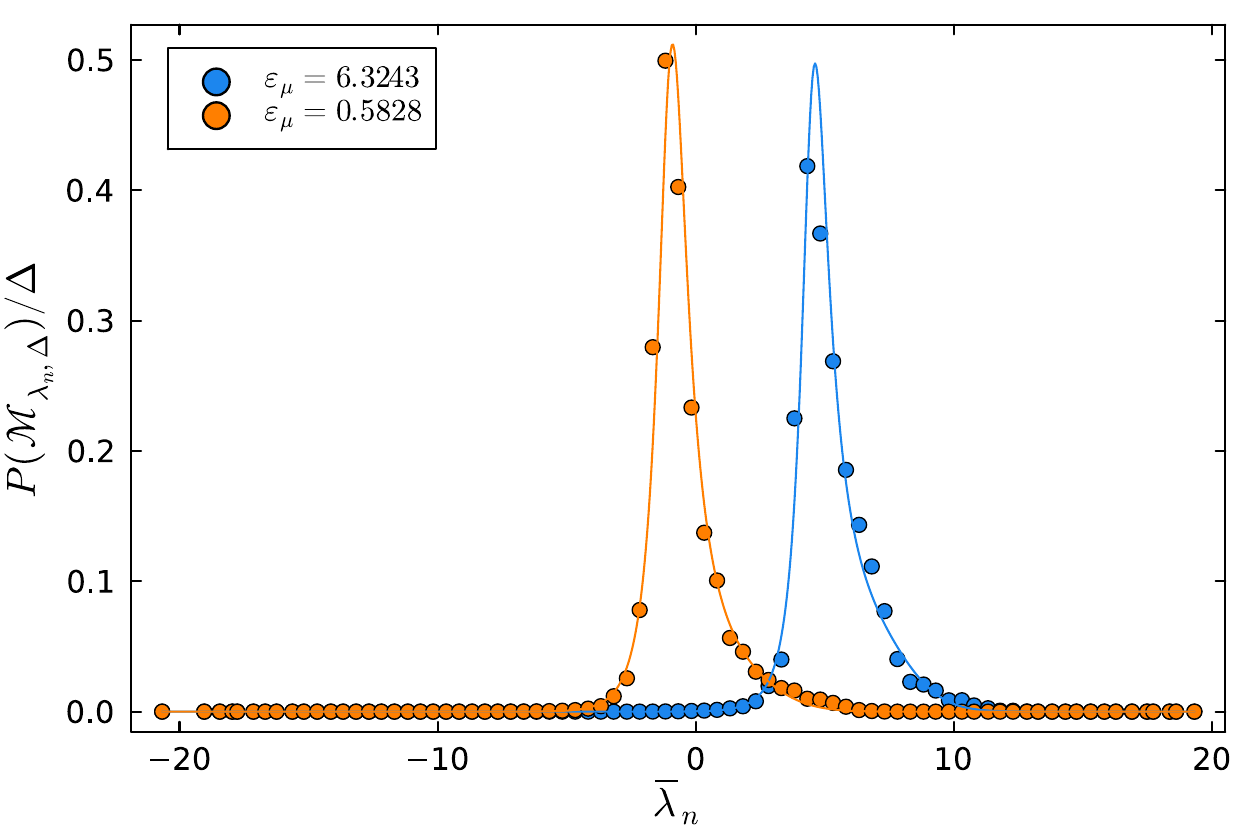}}
    \subfigure{\includegraphics[width=0.48\textwidth]{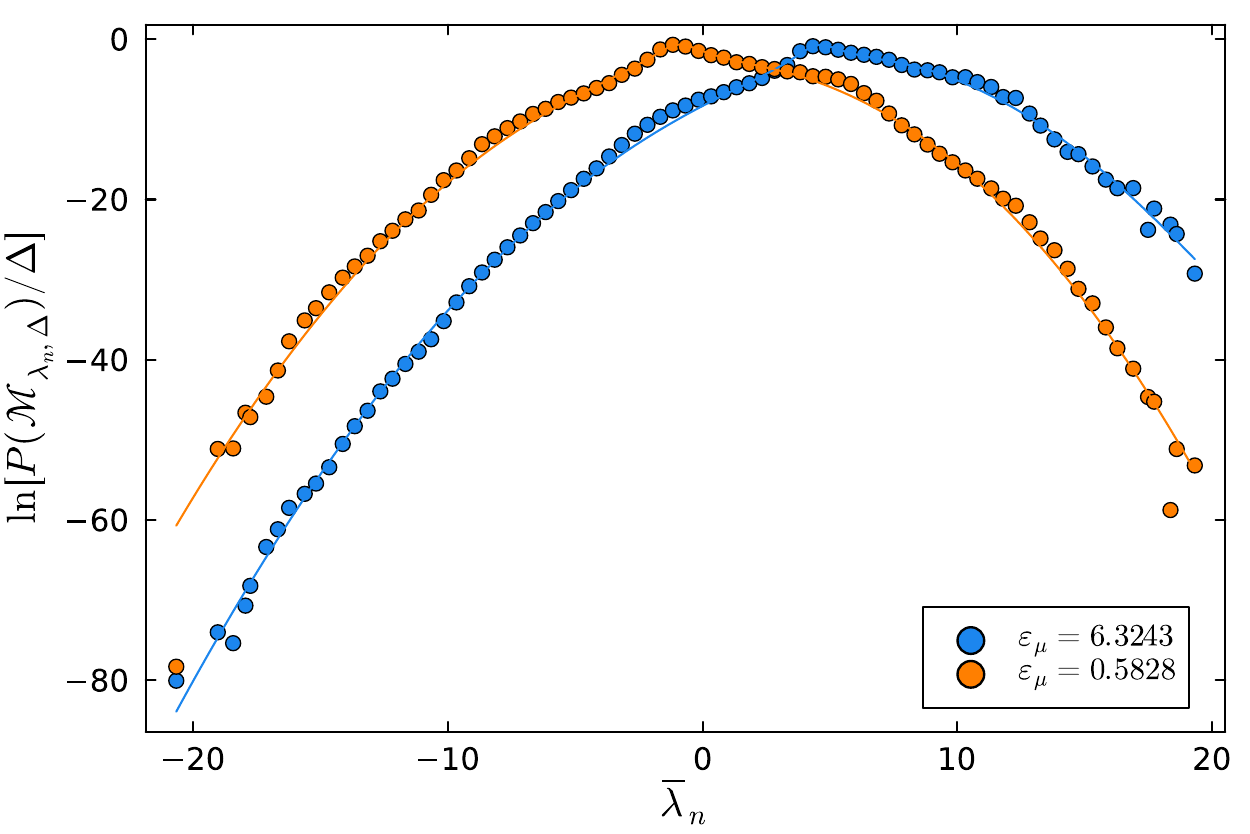}}
    \caption{Binned distribution under the dual-Faddeeva ansatz (\(i=1\), \(\Delta = 0.5\)).
    Fitted parameters: for \(\epsilon_\mu = 6.3243\): \((5.2228, 1.1548, 7.7700, 5.6103, 1.7272, 2.1356)\);
    for \(\epsilon_\mu = 0.5828\): \((-0.4882, 1.0831, 2.4956, -0.5692, 1.7515, 1.9534)\).}
    \label{FIGguifit}
\end{figure*}

A natural generalization employs a linear combination of two Faddeeva functions,
\(f^{\mu i}(\lambda) = w(-\delta\lambda^{(1)}_{\mu i}/(\sqrt{2}\sigma^1_{\mu i})) + w(-\delta\lambda^{(2)}_{\mu i}/(\sqrt{2}\sigma^2_{\mu i}))\),
providing six-parameter flexibility to capture finite-band effects (Fig.~\ref{FIGguifit}).

\subsection{Connection to the Beyond-Mean-Field Hierarchy}

The analysis above is conducted within the mean-field (single-resolvent) sector, where the self-energy is a
linear functional of the resolvent and the spectral function is strictly reflection-symmetric whenever
Eq.~\eqref{GRE} is used self-consistently.
The observed asymmetry \(\Delta'_{\mu i} \neq \Delta_{\mu i}\)---manifest as the difference between Lorentzian
and Gaussian centers---is a clear signature of physics beyond the mean-field level.
As proved in Ref.~\cite{HC26}, spectral skewness originates uniquely from the multi-resolvent hierarchy
\(\mathcal{G}^{\mathrm{CC}}_{\mu i} = \sum_{\ell \ge 3} \mathcal{G}^{(\ell)}_{\mu i}\):
the leading term \(\mathcal{G}^{(3)}_{\mu i}\) involves products of two diagonal resolvents whose
Hilbert-transform structure intrinsically mixes parity sectors (see \cref{cross_structure} and \cite{HC26}).
A complementary mechanism contributing to the observed two-branch Gaussian 
tails, which rooted in the interplay between reflection symmetry and local 
system-bath coupling, is analyzed in \cref{OMTB}, and we discuss its 
connection to the multi-resolvent hierarchy at the end of that appendix.

At the level of entropy scaling, these beyond-mean-field effects produce corrections of order unity to the constant term (reducing the effective constant from \(\ln(4\pi\chi) \approx 2.07\) for the pure Lorentzian to \(H \approx 1.54\) for the LG ansatz), while the dominant logarithmic scaling \(\sim \ln \chi_{\mu i}\) and the extensive density-of-states contribution \(S(a_{\mu i} + \Delta_{\mu i})\) remain robust---the former because \(\chi_{\mu i}\) is controlled by the integrated spectral weight \(\mathcal{V}_{\mu i,j}\), which is invariant under the branch splitting discussed in \cref{OMTB}.
The systematic convergence of the hierarchical ansatz strategy---from Lorentzian to Gaussian tail
to hybrid LG---demonstrates that the complex eigenstate mixing underlying steady-state entropy can be
efficiently captured by a few macroscopic parameters determined self-consistently from the interaction,
with systematically improvable corrections from the full resolvent hierarchy of Ref.~\cite{HC26}.

The leading self-consistent equations used here retain the dominant diagonal
sector of the resolvent hierarchy.  The higher-order sectors developed in
Ref.~\cite{HC26} provide systematic corrections to the self-consistent
envelope parameters---relevant for quantitative parameter reconstruction and
numerical stability---but they do not modify the entropy decomposition
derived below: once the shell-conditioned envelope \(p^{\rm env}(\lambda)\)
is specified, the entropy deficit is determined by the fluctuation sectors of
the same hierarchy.

\subsection{Intra-Shell Fluctuations and the von Neumann Entropy}
\label{IntraShell}

The smoothed estimates above capture the observational entropy but remain
systematically above the von Neumann entropy of the decohered state
(Fig.~\ref{FIG5}).  We now account for this gap: the difference is a computable
intra-shell fluctuation correction carried by the covariance (two-resolvent)
sector of the resolvent hierarchy, whose exact structure has been established
in Ref.~\cite{HOETH}.  This extends the hierarchy from the envelope to the
microscopic fluctuation sector, providing a route toward the full conditional
intensity distribution---and thereby toward the von Neumann entropy
\(S_{\rm vN}=S(\omega)\) itself.

Write the microscopic probability as
\begin{equation}\label{fluct_decomp}
    p_n = p^{\rm env}_n\, x_n, \qquad
    p^{\rm env}_n := \mathbb E_{\rm shell}[\,p_n\,|\,\lambda_n], \qquad
    \mathbb E_{\rm shell}[x_n\,|\,\lambda_n]=1,
\end{equation}
where \(p^{\rm env}_n\) is the \emph{exact} shell-conditional mean.  Here and
throughout this section, \(\mathbb E_{\rm shell}[\cdots|\lambda]\) denotes a
local average over the eigenstates \(n\) within a narrow energy window around
\(\lambda\), at a \emph{fixed} Hamiltonian and channel---not a statistical
average over Hamiltonian realizations; to lighten the notation we write
\(\mathbb E\) for \(\mathbb E_{\rm shell}\) in what follows.  The
empirical support analysis of Appendix~\ref{fluct_entropy} shows the support
to be essentially the full spectrum: only a fraction \(\sim1/(2N)\) of the
eigenstates---the parity-definite ones mismatched with the product
state---has exactly vanishing overlap.  The von Neumann entropy of the
decohered state then decomposes \emph{identically} as
\begin{equation}\label{dS_exact}
    \mathbb E[S_{\rm vN}] = S_{\rm env} + \Delta S,
    \qquad
    \Delta S = -\sum_{n}p^{\rm env}_n\,
    \mathbb E[\,x_n\ln x_n\,|\,\lambda_n],
\end{equation}
with \(S_{\rm env}:=-\sum_n p^{\rm env}_n\ln p^{\rm env}_n\) and no truncation
involved.  The ED verification below uses a single Hamiltonian and averages
over energy shells and channels, exactly as in the definition of
\(\mathbb E_{\rm shell}\).  Equation~\eqref{dS_exact} separates the gap between the
coarse-grained and the exact entropy into its two conceptual origins: the
envelope entropy \(S_{\rm env}\) accounts for the energy-space spreading of
the overlaps, while \(\Delta S\) accounts for the microscopic intra-shell
intensity statistics.  By convexity of \(x\ln x\) and
\(\mathbb E_{\rm shell}[x|\lambda]=1\), Jensen's inequality gives
\(\mathbb E_{\rm shell}[x\ln x|\lambda]\ge0\) in every shell, so
\(S_{\rm vN}\le S_{\rm env}\) follows \emph{exactly}, independently of any
Gaussian or Gamma closure; only the magnitude of the deficit is
model-dependent.

The exact envelope is approximated by the LG profile of Eq.~\eqref{LGdistr},
\(p^{\rm env}_n\approx p^{\rm LG}_n:=e^{-S(\lambda_n)}f^{\mu i}(\lambda_n)\),
so that \(S_{\rm env}\approx S_{\rm LG}\), the LG estimate of
Eq.~\eqref{LGentro}.  The approximation is excellent in practice: the ansatz
error \(|S_{\rm env}-S_{\rm LG}|\lesssim0.1\) is small against the
fluctuation correction \(|\Delta S|\approx1.08\)
(Appendix~\ref{fluct_entropy}).  The resulting three-level decomposition
\begin{equation}\label{threelevel}
    \mathbb E[S_{\rm vN}] = S_{\rm LG}
    + \underbrace{(S_{\rm env}-S_{\rm LG})}_{\text{ansatz error}}
    - \underbrace{\sum_n p^{\rm env}_n\,\mathbb E[x_n\ln x_n|\lambda_n]}_{\text{microscopic fluctuation}},
\end{equation}
therefore isolates the macroscopic envelope description, its own ansatz
error, and the microscopic fluctuation correction as three distinct,
numerically separated contributions.

That statistics is supplied by the covariance sector of the hierarchy of
Ref.~\cite{HOETH}.  The same-channel resolvent covariance
\(\mathcal C_{aa}(z,z')=\mathbb E[\delta\mathcal R_a(z)\,\delta\mathcal R_a(z')]\),
with \(\delta\mathcal R_a=\mathcal R_a-\bar{\mathcal R}_a\), admits the
spectral representation
\(\sum_{nm}\mathbb E[\delta p_n^a\delta p_m^a]/(z-\lambda_n)(z'-\lambda_m)\),
and its diagonal (\(n=m\)) bin is exactly the intensity variance
\begin{equation}\label{diagbin}
    \sum_n \frac{\mathrm{Var}(p_n^a)}{(z-\lambda_n)(z'-\lambda_n)}
    = \sum_n \frac{[q(\lambda_n)-1]\,(p^{{\rm env},\,a}_n)^2}
                  {(z-\lambda_n)(z'-\lambda_n)},
\end{equation}
where
\begin{equation}\label{qdef}
    q(\lambda)
    = \frac{\mathbb E[\,p_n^2\,|\,\lambda_n=\lambda]}{p^{\rm env}(\lambda)^2}
\end{equation}
is the energy-resolved Porter--Thomas factor of Ref.~\cite{HOETH}.
The shell-resolved definition is essential: the global moment ratio
\(\mathbb E[p^2]/\mathbb E[p]^2\) mixes envelope inhomogeneity with intrinsic
amplitude fluctuations, and only the shell-conditioned \(q(\lambda)\)---with
the envelope divided out---is a clean fluctuation measure.
Equation~\eqref{diagbin} is an exact \emph{spectral identification}: the
diagonal ridge of the two-resolvent covariance carries the intra-shell
intensity variance, with coefficient \(q(\lambda)-1\).  The single resolvent
thereby generates the envelope \(S_{\rm env}\), and the two-resolvent
covariance \emph{identifies} the fluctuation correction; what it does not
provide is a microscopic \emph{closure} computing the ridge weight
\(q(\lambda)-1\) from the Hamiltonian.  In the language of
Ref.~\cite{HOETH} that remains the Level-II problem of the covariance
hierarchy, whose ladder resummation is only a small dressing at ETH-scaled
couplings (spectral radius \(0.06\)--\(0.18\) in the models of
Ref.~\cite{HOETH}) while the irreducible vertex
\(\Gamma^{(2)}\) must be closed separately; we return to this distinction in
the outlook.  The complete correction, moreover, is not determined by \(q\)
alone: by Eq.~\eqref{dS_exact} the natural descriptor of the fluctuation
sector is the conditional entropy functional
\(\mathcal H_x(\lambda):=\mathbb E[x\ln x\,|\,\lambda_n=\lambda]\), with
\(\Delta S=-\int d\lambda\,f^{a}(\lambda)\mathcal H_x(\lambda)\), of which
\(q(\lambda)=\mathbb E[x^2|\lambda]\) is only the leading second-moment
diagnostic.  The two-resolvent covariance identifies \(q\); determining the
complete conditional intensity distribution \(P(x|\lambda)\)---and hence
\(\mathcal H_x(\lambda)\)---requires the full covariance and vertex
hierarchy of Ref.~\cite{HOETH}, of which the Gaussian closure below and the
Gamma family of Appendix~\ref{fluct_entropy} are two explicit realizations.
The logical structure of the hierarchy is
\begin{equation}\label{hierdiag}
    \begin{array}{ccccc}
        \mathcal G^{(1)} & \longrightarrow & p^{\rm env}(\lambda)
        & \longrightarrow & S_{\rm env} \\[4pt]
        \mathcal G^{(2)} & \longrightarrow & q(\lambda)
        & & \\[4pt]
        \mathcal G^{(3)},\mathcal G^{(4)},\ldots
        & \longrightarrow & P(x|\lambda)
        & \longrightarrow & \mathcal H_x(\lambda)
        \;\longrightarrow\; \Delta S,
    \end{array}
\end{equation}
with \(\mathcal G^{(r)}\) denoting the successive sectors---the single
resolvent, the two-resolvent covariance, and the higher irreducible vertices
\(\Gamma^{(3)},\Gamma^{(4)},\ldots\) of the covariance hierarchy of
Ref.~\cite{HOETH}: each entropy object in this work is a
functional of a definite level of the hierarchy---the single resolvent for
the envelope, the two-resolvent covariance for the second moment, and the
higher vertices for the complete conditional intensity distribution.
Expanding \(\ln x\) around \(x=1\)
produces the cumulant series
\begin{equation}\label{dS_series}
    \Delta S = \sum_{r\ge2}\Delta S_r, \qquad
    \Delta S_r = \frac{(-1)^{r+1}}{r(r-1)}
    \sum_{n}p^{\rm env}_n\,\kappa_r(x_n),
\end{equation}
whose leading term is the second-order variance correction
\begin{equation}\label{dS2}
    \Delta S^{(2)}
    = -\frac12\sum_{n}\frac{\mathrm{Var}(p_n)}{p^{\rm env}_n}
    = -\frac12\sum_{n}p^{\rm env}_n\,[q(\lambda_n)-1],
\end{equation}
which equals \(-(q-1)/2\) when \(q(\lambda)\) is bulk-constant and
\(-\tfrac12\int d\lambda\,f^{a}(\lambda)[q(\lambda)-1]\) in the continuum
limit (Appendix~\ref{fluct_entropy}).  The linear term of the expansion
vanishes identically in expectation, its global counterpart being the exact
covariance conservation
\(\sum_{nm}\mathbb E[\delta p_n^a\delta p_m^b]=0\) of Ref.~\cite{HOETH}.

In the decorrelated (chaotic) regime the amplitude regularity conditions of
Ref.~\cite{HOETH} reduce the intra-shell amplitudes to Gaussian statistics:
\(c_n=(p^{\rm env}_n)^{1/2}\zeta_n\) with \(\zeta_n\sim\mathcal N(0,1)\) for real
\(H\), so that \(x_n\) follows the Porter--Thomas distribution~\cite{PT56}.
The Porter--Thomas factor is then \(q=3\), and the exact identity
\begin{equation}\label{qk4}
    q(\lambda) = 3 + \frac{\kappa_4(c_n\,|\,\lambda)}
                            {\mathbb E[c_n^2\,|\,\lambda]^2}
\end{equation}
identifies any deviation with the normalized fourth amplitude cumulant
\cite{HOETH}.  More generally, Gaussian amplitudes with Dyson index
\(\beta=1,2,4\) give \(x\sim\chi^2_\beta/\beta\), and
\begin{equation}\label{xlnx_exact}
    \mathbb E[x\ln x] = \psi\!\left(\tfrac\beta2+1\right)+\ln\frac2\beta,
\end{equation}
with \(\psi\) the digamma function (see Appendix~\ref{fluct_entropy} for the
derivation).  \emph{Within the Gaussian amplitude closure} the correction is
thus a symmetry-dependent constant, independent of the envelope parameters
and hence insensitive to the tail-branch structure of \cref{OMTB}:
\begin{equation}\label{SvN_universal}
    \mathbb E[S_{\rm vN}] = S_{\rm env} - c_\beta,
    \qquad
    c_\beta = \psi\!\left(\tfrac\beta2+1\right)+\ln\frac2\beta,
\end{equation}
with \(c_1=2-\gamma-\ln2\approx0.730\), \(c_2=1-\gamma\approx0.423\),
\(c_4=\tfrac32-\gamma-\ln2\approx0.230\) (\(\gamma\) the Euler--Mascheroni
constant).  We stress that Eq.~\eqref{SvN_universal} is a \emph{Gaussian
benchmark} rather than a generic prediction for finite-size
interacting systems: the constant \(c_1\) is universal \emph{within} the
Gaussian-amplitude closure, not a universal constant of generic chaotic
many-body systems, and the ED data below show that the present chain does
\emph{not} realize that limit (\(q\approx4.8\neq3\), \(\Delta S\approx-1.08
\neq-0.73\)).  In the Gaussian-amplitude limit itself,
\begin{equation}\label{mainresult}
    S_{\rm vN} \simeq S_{\rm env} - (2-\gamma-\ln2) \approx S_{\rm env} - 0.73
    \qquad (\beta=1),
\end{equation}
which is the mechanism behind the systematic gap in Fig.~\ref{FIG5}:
observational and von Neumann entropy share the same dominant scaling and
differ by an additive fluctuation constant, computed rather than postulated
\emph{within the Gaussian-amplitude closure}.

The exact diagonalization of the present model places it in the
\emph{mildly non-Gaussian} regime.  The energy-resolved Porter--Thomas factor
is flat across the bulk, \(q(\lambda)\approx4.8\) at \(N=15\) and
\(\approx5.0\) at \(N=14\) (Fig.~\ref{FIGqlambda}), corresponding to a
normalized fourth amplitude cumulant
\(\kappa_4(c)/\mathbb E[c^2]^2\approx1.8\)---between the Gaussian value
\(q=3\) and the strongly structured models of Ref.~\cite{HOETH}
(\(q\sim10^2\)).  The measured entropy correction is
\(S_{\rm vN}-S_{\rm env}=-1.08\,[-1.10,-1.06]\) at \(N=15\) and
\(-1.10\,[-1.15,-1.07]\) at \(N=14\) (median and interquartile range over
bulk channels, Fig.~\ref{FIGgap}), consistent with the exact decomposition
\eqref{dS_exact}, while the second-order truncation
\(-(q-1)/2\approx-1.9\) overshoots it by \(\sim80\%\), consistent with the
asymptotic character of Eq.~\eqref{dS_series} established analytically and
illustrated in Appendix~\ref{fluct_entropy}.  As a one-parameter
\emph{diagnostic} closure constrained by the measured \(q\), a Gamma intensity
model gives \(\Delta S_\Gamma=-\psi(q/(q-1))-\ln(q-1)\approx-1.12\), within
\(4\%\) of the ED correction at \(N=15\) (and \(4\%\)--\(9\%\) at all \(N\),
Appendix~\ref{fluct_entropy}): the three benchmarks---Gaussian
\(c_1\approx-0.73\), Gamma\((q)\approx-1.12\), ED \(-1.08\)---quantify the
accuracy obtained when the measured second moment is used to constrain the
one-parameter Gamma family (the Gaussian value captures only \(68\%\)), the
residual being carried by higher-order intensity statistics.

The finite-size trend of these diagnostics is informative
(Table~\ref{tabfluctN}, Fig.~\ref{FIGscaling}): over \(N=12\)--\(15\) the
bulk-median factor decreases monotonically,
\(q(N)\approx5.6,\,5.3,\,5.0,\,4.8\), and the measured correction as
\(\Delta S\approx-1.19,\,-1.15,\,-1.10,\,-1.08\): both drift systematically
toward their Gaussian-amplitude values \(q=3\) and \(-c_1\), providing
evidence for a finite-size drift of these low-order diagnostics toward their
Gaussian-amplitude values.  A
simple exponential fit gives \(q(N)-3\sim2^{-0.18N}\) and
\(\Delta S(N)+c_1\sim2^{-0.14N}\), but the four available sizes are
insufficient to distinguish this form from algebraic or \(1/N\)-type
finite-size corrections, and the fit is not intended as a determination of
the asymptotic scaling.  We emphasize that \(q\to3\) is only the necessary
\emph{low-order} evidence of amplitude Gaussianization: the full
Gaussianization statement is \(P(x|\lambda)\to P_{\rm PT}(x)\), which the
second moment alone cannot establish.  The measured excess is moreover not a
finite-dimensional Haar-sampling artifact: the GOE finite-size benchmark
\(q_{\rm GOE}(D)=3D/(D+2)\approx3-6/D\) deviates from \(3\) by only
\(\sim10^{-3}\) at the present dimensions---three to four orders of
magnitude below the measured \(q(N)-3\approx1.8\)--\(2.6\)---so the excess
intensity fluctuation is genuine dynamical structure rather than a
finite-\(D\) eigenvector correction (Appendix~\ref{fluct_entropy}).  The
direct comparison in
Fig.~\ref{FIGptscaling} shows that the complete conditional intensity
distribution remains visibly non-Porter--Thomas at both \(N=12\) and
\(N=15\) (heavy tail in both cases, with Kolmogorov--Smirnov distance
\(D_{\rm KS}\approx0.11\)--\(0.15\) at all four sizes), consistent with the
slow approach of the second moment and with the persistent mildly
non-Gaussian regime.  The Gamma diagnostic closure
\(-\psi(q/(q-1))-\ln(q-1)\) tracks the measured \(\Delta S\) to \(4\%\)--\(9\%\)
at every \(N\) (Appendix~\ref{fluct_entropy}).

Two remarks on Eq.~\eqref{dS_series} are essential.  First, its truncation at
\(r=2\) is \emph{not} uniformly accurate.  For Porter--Thomas statistics
\(\kappa_r(x)=(r-1)!\,2^{r-1}\beta^{1-r}\), so the terms grow factorially
(\(\Delta S_2=-1/\beta\), \(\Delta S_3=+4/3\), \(\Delta S_4=-4\) for
\(\beta=1\)) and the series is only asymptotic (Fig.~\ref{FIGfluct}): the
second-order estimate \(-(q-1)/2=-1\) overshoots the exact \(c_1\approx0.730\)
by \(37\%\) (by \(18\%\) and \(9\%\) for \(\beta=2,4\)).  On the present model
the same comparison reads \(-1.9\) versus \(-1.08\) (an \(80\%\) overshoot),
with measured cumulants \(\kappa_2\approx4.0\), \(\kappa_3\approx46\),
\(\kappa_4\approx1.2\times10^3\) and rapidly diverging partial sums
(Appendix~\ref{fluct_entropy}).  The exact
evaluation \eqref{xlnx_exact} is therefore both simpler and quantitatively
superior, mirroring the conclusion of Ref.~\cite{HOETH} that the fluctuation
content resides in the irreducible vertex rather than in ladder resummation:
the Gaussian amplitude closure supplies the full intensity distribution, not
merely its variance.  Second, beyond the Gaussian-amplitude limit the terms of
Eq.~\eqref{dS_series} are organized by the vertex hierarchy of
Ref.~\cite{HOETH}: \(\Delta S_3,\Delta S_4,\ldots\) are determined by the
higher irreducible vertices \(\Gamma^{(3)},\Gamma^{(4)},\ldots\)
(equivalently by the amplitude cumulants \(\kappa_3,\kappa_4,\ldots\)), so the
multi-resolvent hierarchy becomes an entropy-generating cumulant hierarchy.
In structured (non-ergodic) systems the Porter--Thomas factor is strongly
enhanced---up to \(q\sim10^2\) with comparable kurtosis in the non-ergodic
model of Ref.~\cite{HOETH}---and, within the Gamma family, the entropy
deficit grows accordingly; for Gamma-distributed intensities with a
prescribed \(q\), the exact correction is
\(\Delta S=-\psi(q/(q-1))-\ln(q-1)\), growing only logarithmically
(\(\sim-\ln(q-1)+\gamma\)), while the second-order truncation \(-(q-1)/2\)
diverges linearly: the cumulant expansion breaks down precisely where
non-Gaussianity matters most.

The leading estimate of the fluctuation of \(S_{\rm vN}\) around its mean is
supplied by the covariance ridge and is exponentially small,
\(\mathrm{Var}(S_{\rm vN})\approx(q-1)\sum_n (p^{\rm env}_n)^2(1+\ln p^{\rm env}_n)^2
=O(e^{-S}S^2)\) (the ridge estimate evaluates to \(0.185^2\) at \(N=15\) and
\(0.247^2\) at \(N=14\) for the interacting chain,
Appendix~\ref{fluct_entropy}; the synthetic Porter--Thomas benchmark of
Table~\ref{tabfluct} gives \(0.153^2\) versus \(0.164^2\)).  If the envelope
entropy is extensive and the fluctuation factor \(q\) remains subexponential
in \(N\), this estimate implies self-averaging of the additive correction
\eqref{mainresult} as \(N\to\infty\).
This estimate concerns the fluctuations generated by the intra-shell
intensity statistics at fixed envelope and fixed Hamiltonian; it is not the
variance over Hamiltonian ensembles.
The R\'enyi-2 entropy of Sec.~\ref{OscSec} receives the parallel shift
\(S_2\simeq S_2^{\rm env}-\ln q\), confirmed by the data with
\(S_2-S_2^{\rm env}=-1.59\) versus \(-\ln q\approx-1.57\) at \(N=15\)
(\(-1.64\) versus \(-1.62\) at \(N=14\)): the oscillation bound
Eq.~\eqref{upb} is weakened by the factor \(q\approx4.8\) without affecting
the exponential suppression expected for extensive \(S_2(\omega)\).

The direct numerical test of Eq.~\eqref{dS_exact} is reported in
Appendix~\ref{fluct_entropy}: binning the ED overlaps \(p_n^{\mu i}\) in
energy shells yields the exact envelope \(p^{\rm env}_n\), \(q(\lambda)\),
\(\mathbb E[x\ln x]\), and the cumulant
hierarchy, all in agreement with the theory above; the small-scale GOE and
synthetic Porter--Thomas benchmarks reported there verify the Gaussian
benchmark \(c_1\approx0.73\) to high accuracy.

To summarize the status of the argument: (i) the decomposition
Eq.~\eqref{dS_exact} is an exact identity; (ii) Eq.~\eqref{diagbin} is an
exact spectral identification of the fluctuation parameter with the diagonal
ridge of the two-resolvent covariance; (iii) within the Gaussian amplitude
closure \(\Delta S=-c_\beta\) is exact, providing the benchmark
\(c_1\approx0.73\); (iv) the ED data verify (i) and (ii) and measure the
non-Gaussian values \(q(\lambda)\approx4.8\), \(\Delta S\approx-1.08\), and
(v) the microscopic closure \(\mathrm{Hamiltonian}\to\mathcal C^{(2)}\to
q(\lambda)\)---the \emph{prediction} of the ridge weight from the Hamiltonian
rather than its identification---remains the open Level-II problem of the
covariance hierarchy~\cite{HOETH} and is left for future work.
A compact form of the entire structure is
\begin{align}\label{unified}
    p_n = p^{\rm env}_n x_n: 
    S_{\rm env} = -\sum_n p^{\rm env}_n\ln p^{\rm env}_n,\notag\\
    S_{\rm vN} = S_{\rm env} - \sum_n p^{\rm env}_n\,\mathbb E[x\ln x\,|\,\lambda_n],、\notag\\
    S_2 = S_2^{\rm env} - \ln \mathbb E[x^2\,|\,\lambda],
\end{align}
i.e., the different entropy measures probe different functionals of one and
the same conditional intensity distribution \(P(x|\lambda)\).

\section{Oscillations Around Steady State}
\label{OscSec}
Here the steady state denotes the long-time dephased state \(\omega\); the instantaneous state continues to fluctuate around it. For any bounded observable \( A \), the temporal fluctuation amplitude is quantified by:
\begin{equation}
	(\Delta \omega)^2_A := \lim_{t\to \infty} \frac{1}{t}\int_0^{t} d\tau \left\{\Tr[ A \rho(\tau) - A \omega ]\right\}^2.
\end{equation}
Following \cite{S11,WGRE19}, this fluctuation amplitude is upper bounded by the R\'enyi-2 entropy of the steady state \( \omega \):
\begin{equation}\label{upb}
	(\Delta \omega)^2_A \leq \norm{A}^2 e^{-S_2(\omega)},
\end{equation}
where \( S_\alpha \) denotes the R\'enyi entropy:
\begin{equation}
	S_\alpha(\rho) = \frac{1}{1-\alpha}\ln[\Tr(\rho^\alpha)].
\end{equation}
For properties of the R\'enyi entropy in specific systems, see \cite{LLT24}.

Now we analyze the R\'enyi-2 entropy corresponding to the Lorentzian ansatz (\ref{Lorentz}). Substituting \cref{decsta} into the definition of the R\'enyi-2 entropy, and using the smoothed distribution \( p^{\mu i}(\lambda) \) like \cref{aproent}, we get
\begin{align}
	S_2(\omega)=-\ln[\sum_n (p^{\mu i}_n)^2 ]\to  S_2(p^{\mu i}) \notag                                                                                            \\
	=-\ln\left[\int d\lambda \frac{1}{e^{S(\lambda)}} \frac{1}{\pi^2} \frac{\chi_{\mu i}^2}{\left(\delta \lambda_{\mu i}^2 + \chi_{\mu i}^2\right)^2}\right]\notag \\
	\approx-\ln\left[ \int d\lambda  \frac{1}{\pi^2} \frac{\chi_{\mu i}^2}{\left(\delta \lambda_{\mu i}^2 + \chi_{\mu i}^2\right)^2}\right]\notag                  \\
	-\ln(\frac{1}{e^{S( a_{\mu i} +\Delta_{\mu i})}})=S(a_{\mu i} + \Delta_{\mu i}) + \ln(2\pi \chi_{\mu i}),
\end{align}
This R\'enyi-2 entropy is marginally smaller than the Shannon entropy (\ref{finalentro}). The intra-shell fluctuations of Sec.~\ref{IntraShell} shift it further down by an additive constant: writing \(p_n=p^{\rm env}_n x_n\), the participation sum is self-averaging, \(\sum_n p_n^2=q\sum_n (p^{\rm env}_n)^2\,[1+O(e^{-S/2})]\), so \(S_2(\omega)\simeq S_2^{\rm env}-\ln q\); this relation holds for \emph{any} intensity statistics and is confirmed by the data with \(S_2-S_2^{\rm env}=-1.59\) versus \(-\ln q\approx-1.57\) at \(N=15\) (\(-1.64\) versus \(-1.62\) at \(N=14\), Appendix~\ref{fluct_entropy}), serving as an independent consistency check of the same intra-shell fluctuation mechanism, and weakening the bound \cref{upb} by the factor \(q\approx4.8\) without affecting the exponential suppression expected for extensive \(S_2(\omega)\).
If \(S_2(\omega)\) is extensive---as holds for the LG estimate \(S_2(\omega)\simeq S(a_{\mu i}+\Delta_{\mu i})+\ln(2\pi\chi_{\mu i})\), whose density-of-states term is extensive in \(N\)---the upper bound in \cref{upb} decays exponentially with \(N\). This implies that post thermalization, the system executes negligible oscillations about the steady state. Consequently, the steady-state entropy effectively characterizes the system's thermodynamic behavior.


\section{Conclusion and Discussion}\label{CD}

We have developed a self-consistent mechanism for steady-state entropy production in isolated quantum systems governed by interactions with statistically independent phases.
The cornerstone of our theoretical framework is the derivation of self-consistent equations for the probability distribution \(p^{\mu i}_n\), which bridge microscopic interaction details and emergent thermodynamic behavior.
The resolvent-based formulation, whose full nonperturbative hierarchy is developed in Ref.~\cite{HC26}, allows us to quantitatively characterize entropy generation through self-consistent equations, moving beyond phenomenological descriptions.

A central achievement of this work is the solution of these self-consistent equations through a hierarchical ansatz strategy~\cite{HC26}.
We first employed a Lorentzian ansatz for the bulk distribution, reducing the problem to determining the width \(\chi_{\mu i}\) and shift \(\Delta_{\mu i}\) via Eqs.~(\ref{consist}) and (\ref{consist2}).
The agreement between numerically extracted parameters and those reconstructed from the self-consistent equations (Fig.~\ref{FIG3}) provides strong validation.
This framework directly yields the entropy scaling \(S(\omega) \approx S(a_{\mu i} + \Delta_{\mu i}) + \ln(4\pi \chi_{\mu i})\), revealing the logarithmic dependence of entropy on interaction-induced broadening.

To refine the characterization, we extend the self-consistent approach to the distribution tails.
A Gaussian ansatz captures the faster decay at large energy differences, governed by the tail self-consistency condition Eq.~\eqref{tailconsis}.
Numerical validation (Figs.~\ref{FIGgf}, \ref{FIGgp}, \ref{sigdis}) confirms its robustness and reveals a splitting of the distribution into distinct branches, explainable by the symmetry-breaking mechanism analyzed in \cref{OMTB}.
To unify peak and tail descriptions, we proposed an enhanced hybrid Lorentzian-Gaussian ansatz, Eq.~\eqref{LGdistr}, which synthesizes the sharp central peak with the rapidly decaying tail.
The self-consistent equations for the LG parameters, Eqs.~\eqref{EAFC1}--\eqref{EAFCfar}, yield reconstructed parameters (Eq.~\eqref{LGpara}) in good agreement with direct fits (Fig.~\ref{FIGLGconsis}), and the resulting entropy estimate \(S \approx S(\lambda_p) + 1.54\) (with \(H \approx 1.54\)) refines the pure Lorentzian prediction \(S \approx S(a_{\mu i} + \Delta_{\mu i}) + 2.07\) (with \(\ln(4\pi\chi) \approx 2.07\)), bringing the theoretical curve in Fig.~\ref{FIG5} into close alignment with the observational entropy. The Lorentzian estimate systematically overestimates the entropy because it assigns excessive weight to the broad tails, whereas the hybrid profile suppresses the tails and yields values closer to the observational entropy.

The success of these ans\"atze rests on a two-layer structure characteristic of the full resolvent hierarchy of Ref.~\cite{HC26}:
the mean-field layer, built from the random-phase cancellation of cross terms, provides the dominant self-consistent equations for the Lorentzian core; the fluctuation layer, consisting of the systematically expandable 
multi-resolvent contributions \(\mathcal{G}^{\mathrm{CC}}_{\mu i} 
= \sum_{\ell \ge 3} \mathcal{G}^{(\ell)}_{\mu i}\),
accounts for the Gaussian tails and the spectral skewness 
\(\Delta'_{\mu i} \neq \Delta_{\mu i}\).
The observed branch splitting in the tail parameters, meanwhile, 
appears to arise from an interplay between symmetry-related bath 
structures (Appendix~\ref{OMTB}) and the multi-branch solutions 
of the beyond-mean-field self-consistency equations. 
Together, they form a unified framework treating typical behavior and statistical fluctuations on an equal footing.

The covariance sector completes the bridge between the two entropy measures. The smoothed envelope of the hierarchy generates the observational entropy, while the diagonal bin of the same-channel resolvent covariance supplies the intra-shell intensity variance \(V(\lambda)=[q(\lambda)-1]p^{\rm env}(\lambda)^2\), which gives the leading fluctuation correction to the von Neumann entropy. In the Gaussian-amplitude limit this correction approaches the symmetry-dependent Gaussian benchmark \(2-\gamma-\ln2\approx0.73\), explaining why \(S_{\rm obs}\) and \(S_{\rm vN}\) share the same leading scaling yet differ by an \(O(1)\) offset; the exact diagonalization of the present chain measures instead \(q(\lambda)\approx4.8\) and a correction of \(1.08\), a mildly non-Gaussian regime in which the second-order truncation \((q-1)/2\approx1.9\) overshoots by \(\sim80\%\)---a quantitative entropy-side diagnostic of the deviation from eigenstate Gaussianity, which in strongly structured systems amplifies the deficit further. Entropy is intrinsically sensitive to the full intensity distribution rather than to a finite number of its moments: the variance supplied by the two-resolvent sector is not a controlled approximation to the entropy functional, which is precisely why the higher vertices of the hierarchy matter.

Our self-consistent formalism thus provides a unified framework relating observational entropy to entropy measures derived from the decohered steady state: the smoothed distribution captures the former (Fig.~\ref{FIG5}), while the covariance sector yields the latter through the exact fluctuation correction \(S_{\rm vN}=S_{\rm env}-\sum_n p^{\rm env}_n\mathbb E[x_n\ln x_n]\)---the Gaussian benchmark \(2-\gamma-\ln2\approx0.73\) in the Gaussian-amplitude limit, and \(1.08\) for the mildly non-Gaussian statistics of the present chain (Sec.~\ref{IntraShell}, Appendix~\ref{fluct_entropy}), with the remaining corrections organized by the irreducible vertex hierarchy~\cite{HOETH}.

Higher-order single-resolvent corrections refine the envelope parameters,
whereas the entropy deficit itself is governed by the covariance and higher
fluctuation sectors; the present results therefore isolate the latter as the
essential microscopic ingredient.

Our analytical framework fundamentally relies on the random-phase condition, under which off-diagonal matrix elements behave as random variables whose fluctuating phases dominate systematic biases.
This condition allows cross-correlated terms to average out, enabling a closed set of self-consistent equations.
In special cases where this condition is violated---such as in strictly integrable or many-body localized (MBL) systems---these simplifications no longer apply.
Integrable models possess extensive conserved quantities and admit explicit solutions; MBL systems strongly violate the random-phase condition and exhibit anomalously slow thermalization, so the logarithmic \(\ln\chi\) contribution to the entropy predicted here is not expected to hold.
Characterizing entropy production beyond the random-phase paradigm remains an important subject for future research.

Future directions emerging from this work include:
(i) extending the energy-resolved fluctuation diagnostics of Sec.~\ref{IntraShell}---whose \(N=12\)--\(15\) scaling is reported in Appendix~\ref{fluct_entropy}---to larger system sizes and to the strongly structured (non-ergodic) regime, where the cumulant hierarchy predicts a growing entropy deficit;
(ii) exploring multi-scale entanglement dynamics during prethermalization;
(iii) developing experimental protocols to measure the key parameters of our models;
(iv) extending the self-consistent analysis to larger system sizes and a wider class of nonintegrable models to test the universality of the hierarchical ansatz framework;
and (v) employing the multi-resolvent expansion and Lanczos continued-fraction techniques developed in the full nonperturbative hierarchy of Ref.~\cite{HC26} to derive more accurate self-consistent equations beyond the mean-field level; for applications demanding higher precision, these refined self-consistent equations can be used to determine the parameters of the ansatz distributions with greater accuracy;
(vi) establishing the microscopic closure of the irreducible ridge weight \(q(\lambda)-1\)---the Level-II problem of the covariance hierarchy~\cite{HOETH}---which would convert the exact spectral identification of Sec.~\ref{IntraShell} into a first-principles prediction of the entropy correction \(\Delta S\) from the Hamiltonian.

\begin{acknowledgments}
	This work is supported by the National Natural Science Foundation of
	China under Grant No. 12305035 and Innovational Fund of Hainan Province (Grant No. KJRC2023B11).
\end{acknowledgments}

\appendix
\section{Origin and Manifestation of the Two Branches}\label{OMTB}

The emergence of two distinct branches in the Gaussian tail parameters, shown in Figs.~\ref{FIGgp} and \ref{FIGlgp}, originates from the interplay between the reflection symmetry of the unperturbed bath and the local system-bath coupling.
In the absence of coupling, the open-chain bath Hamiltonian \(H_B\) is invariant under reflection about the chain midpoint (\(j \leftrightarrow N-j\) for the bath sites), and its eigenstates have definite reflection parity. This symmetry gives rise to exact or near degeneracies between parity-related pairs of eigenstates whose amplitude is concentrated near the two defect sites.
The local interaction \(V = -\sigma_z^S \otimes (\sigma_z^1 + \sigma_z^{14})\) explicitly breaks the reflection symmetry by coupling the system spin to the two defect sites, effectively imposing an open-boundary-like perturbation.

For a fixed system state \(i\) and bath energy \(\epsilon_\mu\), the bath eigenstates that couple most strongly are those with significant amplitude on the two defect sites.
Reflection symmetry implies that these come in pairs related by reflection about the chain midpoint.
The local perturbation hybridizes these nearly degenerate states into symmetric and antisymmetric combinations with respect to the two defect sites, which couple differently to the system spin, leading to the two observed branches~\cite{CH11,NZH17,PL21}.

An intriguing observation is that while the Gaussian parameters \(\Delta_{\mu i}'\) and \(\sigma_{\mu i}\) exhibit a clear two-branch structure, the Lorentzian parameters \(\Delta_{\mu i}\) and \(\chi_{\mu i}\) remain largely single-valued (Fig.~\ref{FIGlgp}).
This dichotomy admits a direct explanation from the structure of the self-consistent equations.
The fundamental equation for the Lorentzian width, Eq.~\eqref{consist}, depends on the interaction matrix elements exclusively through the integrated spectral weight
\begin{equation}
    \mathcal{V}_{\mu i,j} = \sum_{\nu} |V_{\mu i,\nu j}|^2,
\end{equation}
rather than on the detailed connectivity between individual branches.
When the bath eigenstates are reorganized into symmetric and antisymmetric combinations relative to the defect sites, the total spectral weight \(\mathcal{V}_{\mu i,j}\) within the energy shell is preserved---the sum is merely partitioned differently among the branches.
Consequently, as long as the two branches lie at similar detunings (\(\delta_A \simeq \delta_B\)), the self-consistent equation effectively senses only the total weight, and the Lorentzian core parameters \((\chi_{\mu i}, \Delta_{\mu i})\) remain largely single-valued.
The Gaussian parameters \((\Delta'_{\mu i}, \sigma_{\mu i})\), by contrast, govern the tail behavior described by Eq.~\eqref{tailconsis}, where the distribution of the spectral weight among branches matters because the tail probes energy differences comparable to or larger than the branch separation.
This explains why the tail parameters split into two distinct branches while the core parameters do not.

From the perspective of the self-consistent framework, the central peak parameters satisfy a single set of equations that averages over symmetry classes, while the tail parameters obey two distinct self-consistent relations corresponding to the symmetric and antisymmetric branches.
This also clarifies why the dominant entropy scaling \(\sim \ln\chi_{\mu i}\) is robust against branch splitting: the parameter \(\chi_{\mu i}\) is controlled by the invariant total spectral weight, so the logarithmic scaling is unaffected by the branch reorganization; the branch structure only modifies the constant term of the entropy by order unity.

\section{Cross-Correlated Terms and the Multi-Resolvent Hierarchy}\label{cross_structure}

In the main text, the random-phase condition was invoked to argue that cross-correlated contributions vanish at the ensemble-average level, leading to the closed mean-field equations \eqref{consist}--\eqref{consist2}.
While this approximation accurately captures the bulk distribution and leading entropy scaling, the actual overlaps exhibit fluctuations, tail structure, and the branch splitting discussed in \cref{OMTB}---all of which originate from the very cross terms that average to zero.

A systematic treatment of these cross terms is provided in Ref.~\cite{HC26}, where the projected self-energy \(\mathcal{G}_{\mu i}\) is shown to admit an exact recursive decomposition
\[
\mathcal{G}^{\mathrm{CC}}_{\mu i} = \sum_{\ell \ge 3} \mathcal{G}^{(\ell)}_{\mu i},
\]
with the leading term
\[
\mathcal{G}^{(3)}_{\mu i} = \sum_{\xi k \neq \nu j \neq \mu i} V_{\mu i,\xi k} V_{\xi k,\nu j} V_{\nu j,\mu i} \, \mathcal{R}_{\mu i}(z) \mathcal{R}_{\xi k}(z)
\]
expressed purely in terms of products of diagonal resolvents.
The real and imaginary parts of \(\mathcal{G}^{(3)}_{\mu i}\) involve Hilbert transforms of the probability distributions, generating the spectral skewness (\(\Delta'_{\mu i} \neq \Delta_{\mu i}\)) and the corrections to the tail decay rate that are observed in our numerical data.
We refer the reader to Ref.~\cite{HC26} for the complete recursive construction, the systematic connection to higher-order ETH correlations, and the analysis of how these terms modify the self-consistent equations beyond the mean-field level.

\section{Hilbert Transform of the Hybrid Ansatz}\label{HTEA}

We provide a compact derivation of the principal-value integral identity used in Eq.~\eqref{REGui} of the main text.
Let \(G(x;\sigma) = e^{-x^2/(2\sigma^2)}/(\sqrt{2\pi}\sigma)\) be the normalized Gaussian and \(L(x;\chi) = \chi/[\pi(\chi^2+x^2)]\) the Lorentzian.
Define the normalized Voigt profile \(V(x_0;\sigma,\chi) = \int dx' G(x';\sigma) L(x_0 - x'; \chi)\) and the dispersion profile \(D(x;\sigma,\chi) = \frac{1}{\sqrt{2\pi}\sigma} \Im\, w\!\bigl(\frac{x+\mathrm{i}\chi}{\sqrt{2}\sigma}\bigr)\), where \(w(z) = e^{-z^2}\operatorname{erfc}(-\mathrm{i}z)\) is the Faddeeva function.

With \(x_0 = \mu_2 - \mu_1\), \(x' = \lambda' - \mu_1\), \(\Delta = \lambda' - \mu_2\), and \(a = \mu_2 + \mathrm{i}\chi\), the integral of interest is
\[
S := \operatorname{PV}\! \int_{-\infty}^{\infty} \frac{G(\lambda - \mu_1; \sigma) L(\lambda - \mu_2; \chi)}{\lambda' - \lambda} \, d\lambda.
\]

Using the partial-fraction representation \(L(\lambda-\mu_2;\chi) = \frac{1}{2\pi\mathrm{i}} \bigl(\frac{1}{\lambda-a} - \frac{1}{\lambda-\bar a}\bigr)\) and the algebraic identity
\(\frac{1}{(\lambda'-\lambda)(\lambda-a)} = \frac{1}{\lambda'-a}\bigl(-\frac{1}{\lambda-\lambda'} + \frac{1}{\lambda-a}\bigr)\),
the integral reduces to a linear combination of the kernel \(I(z) = \operatorname{PV}\! \int \frac{G(\lambda-\mu_1;\sigma)}{z-\lambda} d\lambda\).
Evaluating \(I(z)\) via the Faddeeva function---\(I(z) = -\mathrm{i}\pi [V + \mathrm{i}D]\) for \(\Im z > 0\) and \(I(\lambda') = \pi D(x';\sigma,0)\) on the real axis---and simplifying, one obtains the compact result
\begin{equation}\label{eq:final_HT_faddeeva}
    \boxed{
        \frac{S}{V(x_0;\sigma,\chi)}
        = \frac{\Delta + \chi\,\frac{D(x';\sigma,0) - D(x_0;\sigma,\chi)}{V(x_0;\sigma,\chi)}}{\Delta^2 + \chi^2}.
    }
\end{equation}
This identity, which is used in Eq.~\eqref{REGui} of the main text and in the construction of the self-consistent equation \eqref{LGsimpcon}, generalizes the simple Lorentzian Hilbert transform by introducing a dispersive correction proportional to \(\chi\) and to the difference of dispersion profiles.
In the limit \(\sigma \to \infty\), \(D \to 0\) and the standard single-pole result is recovered.
A more detailed step-by-step derivation and an analysis of the Kramers-Kronig structure can be found in Ref.~\cite{HC26}.

\section{Effective Self-Energy Representation of the Hybrid Profile}\label{ESrVN}

The hybrid LG distribution \eqref{LGdistr} can be equivalently represented as a spectral function with an effective energy-dependent self-energy, which is technically advantageous for imposing causality constraints.
Starting from the normalized Voigt-type profile
\[
    f_V(\lambda) = \frac{L(\lambda - \epsilon_L; \chi) G(\lambda - \epsilon_G; \sigma)}{V(\epsilon_L - \epsilon_G; \sigma, \chi)},
\]
one constructs the approximation
\begin{equation}\label{eq:voigt_selfenergy_equiv}
    f_V(\lambda) \simeq \frac{1}{\pi} \Im \frac{1}{\lambda - \epsilon_{\mathrm{eff}} - \mathrm{i}\chi_{\mathrm{eff}} \, w\!\bigl(\frac{-\lambda + \epsilon_G}{\sqrt{2}\sigma}\bigr)},
\end{equation}
where \(w(z)\) is the Faddeeva function.
The effective parameters \(\epsilon_{\mathrm{eff}}\) and \(\chi_{\mathrm{eff}}\) are determined by matching the peak position and peak height of the two representations.

The peak position of \(f_V\) is, to leading order for \(|\epsilon_L - \epsilon_G| \lesssim \chi\),
\begin{equation}\label{eq:voigt_peak}
    \lambda_p \approx \frac{2\sigma^2 \epsilon_L + \chi^2 \epsilon_G}{2\sigma^2 + \chi^2}.
\end{equation}
Equating this with the peak position obtained from Eq.~\eqref{eq:voigt_selfenergy_equiv} and matching the peak heights yields two equations that uniquely fix \(\epsilon_{\mathrm{eff}}\) and \(\chi_{\mathrm{eff}}\).
This procedure ensures that the effective spectral representation reproduces the key features of the hybrid profile while remaining manifestly causal (the Faddeeva function is analytic in the appropriate half-plane).
Numerical comparisons confirm that the agreement extends beyond the immediate peak region.
The broader analytic structure of this representation---including its role in a continued-fraction backbone and as a building block for the full resolvent hierarchy---is analyzed in detail in Ref.~\cite{HC26}.

\section{Derivation of the Entropy Contribution}\label{DAE}

We derive the entropy contribution \(H(\Delta, \sigma, \chi)\) appearing in Eq.~\eqref{LGentro}.
Starting from the kernel
\[
    K(\varepsilon) = \frac{1}{\sqrt{2\pi}\,\sigma} e^{-\varepsilon^2/(2\sigma^2)} \, \frac{1}{\pi} \frac{\chi}{(\varepsilon - \Delta)^2 + \chi^2},
\]
and the normalized density \(p(\varepsilon) = K(\varepsilon)/V\) with \(V = V(\Delta; \sigma, \chi)\), the Shannon entropy is
\[
    S = -\int p(\varepsilon) \ln p(\varepsilon) d\varepsilon
      = -\int p(\varepsilon) \ln K(\varepsilon) d\varepsilon + \ln V.
\]
Expanding and defining the moments
\begin{align}
	    M_0 = \int K(\varepsilon) d\varepsilon, \quad
    M_2 = \int K(\varepsilon) \varepsilon^2 d\varepsilon, \notag\\
    N_{\ln} = \int K(\varepsilon) \ln\!\bigl((\varepsilon - \Delta)^2 + \chi^2\bigr) d\varepsilon,
\end{align}
the entropy identity becomes
\begin{equation}\label{eq:S_exact}
    S = \frac12 \ln(2\pi\sigma^2) - \ln\frac{\chi}{\pi} + \ln M_0 + \frac{M_2}{2\sigma^2 M_0} + \frac{N_{\ln}}{M_0}.
\end{equation}

Introducing the dimensionless variables $s = \sigma/\Delta$, $\eta = \chi/\Delta$, and the Faddeeva function $w(z)$, the moments $M_0$ and $M_2$ can be evaluated exactly in closed form:
\begin{align}
    M_0 &= \frac{1}{\sigma\sqrt{2\pi}} \Re(\wp), \quad \wp = w\!\Bigl(\tfrac{1+\mathrm{i}\eta}{\sqrt{2}\,s}\Bigr), \label{eq:M0}\\
    M_2 &= \frac{1}{\sigma\sqrt{2\pi}} \bigl[(\Delta^2 - \chi^2)\Re(\wp) - 2\Delta\chi\,\Im(\wp)\bigr] + \frac{\chi}{\pi}. \label{eq:M2}
\end{align}

For the logarithmic moment $N_{\ln}$, an elementary closed form is unavailable. We provide an exact one-dimensional integral representation via parametric integration. By employing the integral identity $[(\varepsilon-\Delta)^2+\chi^2]^{\alpha-1} = \frac{1}{\Gamma(1-\alpha)}\int_0^{\infty} t^{-\alpha} e^{-t[(\varepsilon-\Delta)^2+\chi^2]}dt$ and differentiating with respect to $\alpha$ at $\alpha=0$, the spatial integral over $\varepsilon$ can be evaluated analytically. This yields the exact relation
\begin{equation}\label{eq:Nln_exact}
    \frac{N_{\ln}}{M_0} = -\gamma + \ln(2\sigma^2) 
    - \frac{\displaystyle\int_0^{\infty} \frac{\ln u}{\sqrt{1+u}} \, e^{-\frac{u}{2s^2}\left[\eta^2 + \frac{1}{1+u}\right]} du}
           {\displaystyle\int_0^{\infty} \frac{1}{\sqrt{1+u}} \, e^{-\frac{u}{2s^2}\left[\eta^2 + \frac{1}{1+u}\right]} du},
\end{equation}
where $\gamma \approx 0.5772$ is the Euler--Mascheroni constant. Both the numerator and denominator integrals share a well-behaved positive weight and converge exponentially, rendering standard numerical quadrature straightforward and stable to machine precision.

Combining Eqs.~\eqref{eq:M0}--\eqref{eq:Nln_exact}, the total Shannon entropy simplifies to the following exact formulation:
\begin{equation}\label{eq:DAE}
    \boxed{
        S = \ln(\pi\chi) + \ln\Re(\wp) + \frac{M_2}{2\sigma^2 M_0} - \gamma + \ln(2\sigma^2) - \frac{I_1}{I_0},
    }
\end{equation}
where $I_0$ and $I_1$ represent the denominator and numerator integrals in Eq.~\eqref{eq:Nln_exact}, respectively, and the exact ratio for the second moment is given by
\begin{equation}
    \frac{M_2}{2\sigma^2 M_0} = \frac{1-\eta^2}{2s^2} - \frac{\eta}{s^2}\frac{\Im(\wp)}{\Re(\wp)} + \frac{\eta}{s\sqrt{2\pi}\,\Re(\wp)}.
\end{equation}

For the representative parameters of our model (\(\Delta - \Delta' = \pm 1.8\), \(\sigma = 2.0\), \(\chi = 0.75\), cf.\ Fig.~\ref{FIGlgp}), numerical quadrature of Eq.~\eqref{eq:DAE} yields $S \approx 1.54$ for the continuous-distribution entropy \(H\) appearing in Eq.~\eqref{LGentro}. The full entropy is therefore \(S(p^{\mu i}) \approx S(\lambda_p) + 1.54\).
This is significantly smaller than the pure Lorentzian estimate \(\ln(4\pi\chi) \approx 2.07\), reflecting the entropy reduction due to the Gaussian suppression of the tails.

\section{Entropy Correction from the Resolvent Covariance Hierarchy}
\label{fluct_entropy}

This appendix supplies the derivations behind Sec.~\ref{IntraShell} and the
small-scale numerical verification of the fluctuation identities.

\subsection{Exact decomposition and the cumulant expansion}

With \(p_n=p^{\rm env}_n x_n\) and \(p^{\rm env}_n=\mathbb E_{\rm shell}[p_n|\lambda_n]\),
so that \(\mathbb E_{\rm shell}[x_n|\lambda_n]=1\) by construction, the entropy difference
is
\[
    \Delta S = -\sum_{n}p^{\rm env}_n\,\mathbb E[x_n\ln x_n],
\]
since \(-\mathbb E[p_n\ln p_n]=-p^{\rm env}_n\ln p^{\rm env}_n-p^{\rm env}_n\mathbb E[x_n\ln x_n]\)
and the linear term \(p^{\rm env}_n\ln p^{\rm env}_n\,\mathbb E[x_n-1]\)
vanishes.  This is the exact identity of Eq.~\eqref{dS_exact}: the empirical
envelope \(p^{\rm env}_n\) enters exactly, and the LG profile of
Eq.~\eqref{LGdistr} is only its approximation,
\(p^{\rm env}_n\approx p^{\rm LG}_n=e^{-S(\lambda_n)}f^{\mu i}(\lambda_n)\).
In the ED data the two agree closely: for the reference channels the ansatz
error is \(|S_{\rm env}-S_{\rm LG}|\lesssim0.05\), an order of magnitude below
the fluctuation correction \(|\Delta S|\approx1.08\), which validates the
three-level decomposition \eqref{threelevel}.  The
Taylor series \((1+u)\ln(1+u)=\sum_{s\ge2}(-1)^s u^s/[s(s-1)]\) around
\(u=0\) yields the moment expansion
\[
    \Delta S = \sum_{s\ge2} \frac{(-1)^{s+1}}{s(s-1)}
    \sum_{n}p^{\rm env}_n\,\mathbb E[(x_n-1)^s],
\]
which reorganizes into the cumulant series of Eq.~\eqref{dS_series}; the
moment and cumulant forms coincide for \(s=2,3\) and differ at \(s\ge4\) only
by a regrouping of lower-order terms within the same asymptotic series.  The
leading term reproduces Eq.~\eqref{dS2}; in the dense-spectrum (continuum)
approximation, with
\(p^{\rm env}_n=e^{-S(\lambda_n)}f(\lambda_n)\) and
\(\sum_np^{\rm env}_n=\int d\lambda\,f(\lambda)=1\) (the support is essentially
the full spectrum in the present model, see Sec.~\ref{fluct_num}), it reads
\(\Delta S^{(2)}=-\tfrac12\int d\lambda\,f(\lambda)[q(\lambda)-1]\).

The two-resolvent identification follows from the resolvent covariance
\(\mathcal C_{aa}(z,z')=\mathbb E[\delta\mathcal R_a(z)\delta\mathcal R_a(z')]\),
whose spectral representation and diagonal bin are given in
Eq.~\eqref{diagbin}; the same object admits the ridge decomposition
\(\mathcal C_{aa}(\lambda,\lambda')=\delta(\lambda-\lambda')\mathcal W_{aa}(\lambda)
+\mathcal C^{\rm reg}_{aa}\) with ridge weight
\(\mathcal W_{aa}(\lambda)=[q(\lambda)-1]e^{-2S}\rho(\lambda)f^a(\lambda)^2\)
\cite{HOETH}.  The entropy correction is thus a spectral functional of the
two-resolvent covariance, \(\Delta S^{(2)}=-\tfrac12\int d\lambda\,
e^{S(\lambda)}\mathcal W_{aa}(\lambda)/f^a(\lambda)\).  The vanishing of the
linear term is the entropy-side counterpart of the exact covariance
conservation \(\sum_{nm}\mathbb E[\delta p_n^a\delta p_m^b]=0\) of
Ref.~\cite{HOETH}: the diagonal and off-diagonal bins of the covariance carry
compensating total weight.  The leading estimate of the fluctuation of the
entropy itself is supplied by the covariance ridge: the
linear term \(\sum_n(1+\ln p^{\rm env}_n)\delta p_n\) has variance
\((q-1)\sum_n(p^{\rm env}_n)^2(1+\ln p^{\rm env}_n)^2\) from the ridge plus an
off-diagonal contribution bounded by covariance conservation, both
\(O(e^{-S})\), so \(\mathrm{Var}(S_{\rm vN})=O(e^{-S}S^2)\); if the envelope
entropy is extensive and \(q\) remains subexponential, the additive
correction is self-averaging as \(N\to\infty\).

\subsection{Gaussian amplitude closure and the Porter--Thomas constant}

Under the amplitude regularity conditions of Ref.~\cite{HOETH}, the intra-shell
amplitudes are Gaussian: \(c_n\sim\mathcal N(0,p^{\rm env}_n)\) per shell for real
\(H\).  Then \(x_n=c_n^2/p^{\rm env}_n\sim\chi^2_1\), and for general Dyson index
\(\beta\), \(x\sim\chi^2_\beta/\beta=\mathrm{Gamma}(\beta/2,2/\beta)\).
Using \(\mathbb E[x^s]=(2/\beta)^s\Gamma(s+\beta/2)/\Gamma(\beta/2)\) and
differentiating at \(s=1\),
\[
    \mathbb E[x\ln x] = \frac{d}{ds}\mathbb E[x^s]\Big|_{s=1}
    = \psi\!\left(\tfrac\beta2+1\right)+\ln\frac2\beta,
\]
which gives \(c_1=2-\gamma-\ln2\), \(c_2=1-\gamma\), and
\(c_4=\tfrac32-\gamma-\ln2\).  The Porter--Thomas factor
\(q=\mathbb E[x^2]=1+2/\beta\) takes the values \(3,2,3/2\), and the exact
identity \(q=3+\kappa_4(c)/\mathbb E[c^2]^2\) follows from
\(\mathbb E[c^4]=3\mathbb E[c^2]^2+\kappa_4(c)\).  For GOE-type ensembles the
finite-size value is \(q=3D/(D+2)\) (eigenvectors uniform on the sphere), and
\(\mathbb E[x\ln x]=\psi(3/2)-\psi(D/2+1)+\ln D=2-\gamma-\ln2+O(1/D)\).

\begin{figure}[htbp]
    \centering
    \includegraphics[width=0.48\textwidth]{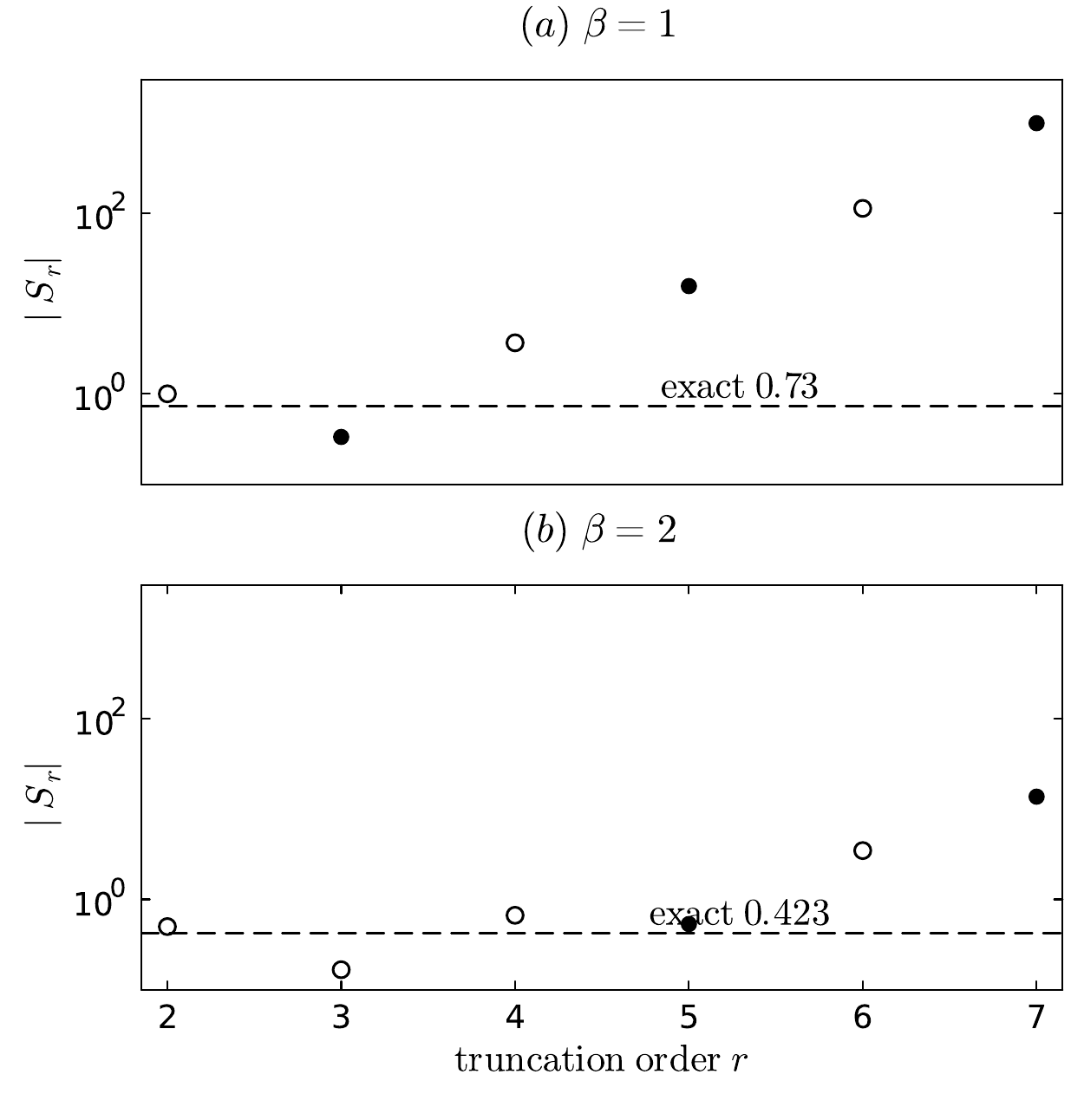}
    \caption{Partial sums \(S_r=\sum_{s=2}^{r}\Delta S_s\) of the cumulant
    expansion \eqref{dS_series} for Porter--Thomas intensities with Dyson
    index \(\beta=1\) (a) and \(\beta=2\) (b), on a logarithmic scale; filled
    (open) symbols denote positive (negative) partial sums.  The dashed lines
    mark the exact corrections \(|c_1|=0.730\) and \(|c_2|=0.423\).  The
    series is asymptotic: the partial sums diverge factorially and no finite
    truncation reproduces the exact value.}
    \label{FIGfluct}
\end{figure}

The cumulants of the Porter--Thomas intensity are
\[
    \kappa_r(x) = (r-1)!\,\alpha\theta^r = (r-1)!\,2^{r-1}\beta^{1-r},
    (\alpha=\beta/2,\ \theta=2/\beta),
\]
giving for \(\beta=1\): \(\kappa_2=2,\kappa_3=8,\kappa_4=48,\ldots\).
The corresponding entropy terms \(\Delta S_2=-1\), \(\Delta S_3=+4/3\),
\(\Delta S_4=-4\) grow factorially, and the partial sums
\(S_r=\sum_{s=2}^r\Delta S_s\) never approach \(-c_1\approx-0.730\)
(Fig.~\ref{FIGfluct}): the series is asymptotic, and no finite truncation
reproduces the exact value.  The closed form \eqref{xlnx_exact}, by contrast,
is the resummation of all orders at once---the Gaussian closure supplies the
complete intensity distribution, not merely its variance.

\subsection{Vertex hierarchy and non-Gaussian corrections}

Beyond the Gaussian-amplitude limit the fluctuation data are organized by the
irreducible vertex hierarchy \(\Gamma^{(2)},\Gamma^{(3)},\ldots\) of the
covariance sector \cite{HOETH}.  The leading vertex \(\Gamma^{(2)}\) controls
\(q-1\), and the higher vertices control the higher entropy terms; for real
amplitudes, e.g.,
\[
    \frac{\kappa_3(p_n)}{(p^{\rm env}_n)^3}
    = \frac{\kappa_6(c_n)}{(p^{\rm env}_n)^3}
      + 12\,\frac{\kappa_4(c_n)}{(p^{\rm env}_n)^2} + 8,
\]
where the constant \(8\) is the Gaussian value and the cumulant terms are the
non-Gaussian corrections, so that \(\Delta S_3\) probes six-point amplitude
correlations invisible at the second-order level.  In the decorrelated regime
\(\kappa_3,\kappa_4\to0\) and the Porter--Thomas values are recovered; in
structured systems \(\kappa_4\) can be large---Ref.~\cite{HOETH} reports
\(q\approx139\) with kurtosis \(123\) in a sparse non-ergodic model at
\(D=1024\)---and the entropy deficit grows beyond \(c_1\).  For Gamma-type
intensity statistics with prescribed \(q\),
\(x\sim\mathrm{Gamma}(1/(q-1),q-1)\), the exact correction is
\begin{equation}\label{eq:gamma}
    \Delta S = -\psi\!\left(\frac{q}{q-1}\right)-\ln(q-1)
    \sim -\ln(q-1)+\gamma \qquad (q\gg1),
\end{equation}
while the second-order truncation \(-(q-1)/2\) diverges linearly: at
\(q=139\) the two estimates are \(-4.36\) and \(-69\), respectively.  The
cumulant expansion therefore breaks down exactly in the regime where
non-Gaussianity is most pronounced, and the closed form \eqref{xlnx_exact}---
or its Gamma generalization above---should be used there.

\subsection{Numerical verification}
\label{fluct_num}
\begin{table*}[htbp]
    \centering
    \caption{Small-scale verification of the entropy-fluctuation identities.
    MC: Monte Carlo with \(n=3\times10^7\) samples; GOE: \(D=512\),
    16 realizations; LG: synthetic envelope at the model's parameters
    (\(\Delta-\Delta'=1.8\), \(\sigma=2.0\), \(\chi=0.75\), Gaussian DOS of
    width \(\sqrt{N}/(2\kappa)\) with \(N=15\)) on \(2^{14}\) support states,
    200 realizations.}
    \label{tabfluct}
    \begin{tabular}{c c c}
    \toprule
    quantity & measured & prediction \\
    \midrule
    \(\mathbb E[x\ln x]\), \(\beta=1\) & 0.7297 & \(2-\gamma-\ln2=0.7296\) \\[2pt]
    \(\mathbb E[x\ln x]\), \(\beta=2\) & 0.4222 & \(1-\gamma=0.4228\) \\[2pt]
    \(\mathbb E[x\ln x]\), \(\beta=4\) & 0.2298 & \(\tfrac32-\gamma-\ln2=0.2296\) \\[2pt]
    \(q\), GOE & 2.975 & \(3D/(D+2)=2.988\) \\[2pt]
    \(\mathbb E[x\ln x]\), GOE & 0.724 & \(\psi(\tfrac32)-\psi(\tfrac D2+1)+\ln D=0.728\) \\[2pt]
    \(\Delta S\), LG + Porter--Thomas & \(-0.731\pm0.011\) & \(-0.7296\) \\[2pt]
    \(\Delta S^{(2)}\), LG + Porter--Thomas & \(-0.999\) & \(-(q-1)/2=-1\) \\[2pt]
    \(\mathrm{Var}(S_{\rm vN})\), synthetic LG + Porter--Thomas & \(0.153^2\) & ridge: \(0.164^2\) \\[2pt]
    \(\mathbb E[x\ln x]\), Gamma \(q=5\) & 1.158 & \(\psi(5/4)+\ln4=1.159\) \\
    \bottomrule
    \end{tabular}
\end{table*}

\begin{figure}[htbp]
    \centering
    \includegraphics[width=0.48\textwidth]{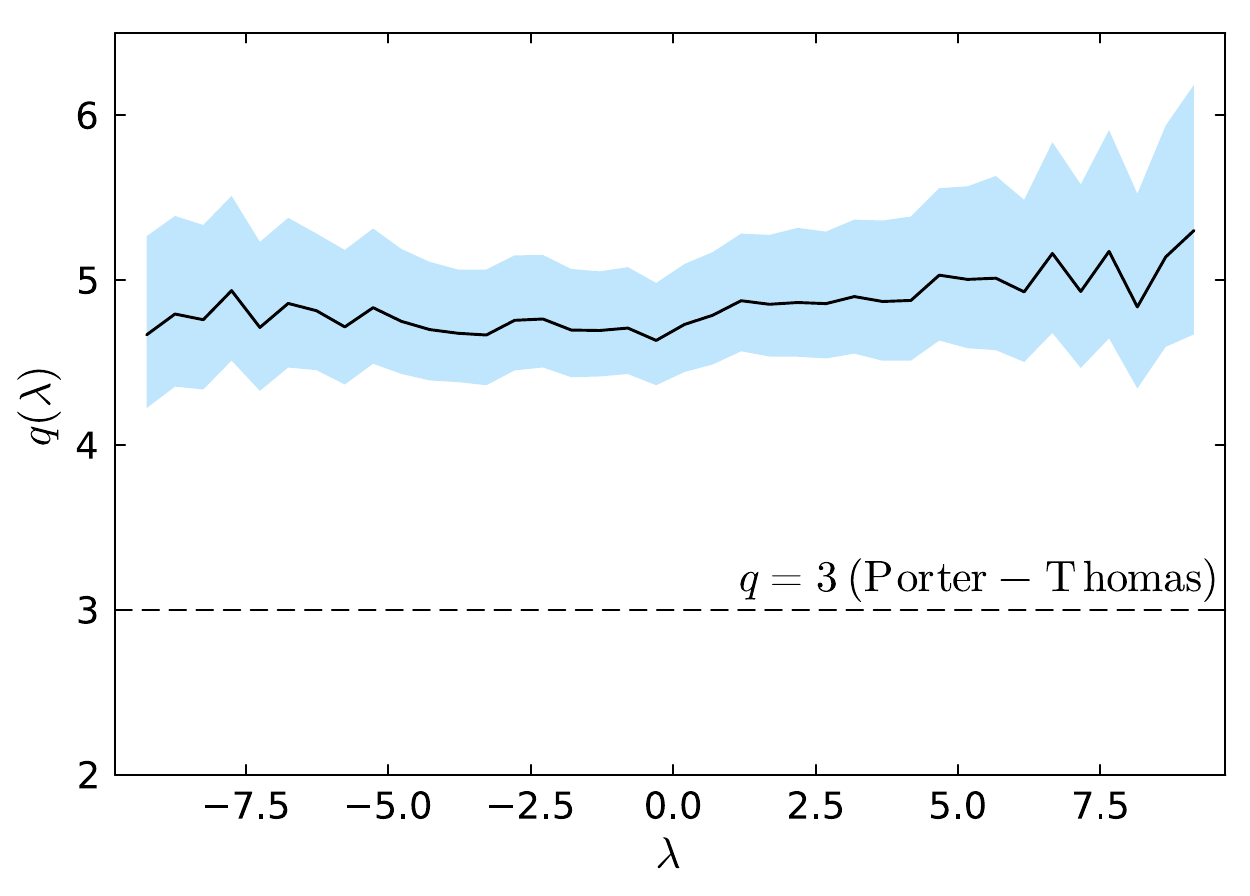}
    \caption{Energy-resolved Porter--Thomas factor \(q(\lambda)\) for
    \(N=15\) (median and 25\%--75\% range over bulk channels, shells of width
    \(0.5\)).  The flat profile \(q(\lambda)\approx4.8\) lies between the
    Gaussian value \(q=3\) (dashed) and the strongly structured models of
    Ref.~\cite{HOETH} (\(q\sim10^2\)): the intra-shell amplitudes are mildly
    non-Gaussian, with \(\kappa_4(c)/\mathbb E[c^2]^2\approx1.8\).}
    \label{FIGqlambda}
\end{figure}

\begin{figure}[htbp]
    \centering
    \includegraphics[width=0.48\textwidth]{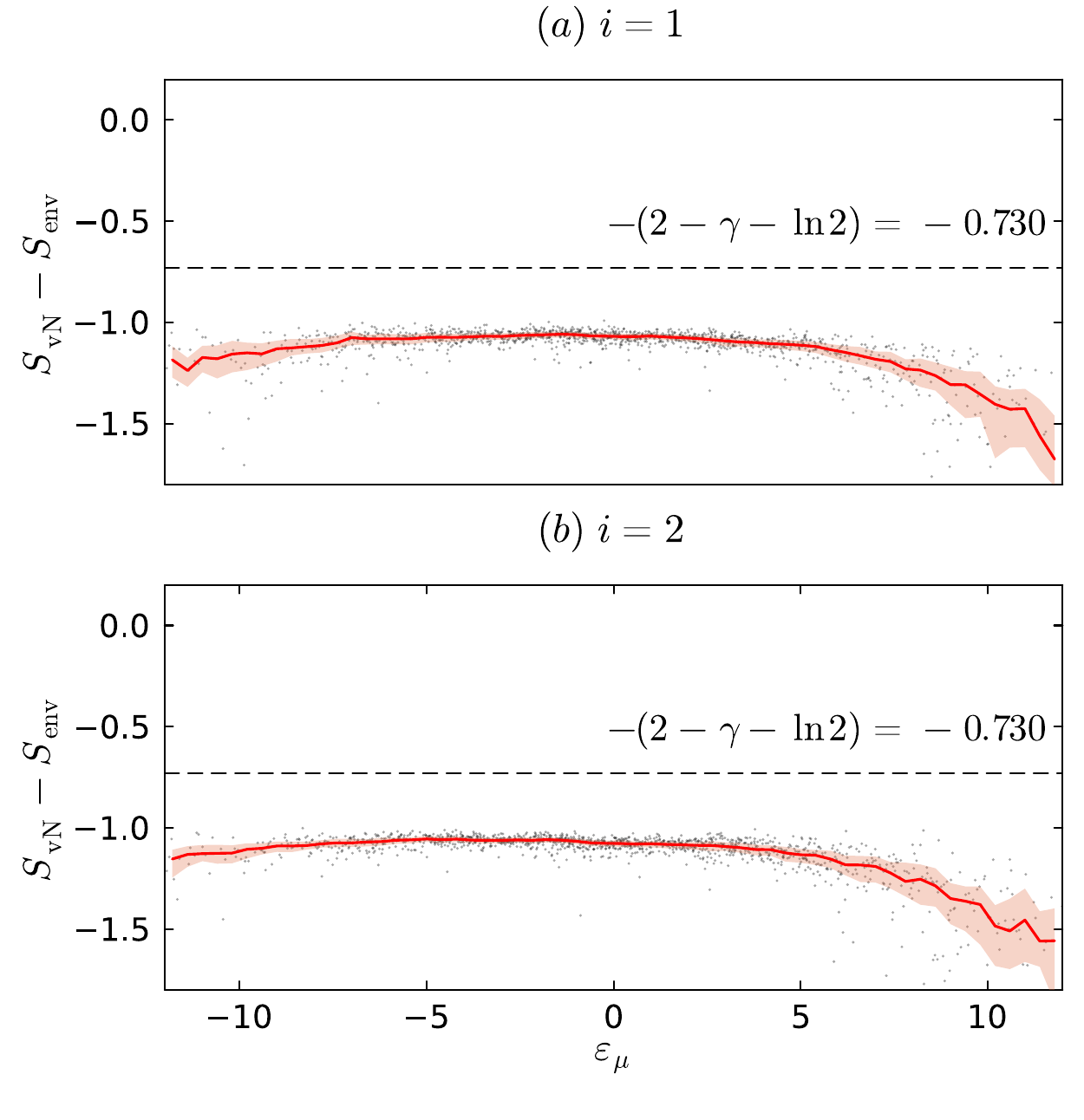}
    \caption{Entropy correction \(S_{\rm vN}-S_{\rm env}\) versus bath energy
    \(\epsilon_\mu\) for \(i=1\) (a) and \(i=2\) (b); scatter: individual
    channels, line and band: binned median and quartiles.  The measured
    plateau \(\approx-1.08\) lies below the Gaussian-limit benchmark
    \(-(2-\gamma-\ln2)=-0.730\) (dashed), and the second-order truncation
    \(-(q-1)/2\approx-1.9\) overshoots the measured value by \(\sim80\%\).}
    \label{FIGgap}
\end{figure}

\begin{figure}[htbp]
    \centering
    \includegraphics[width=0.48\textwidth]{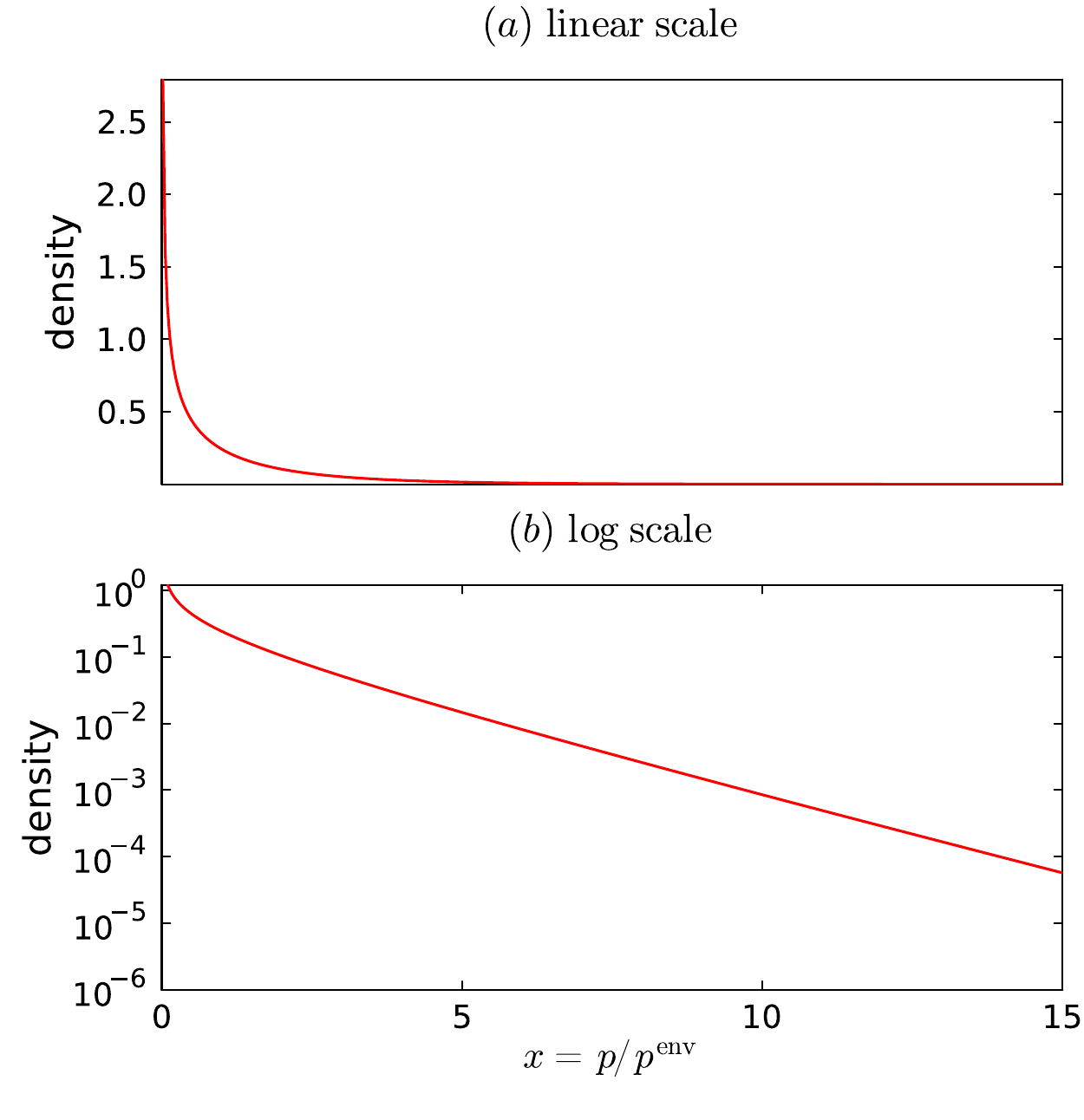}
    \caption{Distribution of the normalized intensity \(x=p/p^{\rm env}\) within
    the envelope-peak shells of the \(\epsilon_\mu=6.324\) channel (\(i=1\)),
    linear (a) and logarithmic (b) scales.  The measured distribution deviates
    from the Porter--Thomas \(\chi^2_1\) law (solid curve) by a pronounced
    heavy tail---the origin of \(q\approx4.8\) and of the failure of the
    second-order truncation.}
    \label{FIGpt}
\end{figure}

All identities of this appendix were verified both by small-scale Monte Carlo
and exact-diagonalization (ED) computations (Julia, seed-averaged; no fit
parameters).  The small-scale synthetic benchmarks are summarized in
Table~\ref{tabfluct}, and the ED results on the interacting chain of
Sec.~\ref{SSCS} in Table~\ref{tabfluctED} and Figs.~\ref{FIGqlambda},
\ref{FIGgap}, \ref{FIGpt}.

\begin{table*}[htbp]
    \centering
    \caption{Bulk-channel diagnostics of the fluctuation sector on the
    interacting chain (median and 25\%--75\% range over channels whose
    envelope peaks in the interior of the density of states; 25\,240 of
    32\,768 channels at \(N=15\), 12\,303 of 16\,384 at \(N=14\)).
    \(q\): energy-resolved Porter--Thomas factor at the envelope peak;
    \(S_{\rm env}\): entropy of the exact shell-conditional envelope
    (Eq.~\eqref{dS_exact}); \(\Delta S=S_{\rm vN}-S_{\rm env}\);
    \(\Delta S^{(2)}\): second-order truncation \eqref{dS2};
    \(S_2^{\rm env}\): R\'enyi-2 entropy of the envelope.  The predictions
    are \(q=3\), \(\Delta S=-(2-\gamma-\ln2)=-0.730\),
    \(\Delta S^{(2)}=-(q-1)/2\) with the measured \(q\), and
    \(S_2-S_2^{\rm env}=-\ln q\).}
    \label{tabfluctED}
    \begin{tabular}{c c c c c}
    \toprule
    & \multicolumn{2}{c}{\(N=15\)} & \multicolumn{2}{c}{\(N=14\)} \\
    \cmidrule{2-3} \cmidrule{4-5}
    quantity & measured & prediction & measured & prediction \\
    \midrule
    \(q\) & \(4.80\ [4.46,5.26]\) & \(3\) & \(5.04\ [4.51,5.86]\) & \(3\) \\[2pt]
    \(\Delta S\) & \(-1.08\ [-1.10,-1.06]\) & \(-0.730\) & \(-1.10\ [-1.15,-1.07]\) & \(-0.730\) \\[2pt]
    \(\Delta S^{(2)}\) & \(-1.95\ [-2.09,-1.86]\) & \(-1.90\) & \(-2.07\ [-2.33,-1.92]\) & \(-2.02\) \\[2pt]
    \(S_2-S_2^{\rm env}\) & \(-1.59\ [-1.66,-1.54]\) & \(-1.57\) & \(-1.64\ [-1.76,-1.57]\) & \(-1.62\) \\[2pt]
    ridge \(\sqrt{\mathrm{Var}(S_{\rm vN})}\) & \(0.185\) & --- & \(0.247\) & --- \\
    \bottomrule
    \end{tabular}
\end{table*}

The exact decomposition Eq.~\eqref{dS_exact} is confirmed by the LG row:
averaging the entropy difference over 200 Porter--Thomas realizations of the
synthetic envelope gives \(-0.731\pm0.011\) against the Gaussian benchmark
\(-(2-\gamma-\ln2)=-0.7296\), while the second-order truncation gives
\(-0.999\approx-(q-1)/2\), confirming the \(37\%\) overshoot discussed in
Sec.~\ref{IntraShell}.  The measured entropy fluctuation
\(\mathrm{Var}(S_{\rm vN})=0.153^2\) matches the ridge estimate
\(0.164^2\) computed from \((q-1)\sum_n(p^{\rm env}_n)^2(1+\ln p^{\rm env}_n)^2\), and both
scale as \(e^{-S(\lambda_p)}\sim4\times10^{-4}\) times an \(O(S^2)\) factor,
as expected for a self-averaging correction when the envelope entropy is
extensive.  The GOE runs verify the
fluctuation sector of Ref.~\cite{HOETH} at the level relevant here:
\(q=2.975\) against the finite-size value \(3D/(D+2)=2.988\), and
\(\mathbb E[x\ln x]=0.724\) against the finite-size prediction \(0.728\)
(\(D\to\infty\): \(0.7296\)); the small discrepancy is within the
finite-sample fluctuations of the 16 GOE realizations.  Finally, the Gamma-type example with \(q=5\)
confirms the exact correction \(-\psi(q/(q-1))-\ln(q-1)=-1.159\) against the
second-order \(-2\), and the \(q=139\) case quoted in the text evaluates to
\(-4.362\) versus \(-69\) and the logarithmic asymptote \(-4.350\).

\emph{Exact-diagonalization verification.} The full program was also run on
the interacting chain of Sec.~\ref{SSCS} (\(N=15\) spins, \(g=1.05\),
\(h=0.1\), with the system spin closing the ring,
\(V=-\sigma_z^S(\sigma_z^1+\sigma_z^{14})\)).  The complete overlap matrix
\(p_n^{\mu i}=|\braket{\psi_n|\phi_{\mu i}}|^2\) over all \(2\times2^{14}\)
channels was obtained from the stored eigenpairs of the closed \(15\)-spin
chain and of the \(14\)-spin open-chain bath by two blocked matrix
multiplications, accumulating the shell statistics channel by channel (peak
memory \(\sim20\)\,GB).  The pipeline was validated by reproducing the
Lorentzian parameters of Fig.~\ref{FIGs1} to three digits
(\(\chi=0.754\), \(a+\Delta=4.661\) for the \(\epsilon_\mu=6.324\) channel;
\(\chi=0.609\), \(a+\Delta=-1.005\) for \(\epsilon_\mu=0.583\)), and by the
normalization \(\sum_n p_n^{\mu i}=1\) to \(10^{-12}\).  The same statistics
were collected for the \(N=14\) chain, whose \(13\)-spin bath was
diagonalized on the fly.  The results are summarized in
Table~\ref{tabfluctED}.

\begin{table*}[htbp]
    \centering
    \caption{Finite-size trend of the bulk fluctuation diagnostics on the
    interacting chain, \(N=12\)--\(15\) (median and 25\%--75\% range over bulk
    channels; \(N=14,15\) from Table~\ref{tabfluctED}, \(N=12,13\) from the
    same pipeline with the chain Hamiltonians diagonalized on the fly).
    \(q\): Porter--Thomas factor at the envelope peak; \(\Delta S=S_{\rm vN}-
    S_{\rm env}\); \(\Delta S^{(2)}\): second-order truncation (predicted
    \(-(q-1)/2\) in parentheses); \(\Delta S_\Gamma\): Gamma diagnostic
    closure \eqref{eq:gamma} at the measured \(q\); \(S_2-S_2^{\rm env}\)
    \(S_2-S_2^{\rm env}\)
    versus \(-\ln q\) in parentheses.  Both \(q(N)-3\) and
    \(\Delta S(N)+c_1\) decrease monotonically toward the Gaussian values
    \(0\) and \(0\); a simple exponential fit gives \(2^{-0.18N}\) and
    \(2^{-0.14N}\), but the four sizes cannot distinguish exponential,
    algebraic, or \(1/N\)-type finite-size corrections, and the fit is not
    intended as a determination of the asymptotic scaling.
    \(D_{\rm KS}\): Kolmogorov--Smirnov distance between the peak-shell
    \(P(x|\lambda)\) of a representative channel (bath energy nearest to
    zero, \(i=1\)) and the Porter--Thomas distribution.}
    \label{tabfluctN}
    \begin{tabular}{c c c c c c c}
    \toprule
    \(N\) & \(q\) & \(\Delta S\) & \(\Delta S^{(2)}\) & \(\Delta S_\Gamma\) & \(S_2-S_2^{\rm env}\) & \(D_{\rm KS}\) \\
    \midrule
    12 & \(5.63\ [4.59,7.44]\) & \(-1.19\ [-1.30,-1.12]\) & \(-2.35\ (-2.31)\) & \(-1.26\) & \(-1.76\ (-1.73)\) & \(0.114\) \\[2pt]
    13 & \(5.34\ [4.58,6.69]\) & \(-1.15\ [-1.22,-1.10]\) & \(-2.23\ (-2.17)\) & \(-1.22\) & \(-1.71\ (-1.68)\) & \(0.147\) \\[2pt]
    14 & \(5.04\ [4.51,5.86]\) & \(-1.10\ [-1.15,-1.07]\) & \(-2.07\ (-2.02)\) & \(-1.17\) & \(-1.64\ (-1.62)\) & \(0.113\) \\[2pt]
    15 & \(4.80\ [4.46,5.26]\) & \(-1.08\ [-1.10,-1.06]\) & \(-1.95\ (-1.90)\) & \(-1.12\) & \(-1.59\ (-1.57)\) & \(0.124\) \\
    \bottomrule
    \end{tabular}
\end{table*}

Across the four sizes the Gamma closure tracks the measured correction to
\(4\%\)--\(9\%\) and the R\'enyi-2 relation \(S_2=S_2^{\rm env}-\ln q\) holds
to \(2\%\) at every \(N\), so the second moment \(q\) provides an effective
low-dimensional diagnostic for the entropy correction within the Gamma
family---without uniquely determining the full intensity distribution, as the
Kolmogorov--Smirnov distance \(D_{\rm KS}\approx0.11\)--\(0.15\) shows, and the four values of \(D_{\rm KS}\) (\(0.114\), \(0.147\), \(0.113\),
\(0.124\)) exhibit no clear monotonic decrease: full-distribution
Gaussianization is not yet numerically established.  The monotonic drift of
\(q(N)\) and \(\Delta S(N)\) toward their Gaussian values provides evidence
for a finite-size flow of the low-order diagnostics toward Gaussian
amplitude statistics, whose slowness connects the entropy-side diagnostics
to the fluctuation-sector scaling of Ref.~\cite{HOETH}.

Three conclusions follow.  First, the exact decomposition
Eq.~\eqref{dS_exact} is confirmed on real data: the measured
\(S_{\rm vN}-S_{\rm env}\) equals the directly measured \(-\mathbb E[x\ln x]\) to
numerical precision, and the R\'enyi-2 shift \(-\ln q\) is reproduced by
\(S_2-S_2^{\rm env}\) at both system sizes.  Second, the interacting chain
sits in the mildly non-Gaussian regime \(q\approx4.8\): its intra-shell
intensities carry a heavy tail (Fig.~\ref{FIGpt}) with
\(\kappa_4(c)/\mathbb E[c^2]^2\approx1.8\), so the Gaussian benchmark
\(-(2-\gamma-\ln2)\approx-0.73\) is \emph{not} attained and the measured
correction is \(-1.08\); as a one-parameter \emph{diagnostic} closure
constrained to reproduce the measured \(q\), the Gamma model of
Eq.~\eqref{eq:gamma} gives
\(-\psi(q/(q-1))-\ln(q-1)=-1.12\), within \(4\%\) of the data.  Third, the
measured cumulants \(\kappa_2\approx4.0\), \(\kappa_3\approx46\),
\(\kappa_4\approx1.2\times10^3\) produce partial sums
\(-2.03,+5.69,-96.4,\ldots\) that diverge rapidly: the asymptotic character
of Eq.~\eqref{dS_series} is quantitative, and the second-order truncation
\(-(q-1)/2\approx-1.9\) overshoots the exact value by \(\sim80\%\)---exactly
the failure mode anticipated by the vertex-hierarchy discussion of
Sec.~\ref{IntraShell}.  This failure is methodological rather than physical:
in the strongly non-Gaussian regime the entropy functional should be
evaluated from the resummed intra-shell distribution \(P(x|\lambda)\)---as in
Eq.~\eqref{xlnx_exact}---rather than from a finite-order cumulant truncation,
which is not a controlled approximation there.

\begin{figure}[htbp]
    \centering
    \includegraphics[width=0.48\textwidth]{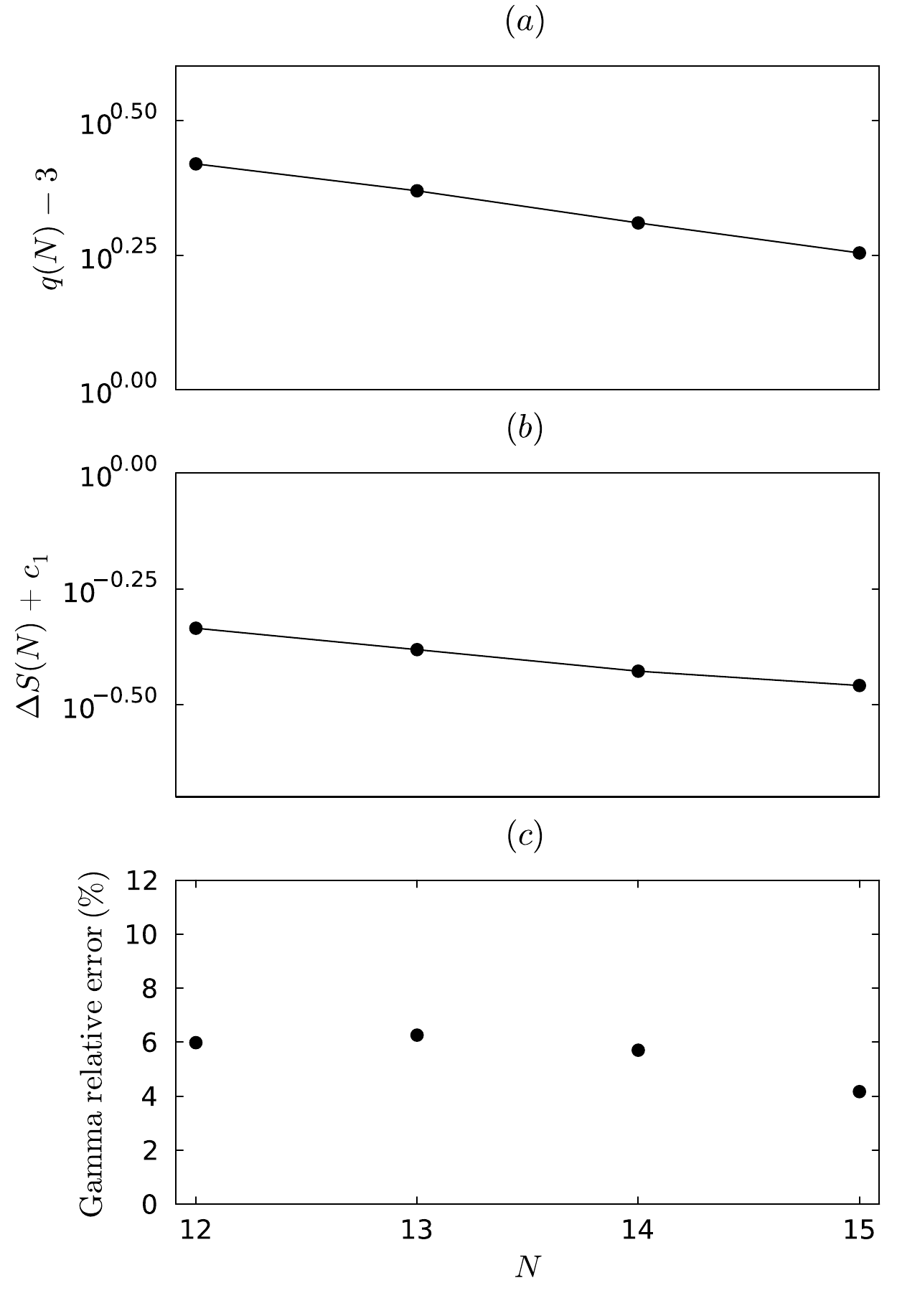}
    \caption{Finite-size flow of the fluctuation diagnostics,
    \(N=12\)--\(15\).  (a) \(q(N)-3\), (b) \(\Delta S(N)+c_1\), both on
    logarithmic scales (the Gaussian-amplitude-limit value \(0\) is approached but not
    attained at the accessible sizes); (c) relative error of the Gamma
    diagnostic closure \eqref{eq:gamma} versus the measured \(\Delta S\).
    The monotonic drift toward the Gaussian-amplitude values is evidence of
    a finite-size flow, but the four sizes do not determine its asymptotic
    form.}
    \label{FIGscaling}
\end{figure}

\begin{figure}[htbp]
    \centering
    \includegraphics[width=0.48\textwidth]{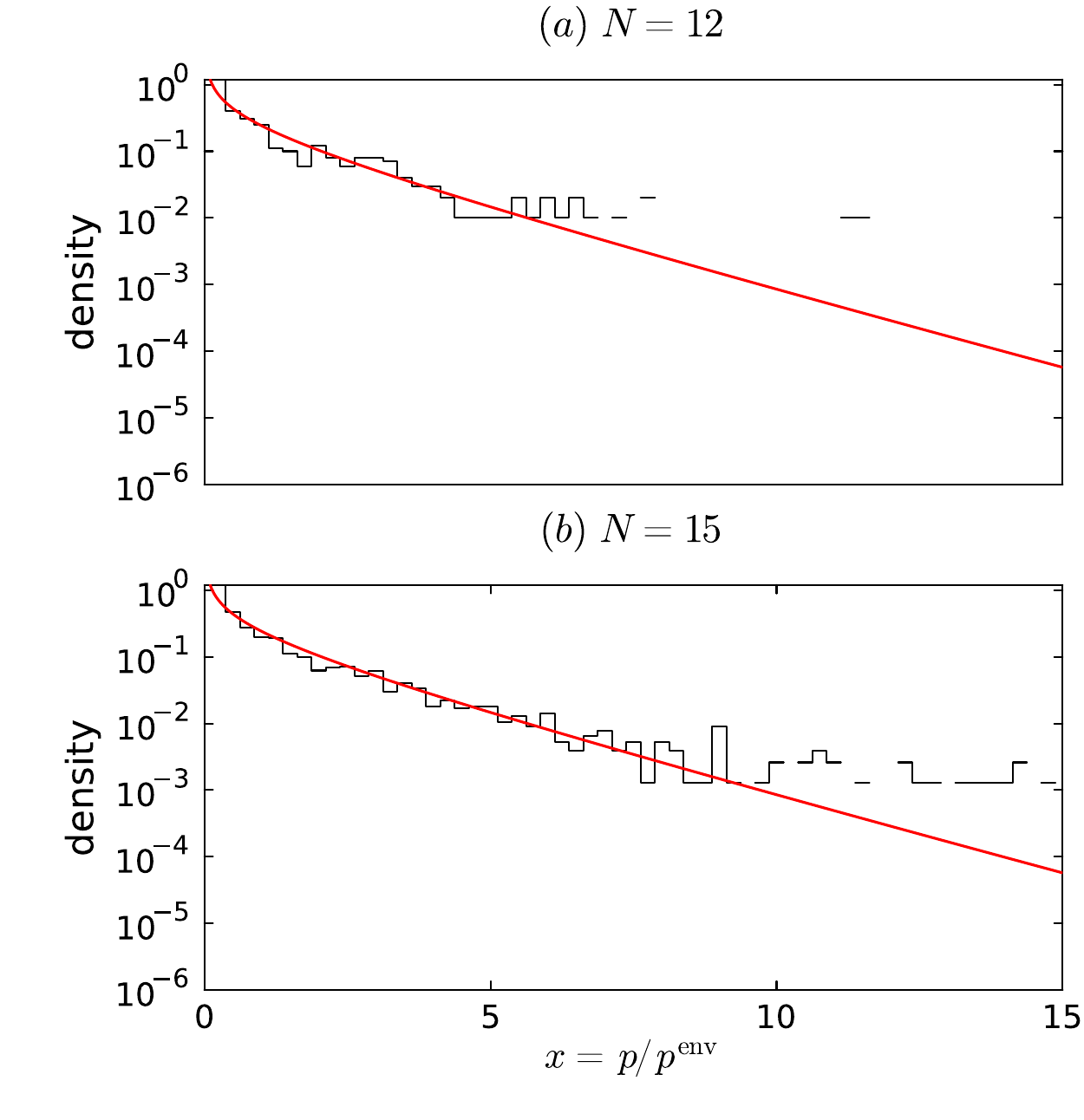}
    \caption{Conditional intensity distribution \(P(x|\lambda)\) within the
    envelope-peak shells of a representative bulk channel (bath energy
    nearest to zero, \(i=1\)) for \(N=12\) (a) and \(N=15\) (b), versus the
    Porter--Thomas \(\chi^2_1\) law (solid curve).  The heavy tail persists
    at both sizes: the drift of the second moment \(q(N)\) toward \(3\) is
    not yet accompanied by convergence of the full distribution, as expected
    from the slow power-law approach.}
    \label{FIGptscaling}
\end{figure}

Fig.~\ref{FIGscaling} displays the monotonic drift of the low-order
diagnostics \(q(N)\) and \(\Delta S(N)\) toward their Gaussian-amplitude
values, while Fig.~\ref{FIGptscaling} shows why this drift must not be read
as Gaussianization of the full distribution: the heavy tail of
\(P(x|\lambda)\) persists at both \(N=12\) and \(N=15\), in line with the
non-monotonic \(D_{\rm KS}\) of Table~\ref{tabfluctN}.  Together the table
and the two figures separate the finite-size drift of the low-order
diagnostics---which the data support---from the convergence of the full
intensity distribution, which the available sizes do not yet establish.

The finite-size benchmark sharpens this statement.  For Haar-typical real
eigenvectors of dimension \(D=2^N\),
\(q_{\rm GOE}(D)=3D/(D+2)=3-6/D+O(D^{-2})\) and
\(\mathbb E[x\ln x]_{\rm GOE}=c_1-1/D+O(D^{-2})\); over \(N=12\)--\(15\)
these finite-dimensional corrections are of order \(10^{-3}\)--\(10^{-4}\),
three to four orders of magnitude below the measured deviations
\(q(N)-3\approx1.8\)--\(2.6\) and \(\mathbb E[x\ln x]-c_1\approx0.35\)--\(0.46\).
The excess intensity fluctuation of the interacting chain is therefore
genuine dynamical (non-Gaussian) structure rather than a finite-dimensional
Haar-sampling artifact, and its slow decay with \(N\) reflects the approach
to amplitude Gaussianization.

\emph{Remark on the support of the overlaps.} The empirical support analysis
underlying Sec.~\ref{IntraShell} is as follows.  Only a fraction
\(\approx1/(2N)\) of the eigenstates (\(4\%\) at \(N=15\), \(8\%\) at
\(N=14\)) has exactly vanishing overlap with a given product
state---precisely the parity-definite eigenstates mismatched with the parity
of \(|\phi_{\mu i}\rangle\), the remainder being members of nearly degenerate
reflection pairs whose stored eigenvectors mix the two parities.  Nor is any
halving of the support recovered in a weaker sense.  In absolute terms about
half of the states do have tiny overlaps (\(p<10^{-8}\) for \(48\%\) and
\(28\%\) of the states of the two reference channels), but they carry only
\(10^{-5}\)--\(10^{-3}\) of the total probability: this is the Gaussian tail
of the envelope itself, already contained in \(S_{\rm env}\).  Relative to the
envelope there is no half-zero structure either: within the peak shells the
states with \(x=p/p^{\rm env}<0.1\) amount to \(36\%\)--\(39\%\) of the spectrum
but carry only \(\sim1\%\) of the weight (the Porter--Thomas law predicts
\(25\%\) and \(1.1\%\)), the lower half of the states (\(x\le\) median)
carries only \(\sim3\%\) of the weight (Porter--Thomas: \(10\%\); the heavy
tail of Fig.~\ref{FIGpt} pushes the weight toward the top percentile, which
carries \(\sim14\%\)), and discarding that lower half and renormalizing
changes \(S_{\rm env}\) by \(\lesssim0.2\), far less than \(\ln2\).  The support of
the overlaps is therefore the full spectrum to an excellent approximation:
the smoothed envelope entropy \(S_{\rm env}\) coincides with the LG estimate of
Eq.~\eqref{LGentro} (within \(0.05\) for the reference channels), the
standard observational entropy (macrostate volumes given by the full shell
dimension, Sec.~\ref{SSCS}) satisfies \(S_{\rm env}\approx S_{\rm obs}\) (within \(0.1\) for the
reference channels), and \(S_{\rm vN}\) lies below both by
\(\approx1.1\)--\(1.2\).


\end{document}